\documentclass[twocolumn]{aa}
\usepackage[switch]{lineno}
\usepackage{lineno}
\usepackage{graphicx}
\usepackage[normalem]{ulem} 
\usepackage{txfonts}
\usepackage{amssymb}
\usepackage{enumerate}
\usepackage{subcaption}
\usepackage{comment}
\usepackage[colorlinks=true,allcolors=blue]{hyperref}%
\usepackage{url}
\newcommand{\degree}{\ensuremath{^\circ}}

\newcommand{\llnr}[1]{{\bf \color{magenta}{[]}} \color{black}} 

\usepackage{xcolor}
\usepackage{soul}
\definecolor{mydarkgreen}{RGB}{0,100,0}

\usepackage{academicons}
\definecolor{orcidlogocol}{HTML}{A6CE39}

\newcommand{\orcid}[1]{\href{https://orcid.org/#1}{\textcolor[HTML]{A6CE39}{\aiOrcid}}}
\begin{document}

 \title{Toroidal modified Miller-Turner CME model in EUHFORIA: II. Validation and comparison with flux rope and spheromak}

 \author{A. Maharana\href{https://orcid.org/0000-0002-4269-056X}{\includegraphics[scale=0.05]{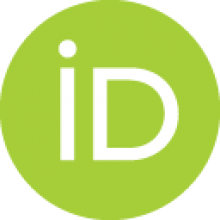}}\inst{1,2} \and L. Linan\href{https://orcid.org/0000-0002-4014-1815}{\includegraphics[scale=0.05]{orcid-ID.png}}\inst{1} \and S. Poedts\href{https://orcid.org/0000-0002-1743-0651}{\includegraphics[scale=0.05]{orcid-ID.png}}\inst{1,3} \and J. Magdaleni\'c \inst{1,2}}

 \institute{Centre for mathematical Plasma-Astrophysics, Department of Mathematics, KU Leuven, Celestijnenlaan 200B, 3001 Leuven, Belgium \\
 \and Solar-Terrestrial Centre of Excellence—SIDC, Royal Observatory of Belgium, 1180 Brussels, Belgium\\
  \and Institute of Physics, University of Maria Curie-Sk{\l}odowska, ul.\ Radziszewskiego 10, 20-031 Lublin, Poland}

 \date{Received: ?; Accepted: ?}

 
 \abstract
 {Rising concerns about the impact of space weather-related disruptions demand modelling and reliable forecasting of coronal mass ejection (CME) impacts.}
 {In this study, we demonstrate the application of the modified Miller-Turner (mMT) model implemented in EUropean Heliospheric FORecasting Information Asset (EUHFORIA), to forecast the geo-effectiveness of observed coronal mass ejection (CME) events in the heliosphere. The goal is to develop a model that not only has a global geometry to improve overall forecasting but is also fast enough for operational space weather forecasting.}
 {We test the original full torus implementation and introduce a new three-fourth Torus version called the Horseshoe CME model. This new model has a more realistic CME geometry, and it overcomes the inaccuracies of the full torus geometry. We constrain the torus geometrical and magnetic field parameters using observed signatures of the CMEs before, during, and after the eruption. We perform EUHFORIA simulations for two validation cases -- the isolated CME event of 12 July 2012 and the CME-CME interaction event of 8-10 September 2014. The assessment of the model's capability to predict the most important $B_z$ component is performed using the advanced Dynamic Time Warping (DTW) technique. }
 {The Horseshoe model prediction of CME arrival time and geo-effectiveness for both validation events compare well to the observations and are weighed to the results obtained with the spheromak and FRi3D models that were already available in EUHFORIA.}
 {The runtime of the Horseshoe model simulations is close to that of the spheromak model, which is suitable for operational space weather forecasting. Yet, the capability of the magnetic field prediction at 1~AU of the Horseshoe model is close to that of the FRi3D model. In addition, we demonstrate that the Horseshoe CME model can be used for simulating successive CMEs in EUHFORIA, overcoming a limitation of the FRi3D model.}


 \maketitle
%
\section{Introduction} \label{sec:Introduction}
The changing physical conditions in the heliosphere induced by the dynamical processes on the Sun and in the interplanetary environment, appearing in the solar wind, magnetosphere, ionosphere, and thermosphere, are called space weather. The main drivers of disturbed space weather conditions are coronal mass ejections \citep[CMEs,][]{Howard2011,Webb2012}, the giant blobs of magnetised plasma erupting from the Sun into the heliosphere. These large-scale structures propagate through the heliosphere, interacting with the solar wind and often causing geomagnetic disturbances on Earth and other planets and spacecraft. 

The interplanetary CME (ICME) is the term used to refer to the CME propagating beyond the corona in the interplanetary medium. ICMEs manifest lucid magnetic field characteristics that can be explained by a flux rope structure. Flux ropes are organised bundles of axially twisted magnetic field lines confining plasma within them \citep{antiochos1999,Torok2005}. Flux ropes are ubiquitous, i.e., in addition to CMEs, they are also associated with magnetic structures in the heliosphere, streamer blow-outs within the heliospheric current sheet, small-scale structures called plasmoids in the heliosphere and in solar flares \citep{Nieves-Chinchilla2023}. The CME flux rope is often referred to as a magnetic cloud (MC) or magnetic ejecta (ME), which is characterised by a strong magnetic field, clear rotations in the magnetic field vector, and a low proton temperature \citep{Burlaga1981,Klein1982,Kilpua2017}.

The majority of the interplanetary CME reconstruction techniques are built on the principle of a force-free magnetic field configuration of the magnetic cloud, i.e., $\nabla \times \mathbf{B} = \alpha \mathbf{B}$, so that $\mathbf{J}$ is parallel to $\mathbf{B}$ and $\mathbf{J}\times\mathbf{B}$ vanishes ($\mathbf{B}$ and $\mathbf{J}$ are the magnetic field and current density respectively, and $\alpha$ is the force-free constant). This condition implies that the electric current is parallel to the magnetic field only in static conditions. This criterion was incorporated into the first intuition of cylindrically symmetrical solution and a constant $\alpha$ across the cloud \citep{Lundquist1951}. Another generally assumed characteristic of CME propagation is a self-similar expansion \citep{Poomvises2010,Davies2012,Subramanian2014} in the initial phase of their evolution. However, incorporating force-free and self-similarity characteristics together was found to be not straightforward. The cylindrical configuration did not maintain the force-free state once it started expanding. Some models changed the geometrical cross-section of the cylinder from circular to elliptical \citep[][]{Hidalgo2002b,Hidalgo2003}, or kinematically distorted the cylindrical flux rope geometry during propagation to conserve the force-free assumption during self-similar expansion. \citep[][]{Owens2006,Vandas2003,Vandas2006}. Others proposed non-cylindrical flux rope models, e.g.\ with toroidal geometry \citep[][]{Romashets2003b,Marubashi2007}. The orientation of CMEs fitted with torus geometry matches the in situ observations better than the cylindrical geometry and is more consistent with their corresponding source region orientations \citep{Marubashi2009}. \\

The static CME model must be self-consistently evolved in a realistic solar wind, and its interaction with the large-scale structures in the solar wind and with other transients must be modelled accurately for reliable space weather forecasting. When designing a numerical model of the CME to be propagated in the heliosphere, the model should have the following requirements. It should be consistent with the definition of flux rope emerging from the solar corona and, upon evolving in the heliosphere, it should be able to reproduce the characteristics of the flux ropes observed in the heliosphere. CME evolution and propagation models like AWSoM \citep{vanderholst2014} or CORHEL-CME \citep{Linker2023} self-consistently erupt the CME from the corona and evolve them to Earth and beyond. However, such modelling is computationally expensive as it covers multiple physical scales and processes. Models like ENLIL \citep{Odstrcil2003}, EUHFORIA \citep{Pomoell2018}, SUSANOO-CME \citep{Shiota2016}, MS-FLUKSS \citep{Singh2018} etc., start evolving the CMEs from 0.1 au. Although such an initialisation restricts the modelling of the scales and physical processes close to the Sun, it is still efficient in understanding the heliospheric processes involved in CME propagation. In this work, we use the solar wind and CME evolution model EUropean Heliospheric FORecasting Information Asset (EUHFORIA) model, which has functional magnetised FR models. The linear force-free (LFF) spheromak model has shown potential in fitting asymmetric magnetic field observations in situ \citep{Vandas1991,Vandas1993a,Vandas1993b}. Although the spheromak model can fit the central part of the magnetic cloud as well as any flux rope (cylindrical or toroidal models), the ``edges" of the cloud cannot be modelled by it \citep{Farrugia1995}. Due to the compact spherical shape and lack of leg-like structures connecting the CME to the surface of the Sun, it is incapable of modelling the flank encounters, i.e., CMEs where the Earth (or a satellite) is impacted by one of these legs. Moreover, the spheromak model tilts in the heliosphere to align its symmetry axis with the ambient field to an extent not actually observed in the heliosphere, and hence, leads to erroneous predictions at Earth \citep{Asvestari2022,Sarkar2024}. To overcome the drawbacks of the spheromak model, the Flux Rope in 3D model \citep[FRi3D; ][]{Isavnin2016} was implemented in EUHFORIA \citep{Maharana2022}. FRi3D has an extended flux-rope geometry with flexible variable cross-sections that incorporate deformations like pancaking, flattening, and skewing. This model significantly improved the prediction of CMEs at 1~au compared to spheromak CME predictions \citep{Maharana2022,Maharana2023,Palmerio2023}. However, the numerical implementation of the deformations and the permanent connection of the CME legs to the Sun gives rise to a higher expenditure of numerical resources. In addition, keeping the legs of one CME attached to the Sun makes it complicated for the numerical injection of the following CME when the legs overlap. Hence, modelling successive CME evolution with the FRi3D model is presently challenging in the framework of EUHFORIA. Therefore, toroidal CME models were developed by \citet{linan2024}, using the analytical magnetic field configurations of the modified Miller-Turner (mMT) and the Soloviev solution of the Grad-Shafranov equation. The aim of that work was to simplify the FRi3D geometry while still overcoming the limitations of the spheromak geometry and shortening the wall-clock time of the simulations.  \\



In this study, building on the work by \citep{linan2024}, we focus on validating the modified Miller-Turner (mMT) toroidal CME model and suggest the changes to upgrade it into the ``Horseshoe model''. Due to the multiple non-trivial geometrical parameters of the Soloviev solution, constraining them from observations is not straightforward. Hence, the validation of the Soloviev toroidal CME model will be carried out in a future publication. Section~\ref{sec:Models} introduces the Horseshoe model and its implementation in EUHFORIA. In Section~\ref{sec:constrain_params}, we explain the methodologies to constrain the geometrical and magnetic field parameters of the Horseshoe model from observations. The details of the validation events are provided in Section~\ref{sec:validation_events}. In Section~\ref{sec:results_discussion}, we assess the performance of the Horseshoe model and compare it to the other magnetised CME models in EUHFORIA. We summarise the work and discuss the outlook in Section~\ref{sec:conclusion}.

\section{Models} \label{sec:Models}
In this section, we introduce the Horseshoe model, EUHFORIA, and the numerical implementation of the Horseshoe model in EUHFORIA. The comparison of the full torus \citep{linan2024} and the Horseshoe implementation of the mMT model is presented.

\subsection{Horseshoe CME model} \label{subsec:CME_Model}
The Horseshoe CME model is a modification of the toroidal CME model introduced by \citet{linan2024} in EUHFORIA. The geometry of the Horseshoe model is a full torus devoid of the rear part of the torus, hence resembling a ``Horseshoe". The rationale behind the new implementation is two-fold: (1)~To resemble a realistic flux rope structure with legs, and (2)~to avoid reproducing inaccurate magnetic field configuration in the heliosphere because a real CME does not have the geometry of a full torus. The magnetic field topology, the same as in \citet{linan2024}, is defined using the the modified Miller-Turner model \citep[mMT,][]{Romashets2003b}:
\begin{eqnarray} 
B_{\rho_{l}}&=& B_{0}\frac{R_{0}-2\rho_{l}\cos\theta_{l}}{2\alpha R_{0}(R_{0}+\rho_{l}\cos\theta_{l})}J_{0}(\alpha\rho_{l})\sin\theta_{l} \,,\label{eq:modMT_1} \\
B_{\phi_{l}}&=& B_{0}\left(1-\frac{\rho_{l}}{2 R_{0}}\cos\theta_{l}\right)J_{0}(\alpha\rho_{l}) \,,\label{eq:modMT_2} \\
B_{\theta_{l}}&=& B_{0}\frac{R_{0}-2\rho_{l}\cos\theta_{l}}{2\alpha R_{0}(R_{0}+\rho_{l}\cos\theta_{l})}J_{0}(\alpha\rho_{l})\cos\theta_{l} \nonumber\\ 
&&-B_{0}\left(1-\frac{\rho_{l}}{2 R_{0}}\cos\theta_{l}\right)J_{1}(\alpha\rho_{l}) \,.\label{eq:modMT_3}
\end{eqnarray}
\noindent where ($\rho_l,\phi_l,\theta_l$) are the local toroidally curved cylindrical coordinates \citep[for more information, see Fig.~5 in][]{linan2024}. Here, $R_0$ and $a$ are the major and minor radii of the torus, respectively, while $B_0$ is the axial magnetic field strength. In the local coordinate system, $\rho_l$ extends from [0,~$a$], and $\phi_l$ and $\theta_l$ both cover the [0,~2$\pi$] interval. $J_m$ is the Bessel function of m-th order and $\alpha$ is the force-free constant. The advantage of this magnetic field topology is its fully analytical form of the force-free magnetic field ($\nabla\times \mathbf{B} = \alpha \mathbf{B}$) inside the torus. The magnetic field topology is axisymmetric about $\phi_l$. To confine the magnetic field inside the torus with a circular cross-section, the magnetic field must be completely poloidal at the torus boundary. $B_{\rho_{l}}$ and $B_{\phi_{l}}$ are maximal at the centre of the torus cross-section and monotonically drop to zero when $\alpha \rho_{l} \approx 2.405$, i.e., the first zero of $J_0$. The parameter $\alpha$ is related to the flux rope minor radius and chirality ($C$) by:
\begin{equation}
    \alpha = C~ \frac{2.405}{a}.
\end{equation}
The twist varies from zero at the centre to infinity at the outer boundary of the torus. The mMT configuration is a generalised version of the Miller-Turner topology \citep[MT,][]{Miller1981}, as it satisfies the exact solenoidality ($\nabla\cdot \mathbf{B} = 0$) for all aspect ratios ($R_0/a$) of the torus, while MT is solenoidal only for higher aspect ratios ($>10$), and coincides asymptotically with the mMT and \citet{Lundquist1951} solutions in that regime. For low aspect ratios ($<3$), the magnetic field profiles for the mMT model match the numerical solutions better. Both MT and mMT solutions are approximately force-free for smaller aspect ratios, but the MT solution performs better than the mMT \citep{Vandas2015}. 
The interplanetary flux ropes fitted at 1~au with toroidal CME models point to an aspect ratio greater than 3 \citep{Marubashi2015}. 
However, the flux ropes close to the Sun or the locally distorted loop-like flux ropes \citep{Vandas2002} can have aspect ratios less than $2$ \citep{Romashets2003a}. 
The CMEs initialised at 0.1~au have a low aspect ratio, and hence, ideally, we can use both the MT and mMT solutions in our numerical modelling. Reaffirming our claim, \citet{linan2024} demonstrated the numerical stability of using tori (with the mMT configuration) of low aspect ratios (between 1 and 3) in the framework of EUHFORIA.

Previous studies \citep{Farrugia1995,Mulligan2001,Hidalgo2002a,Mostl2009b} suggest that the force-free assumption is mostly not valid to fully explain the pressure gradients in the core and boundary of the interplanetary CMEs. Similarly, \citet{Isavnin2016} does not consider the non-force-free nature of the FRi3D model as a disadvantage. Although mMT is not force-free for all aspect ratios, as it satisfies the divergence-free condition better, we consider adopting it in our study. \citet{Romashets2003b} suggest the validity of the mMT solution for locally treating a part of an extended flux rope rooted at the Sun as a toroid (a torus is a toroid with a circular cross-section). Hence, even if we deviate from the full torus geometry in the novel horseshoe model, the magnetic field topology of mMT is still applicable. 

\subsection{EUHFORIA}
EUHFORIA is a physics-based, data-driven magnetohydrodynamic (MHD) model of solar wind and CME evolution used for both space weather research and forecasting. It has a modular design enabling the combination of different coronal and heliospheric models. The coronal model extends from the photosphere to the low corona up to 0.1~au and provides the boundary condition for the heliospheric domain. The heliospheric domain normally covers the region between 0.1~au and 2~au (expandable) and solves the MHD equations on the 3D grid. We use the semi-empirical, modified Wang–Sheeley–Arge model \citep[WSA, ][]{Arge2004,Pomoell2018} as the coronal model. The WSA model incorporates the Potential Field Source Surface \citep[PFSS, ][]{Altschuler1969} extrapolation initialised by photospheric magnetograms from sources, e.g., the Global Oscillation Network Group \citep[GONG, ][]{Harvey1996} and SDO’s Helioseismic and Magnetic Imager \citep[HMI, ][]{Schou2012} in the low corona up to 2.6~R$_\odot$. It is followed by the Schatten Current Sheet \citep[SCS, ][]{Schatten1969} model, which then radially extends the magnetic field lines up to 21.5~R$_\odot$ (0.1~au). Finally, the plasma properties (speed, density, temperature) are computed at 0.1~au using empirical functions of the flux tube areal expansion factor and the distance of the foot point of the open field lines to the coronal hole boundary \citep[more details in ][]{Pomoell2018,Asvestari2019}. After the solar wind relaxation phase (i.e., filling the heliospheric domain with steady co-rotating solar wind plasma and magnetic field), CMEs are injected by the means of time-dependent boundary conditions at 0.1~au and then self-consistently evolved in the heliosphere by solving the MHD equations. In this work, the simulations with the Horseshoe model in EUHFORIA are performed using a spherical grid in the Heliocentric Earth EQuatorial (HEEQ) coordinate system, with a spatial resolution of 1.6~R$_\odot$ in the radial direction (0.1 -- 2~au), and 4$\degree$ angular resolution in the latitudinal (-70$\degree$ to 70$\degree$), and longitudinal (0$\degree$ to 360$\degree$) directions. In this work, the resolution of the EUHFORIA time series output is 10~minutes.

\subsection{Numerical implementation Horseshoe model in EUHFORIA}
The Horseshoe CME is injected into the heliospheric domain of EUHFORIA as a time-dependent boundary condition starting from a predetermined initiation time (based on observations of the CME speed). {There are two layers of ghost cells before the actual computational domain of the heliosphere. The CME values are first assigned at the interface between the ghost cells and the in-domain cells of the heliosphere and are then interpolated to the in-domain cells \citep[see details in][]{Pomoell2018}}. The implementation follows the same methodology as the torus models as detailed in Section~3.2 of \citet{linan2024}. A mask function is defined to identify the grid cells on the inner boundary (a spherical surface at 0.1~au), which intersect with the CME upon its insertion. Then, $\rho_l$ is computed, and all points on the boundary, where this radius is less than the minor radius of the torus, are considered part of the CME. As the magnetic field for the mMT configuration within the torus is defined in the local curved cylindrical coordinates $(\rho_{l},\phi_{l}, \theta_{l})$, the magnetic field is derived in the local spherical system (${\rho},{\theta},{\phi}$) where $\rho$ is the radius, $\theta$ is the co-latitude, and $\phi$ is the longitude. The two sets of coordinates are related to each other by the relations:
\begin{eqnarray}
    \rho_{l}&=&\sqrt{\rho^{2}+R_{0}^{2}-2\rho  R_{0}\sin\theta} \label{eq:rhol}\\
     \theta_{l}&=&
\begin{cases}
    \sin^{-1}{\frac{\rho\cos\theta}{\rho_l}},& \text{if } \rho\sin\theta-R_{0}\geq 0\\
    \pi-\sin^{-1}{\frac{\rho\cos\theta}{\rho_l}},              & \text{otherwise}.
\end{cases}
\end{eqnarray}
More details on the coordinate systems and the transformations can be found in the Appendix of \citet{linan2024}.

\begin{figure*}[ht!]
    \centering
    \includegraphics[width=0.49\textwidth,trim={0.0cm 0.0cm 9.0cm 0.0cm},clip=]{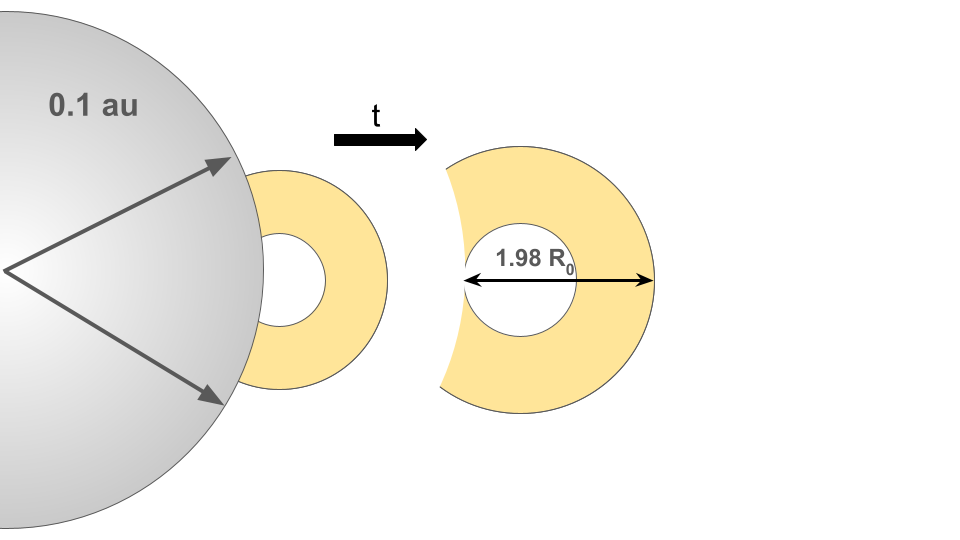}
    \includegraphics[width=0.49\textwidth,trim={0.0cm 0.0cm 0.0cm 0.0cm},clip=]{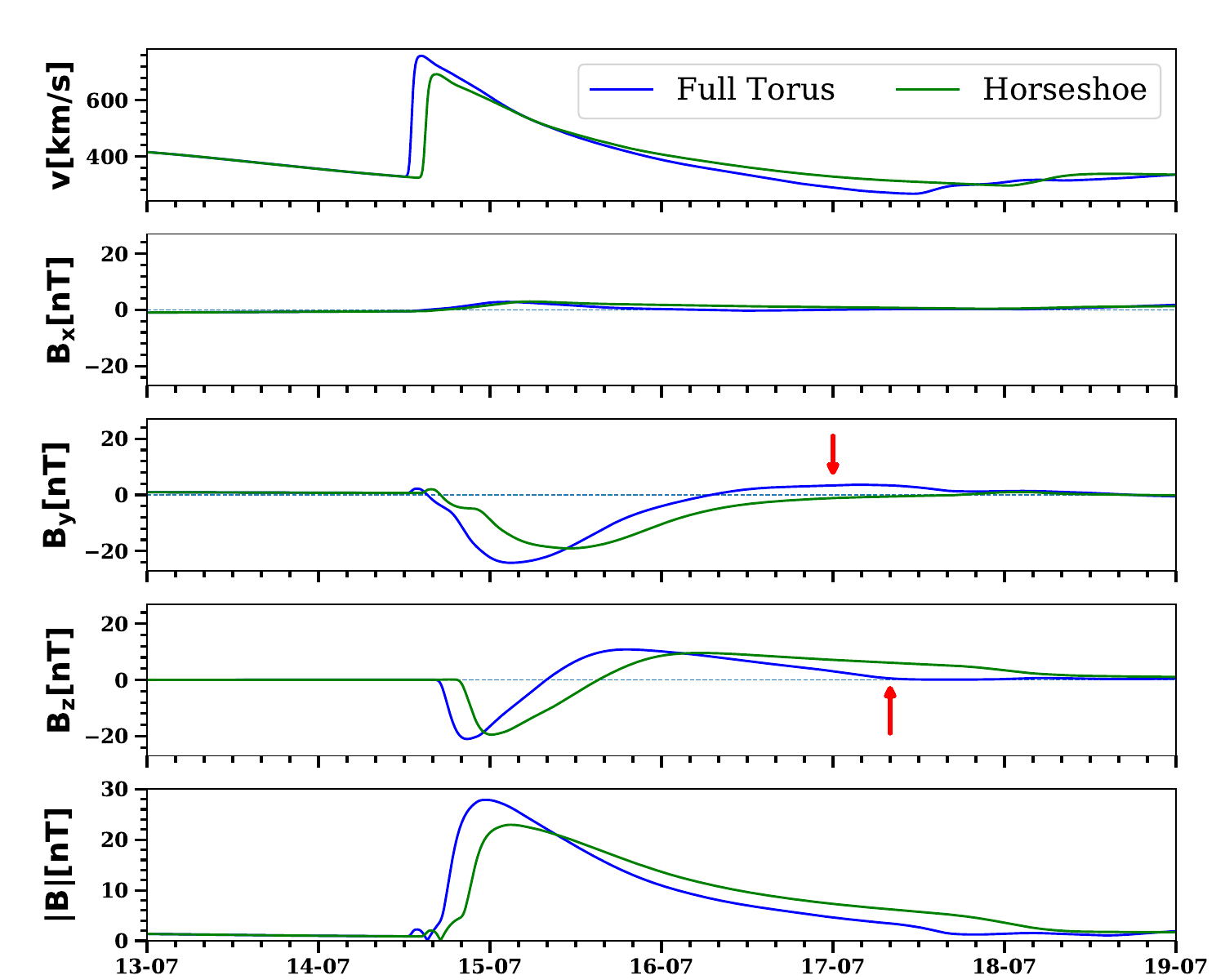}
    \caption{Horseshoe model implementation. Left: A schematic representation (not-to-scale) of the Horseshoe model propagating into the EUHFORIA heliosphere domain in the equatorial plane ($x-y$); Right: Comparison of the EUHFORIA predictions at 1~au between the Horseshoe and the full torus geometry with mMT magnetic field configuration. The red arrow in the $B_y$ profile shows the double polarity signatures of the full torus. The $B_z$ profile of the full torus is compressed and merged with the trailing part of the torus.}
    \label{fig:horseshoe_geo}
\end{figure*}

\begin{figure*}[ht!]
    \centering
    \includegraphics[width=0.49\textwidth,trim={0.0cm 0.0cm 0.0cm 0.0cm},clip=]{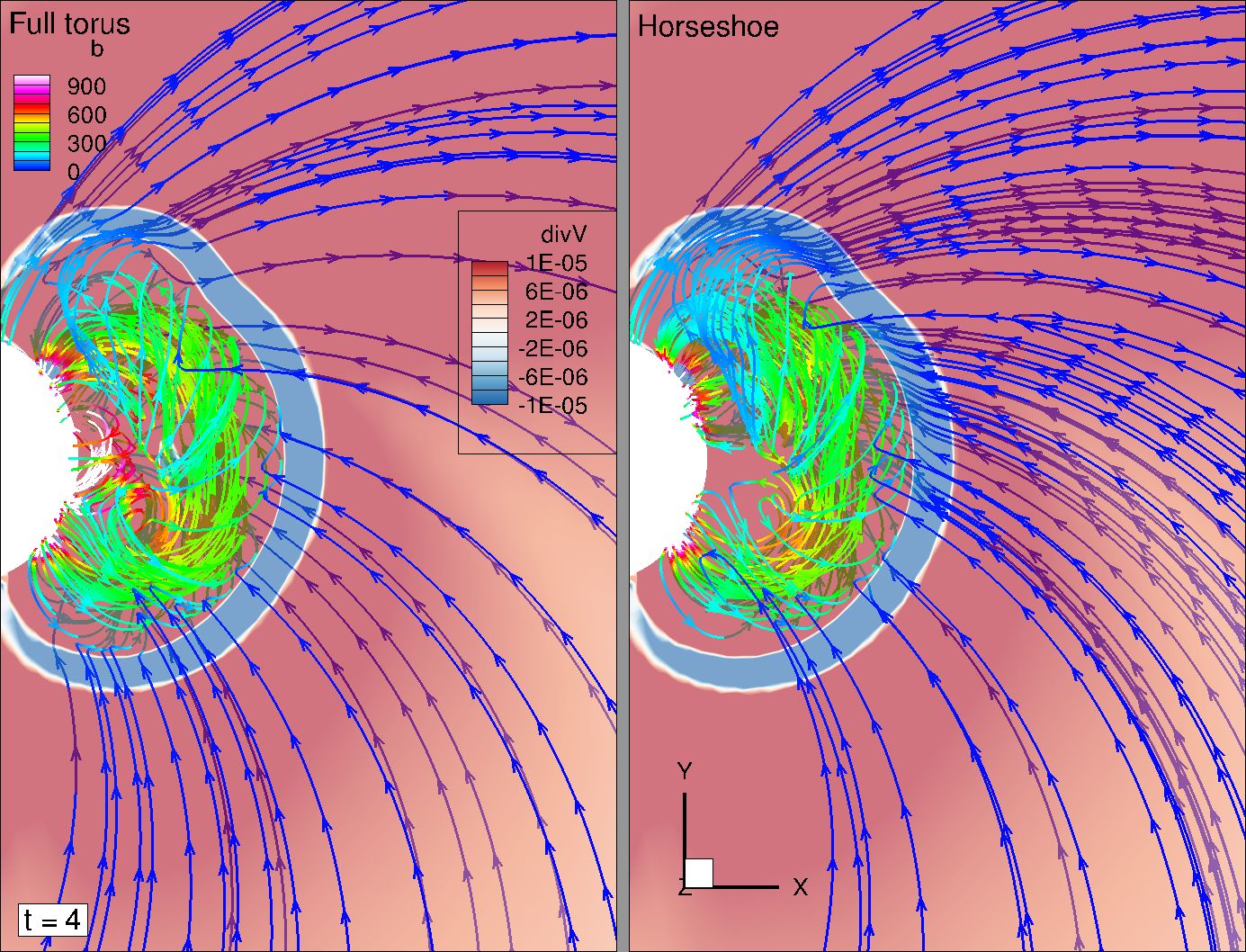}
    \includegraphics[width=0.49\textwidth,trim={0.0cm 0.0cm 0.0cm 0.0cm},clip=]{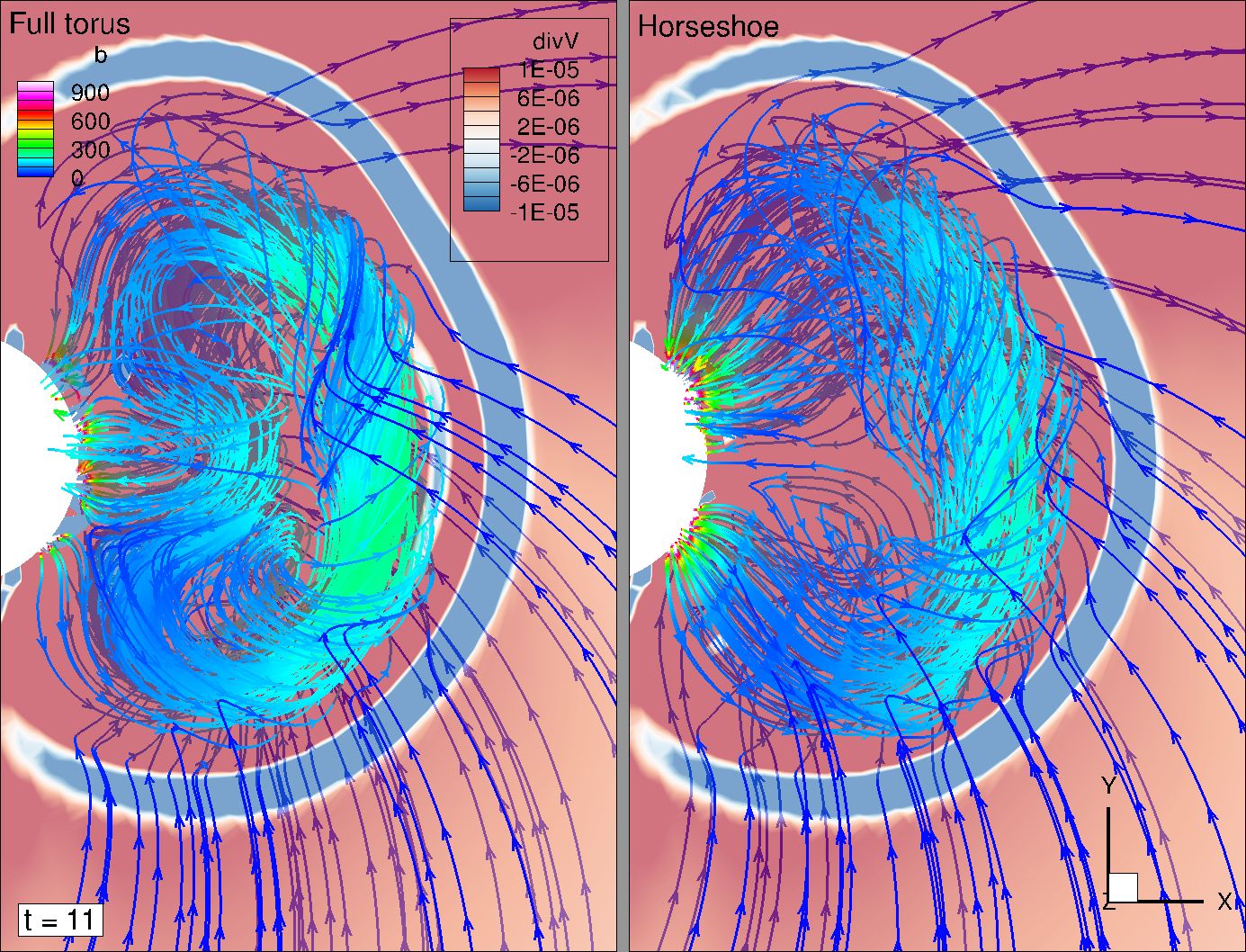}
    \caption{Equatorial snapshot of EUHFORIA simulations of the full torus and the Horseshoe implementation of the mMT magnetic field configuration. The evolution of the internal magnetic field lines of the CMEs (colour-coded with the magnetic field strength), 4 hours (panels 1--2), and 11 hours (panels 3--4), respectively, after the start of the injection at 0.1~au, is illustrated. In the background, the divergence of the speed is plotted. The sheath ahead of the magnetic cloud is depicted by the negative divergence (accumulation of plasma) region, showing a clear envelope around the CME. In panels 1 and 3, the rear part of the full torus is seen injected, whereas panels 2 and 4 show the horseshoe-like geometry creating CME leg-like structures connected to the inner boundary.}
    \label{fig:fulltorus_horseshoe_comp3D}
\end{figure*}

Instead of pushing the full torus of radial size $2(a+R_0)$, we cut the torus at $1.98\;R_0$ which is slightly less than $2\;R_0$ ($a+R_0$ (front half of the torus) + $R_0-a$ (from centre until the last half of the torus)) in order to exclude the injection of the trailing part of the torus. The modified torus of size $1.98\;R_0$, defined as the Horseshoe model, is illustrated schematically in Fig.\ref{fig:horseshoe_geo}. At each time step, the centre of the torus is advanced with a uniform radial speed and the 3D mask region is updated. In the mask region, the solar wind magnetic field values are substituted with the mMT magnetic field. We incorporate a uniform mass density and temperature in the Horseshoe model, similar to the full torus model. We performed a comparison of the EUHFORIA profiles at 1~au for the full torus and the Horseshoe models. We launched the CME along the Sun-Earth line with a zero tilt to get a simplified magnetic field profile at Earth. The time series plot in Fig.\ref{fig:horseshoe_geo} shows the contrasting features in the magnetic field components. The double polarity feature resulting from the front and the back part of the full torus is reflected in the $B_y$ component. The 3D visualisation in Panels 1--3 of Fig.~\ref{fig:fulltorus_horseshoe_comp3D} shows the evolution of the full torus in the equatorial plane, 4 and 11 hours after the start of the CME injection into the heliospheric domain, respectively. The magnetic field lines are colour-coded with the total magnetic field strength, hence distinguishing the background solar wind from the strong field inside the CME. The divergence of the velocity ($\nabla \cdot \mathbf{v}$) is plotted in the background to show the flow of plasma -- the blue spectrum corresponds to the accumulation of matter in the sheath ahead of the CME, and the red region inside the CME is depicting the outflow of mass. Due to the uniform injection speed, the torus is deformed during the injection. More precisely, the torus gets flattened, and the rear part becomes bean-shaped. Moreover, the rear part propagates much faster than the front part as the front part cleans up the background wind so that the density behind it is much lower. Consequently, the Alfv\'en speed increases and the rear part evolves faster than the front part and catches up with it. These features suggest the possibility of erroneous space weather predictions if a full torus model is used. Panels 2--4 of Fig.~\ref{fig:fulltorus_horseshoe_comp3D} depict the evolution of the Horseshoe CME and the mitigation of the double $B_y$ polarity signature. The $B_z$ profile at 1~au in the full torus lasts less long as a result of the trailing part of the torus compressing the front part and merging with it, whereas in the Horseshoe, CME expands more freely. 

In the current version of the Horseshoe model in EUHFORIA, we also implemented the MT magnetic field topology. MT is as stable as mMT, and qualitatively, it gives similar results for low aspect ratios. We also noticed that the aspect ratios obtained from the 3D reconstruction of CMEs from white light images within 0.1~au are usually low ($<2$). In that case, both mMT and MT will perform similarly. However, due to the more general divergence-free nature of the mMT solution, we develop further the methodologies to constrain the magnetic field parameters and optimise our simulations for real events for the mMT implementation in the Horseshoe geometry.

\section{Constraining the CME parameters} \label{sec:constrain_params}
\subsection{Magnetic field parameters}\label{subsec:constrain_mag_params}
To provide an early warning of CME impact, we constrain the CME parameters from remote sensing observations of the photosphere and corona before, during, and after the time of eruption. Estimating these parameters is not straightforward for the CMEs without low-coronal signatures like the streamer blow-out CMEs \citep{Robbrecht2009}. However, for the classical CMEs associated with flares, the source region features can allow the estimation of magnetic configuration of the flux rope \citep{Hudson2001,Palmerio2017}. The magnetic field input parameters required for the initialisation of the CME at 0.1~au are the axial magnetic field strength ($B_0$), the chirality ($C$), and the orientation of the magnetic axis (tilt). 

\textit{Axial magnetic field strength}: The field strength is quantified through the amount of the magnetic flux released during the eruption. Hence, we determine the toroidal and poloidal flux as a function of $B_0$. The toroidal flux, $\phi_t$, is the magnetic flux passing across the cross-section of the torus. Considering flux conservation and substituting with the mMT magnetic field equations from Eq.~\ref{eq:modMT_1}-\ref{eq:modMT_3},

\begin{eqnarray}
    \centering
    \varphi_t &=& \int_S \mathbf{B}\cdot\mathbf{dS} = \int_{0}^{a}\!\!\!\int_{0}^{2\pi} B_{\phi_l} \ \rho_l \ {d\rho_l} \ {d\theta_l} \\
    &=& \int_{0}^{a} \int_{0}^{2\pi} B_0\bigg(1 - \frac{\rho_l \cos\theta_l}{2R_0}\bigg) \ J_0(\alpha \rho_l) \ \rho_l \ {d\rho_l} \ {d\theta_l},
    \label{eq:phi_t_final}
\end{eqnarray}
\noindent where $S$ is the cross-section of the torus and magnetic field $\mathbf{B}$ is integrated over the crossing area. The radius of the surface $S$ is the minor radius of the torus ($a$). Here, $\alpha = \pm\frac{a_0}{a}$, where $a_0$ is the first root of $J_0$, i.e., $J_0(a_0)=0 \implies a_0=2.41\,a$. The sign of $\alpha$ represents the chirality of the flux rope. Inverting the above integral yields,
\begin{equation}
    \centering
    B_0 = \frac{\varphi_t \alpha}{2\pi a J_1(\alpha a)}.
\end{equation}

The poloidal flux, $\varphi_p$, is computed by integrating over the magnetic field lines crossing the $\theta_l = 0$ plane of the torus. Hence, we integrate $B_{\theta_l}$ (${\theta_l} = 0$) over $r$ and $\phi$:

\begin{eqnarray}
    \varphi_p &=& \frac{1}{2\pi}\int_0^{a}\int_0^{2\pi}B_{\theta_l} \rho_l d\rho_l d\phi_l \\
            &=& B_{0}\int_0^{a}\Bigg(\frac{R_{0}-2\rho_l}{2\alpha R_{0}(R_{0}+\rho_l)}\Bigg)J_{0}(\alpha \rho_l) \nonumber\\
            &&-\Bigg(1-\frac{\rho_l}{2 R_{0}}\Bigg)J_{1}(\alpha \rho_l)\rho_l d\rho_l. \label{eq:phi_p_final}
\end{eqnarray}
This integral can be computed numerically. The value of $\varphi_p$ is obtained from observations, and hence, the value of $B_0$ can be obtained by inverting Eq.~\ref{eq:phi_p_final}. The values of the geometrical parameters $a$ and $R_0$ obtained when the torus is self-similarly expanded up to 0.1~au, are substituted in Eq.~\ref{eq:phi_p_final}.

One of the fastest and most accessible methods to obtain $\phi_p$ from observations is using the empirical relation, i.e., using the peak X-ray intensity of the flare. \citet{Kazachenko2017} from a catalogue of 3000 flares (RibbonDB catalog, \url{http://solarmuri.ssl.berkeley.edu/~kazachenko/RibbonDB/}) found a correlation between the reconnection flux ($\phi_r$ in Mx) and the flare peak intensity ($I_{SXR}$, in W~m$^{-2}$). The relation they obtained was $\text{log}_{10}(\phi_r) = 24.42 + 0.64 \text{log}_{10}(I_{SXR})$, with a Spearman’s rank correlation coefficient 0.66. We set the value of $\phi_p$ to half of the total unsigned ribbon reconnection flux provided by the catalogue. The catalogue of \citep{Kazachenko2023} provides for 479 flares, the observed $\phi_p$ (hereafter, $\phi_{p,o}$) and the error associated with each event ($\Delta \phi_{p,e}$). The relative error, $RE=\Delta \phi_{p,e}/\phi_{p,o}\times100$, of the sample set ranges between $7 - 60\%$. The average $RE$ ($RE_{avg}$) from the catalogue is around $23\%$. We derive the absolute average error, $\Delta \phi_{p,avg}$, from $RE_{avg} = \Delta \phi_{p,avg}/\phi_{p,o}\times100$. While performing EUHFORIA simulations, the error estimates will be considered for making ensemble runs. The flares associated with the validation events (Section~\ref{sec:validation_events}) considered in this study are available in the RibbonDB catalogue. Hence, we use the observed poloidal flux to constrain $B_0$. 

\textit{Chirality}: With this parameter, we incorporate the sign of the helicity. The orientation of the magnetic flux rope can be interpreted from the hemisphere rule \citep{Pevtsov2003} or EUV/soft-X-ray sigmoids \citep{Titov1999}. A comprehensive list of different observational proxies of chirality can be found in \citet{Palmerio2017}. A right-handed sigmoid is associated with a positive chirality, and a left-handed one with a negative chirality.

\textit{Tilt}: The flux rope orientation is inferred to be parallel with the polarity inversion line (PIL). We define the east-west pointing flux rope, placed parallel to the solar equator, as the reference zero tilt for the Horseshoe model. The acute angle is assigned a negative (positive) sign when rotated anticlockwise (clockwise). 
\begin{figure*}[ht!]
    \centering
    \subfloat[]{\includegraphics[width=0.25\textwidth,trim={0.5cm 0.6cm 0.0cm 1.5cm},clip=]{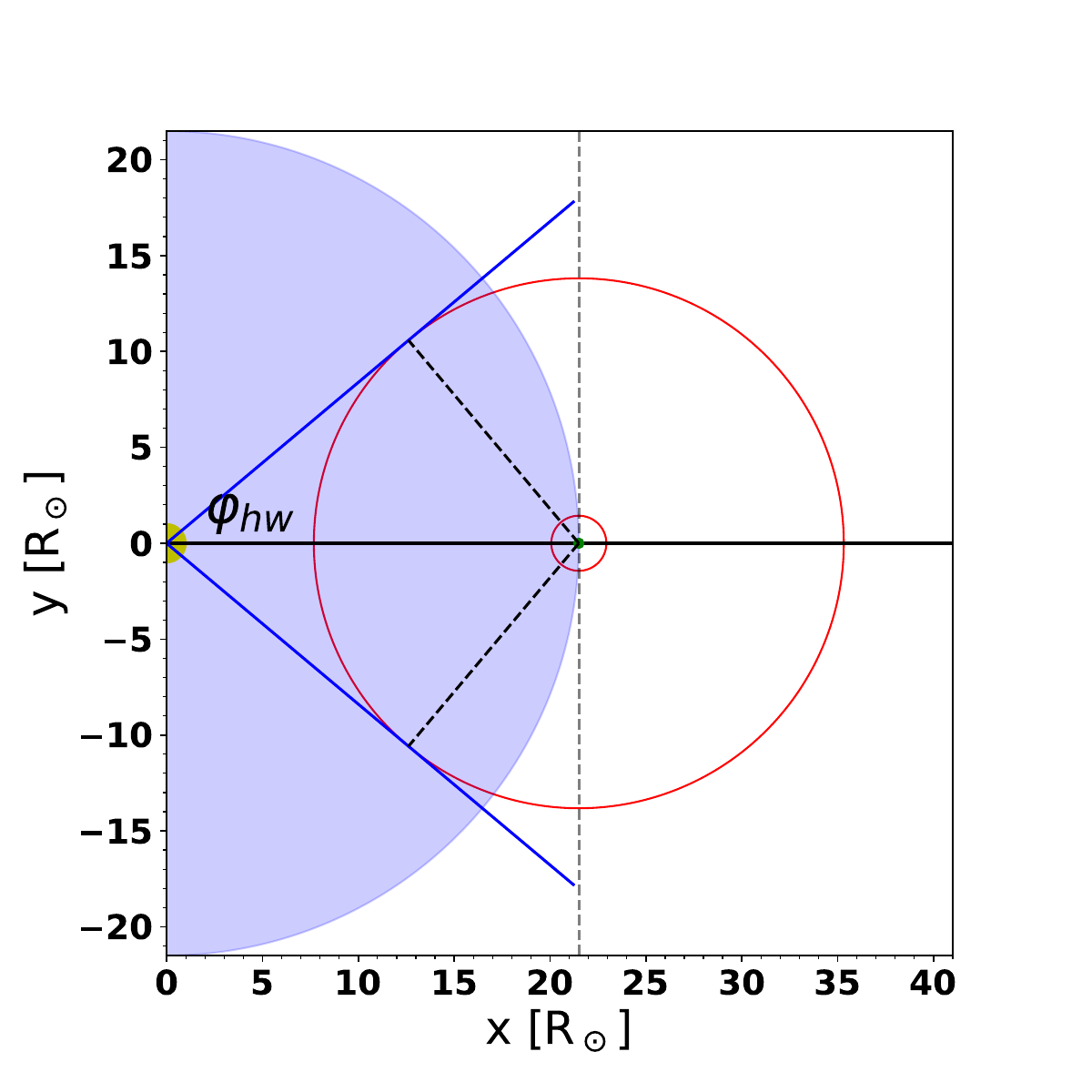}}
    \hspace{-1.5\baselineskip}
    \subfloat[]{\includegraphics[width=0.25\textwidth,trim={0.5cm 0.6cm 0.0cm 1.5cm},clip=]{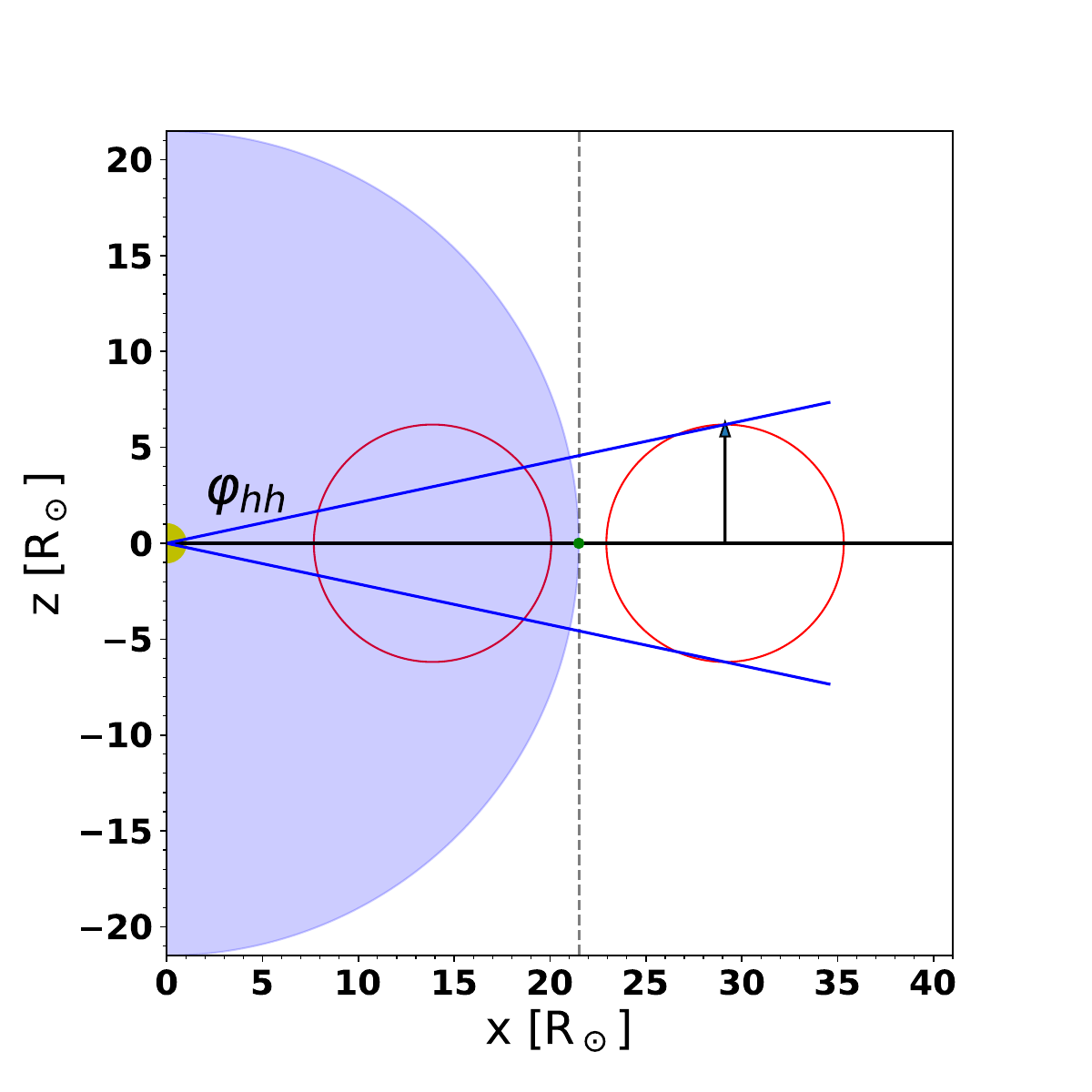}} 
    \subfloat[]{\includegraphics[width=0.25\textwidth,trim={0.5cm 0.6cm 0.0cm 1.5cm},clip=]{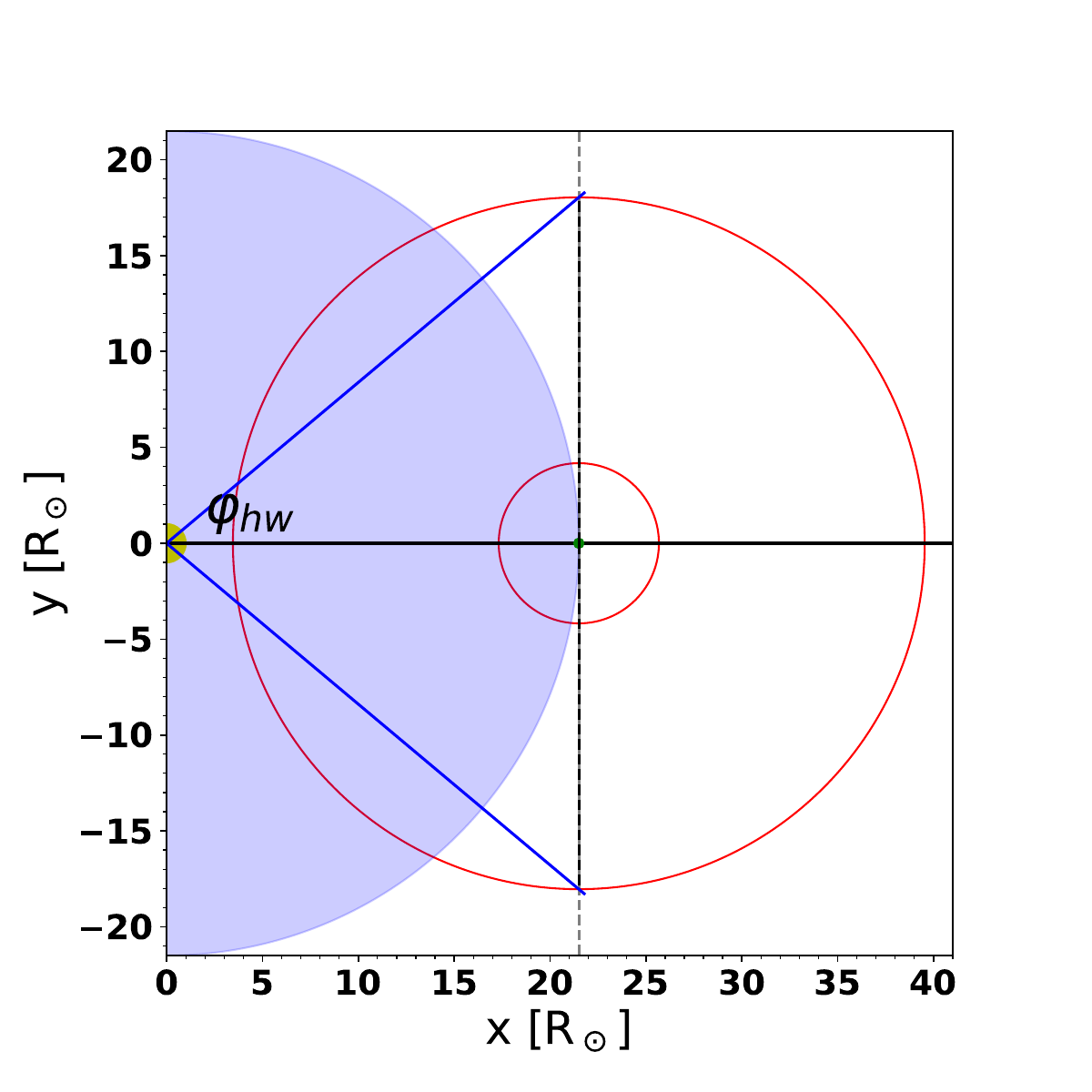}}
    \hspace{-1.5\baselineskip}
    \subfloat[]{\includegraphics[width=0.25\textwidth,trim={0.5cm 0.6cm 0.0cm 1.5cm},clip=]{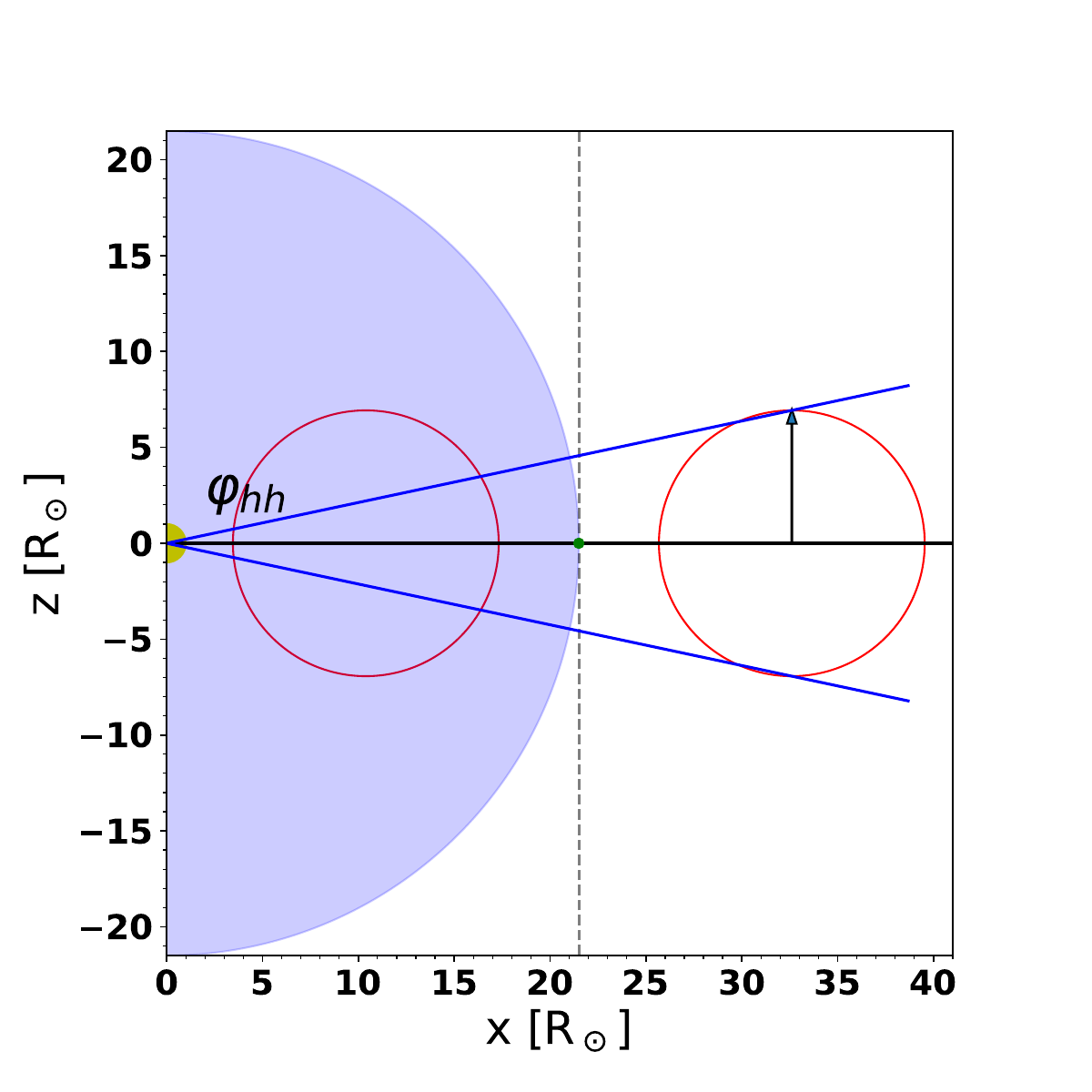}}
    \caption{Torus geometry in the equatorial and meridional planes for Case 1 (a, b) and Case 2 (c, d) geometries are represented with red curves. The horizontal black solid line is the Sun-Earth line. The blue lines on the equatorial and meridional plots have an angular extent of $\varphi_{hw}$ and $\varphi_{hh}$, respectively, about the Sun-Earth line. In this example, $\varphi_{hw}$ = 40$^\circ$ and $\varphi_{hh}$ = 12$^\circ$. In Case 1, $R_0$ = 7.6 and $a$ = 6.2, making the aspect ratio 1.23. In Case 2, $R_0$ = 11.1 and $a$ = 6.9, making the aspect ratio 1.6.}
    \label{fig:torus_geo}
\end{figure*}
\subsection{Geometry}\label{subsec:constrain_geo_params}
The geometrical parameters for the Horseshoe model can be constrained equivalently to the full torus, as its geometry is not altered during the numerical implementation; rather, only a part of the full torus is injected. The simplest CME geometry is generally defined using the angular half-width ($\varphi_{hw}$) and angular half-height ($\varphi_{hh}$), which define the maximum extent of the CME (relative to the plane containing its magnetic axis) in the azimuthal and polar directions, respectively. This approach has been adopted in 3D reconstruction techniques of the Graduated Cylindrical Shell \citep[GCS,][]{Thernisien2006} and Flux Rope in 3D \citep[FRi3D,][]{Isavnin2016} models. For constraining the geometrical parameters to initialise CMEs in heliospheric models like EUHFORIA, the 3D reconstruction of a CME is performed as long as the CME structures can be identified distinctly in white light coronagraph images. Multi-point observations provide the view of the CME in the upper corona until somewhere below 21.5~R$_\odot$. Such data is currently available from the C2 and C3 instruments of the Large Angle and Spectrometric COronagraph \citep[LASCO,][]{Brueckner1995} on board the Solar and Heliospheric Observatory (SOHO), and the COR-2 instrument on board the Sun-Earth Connection Coronal and Heliospheric Investigation (SECCHI) package of the twin-spacecraft Solar Terrestrial Relations Observatory \citep[STEREO,][]{Kaiser2008}, the Metis instrument  \citep{Ester2020} on the Solar Orbiter \citep{Muller2020} etc. So, we determine the geometry of the torus assuming a self-similar expansion of the CME beyond the last observed time frame until 21.5~R$_\odot$. We inject the already expanded torus directly at 21.5~R$_\odot$. The time of arrival of the CME at 21.5~R$_\odot$ is calculated using the speed obtained in the last frame and assuming a uniform propagation of the CME up to 21.5~R$_\odot$. In this section, we define two geometries (Case~1 and Case~2) for establishing the association between the torus parameters $a$, $R_0$, and the centre of the torus ($T_c$) with the observed angular CME parameters, $\varphi_{hw}$ and $\varphi_{hh}$.  
\begin{itemize}
    \item {Case 1:} This geometry is represented by the red curve in the equatorial and meridional planes as shown in Fig.~\ref{fig:torus_geo}(a) and (b), respectively, when $T_c=21.5~R_{\odot}$. The $\varphi_{hw}$ and $\varphi_{hh}$ are defined by:
\begin{equation}
    \sin(\varphi_{hw}) =  \frac{R_0 + a}{T_c},
    \label{eq:case1_hw}
\end{equation}
and
\begin{equation}
    \tan(\varphi_{hh}) =  \frac{a}{T_c + R_0}.
    \label{eq:case1_hh}
\end{equation}
Using Eqs.~(\ref{eq:case1_hw}) and (\ref{eq:case1_hh}), we derive $R_0$ and $a$. 
\begin{equation}
    R_0 = T_c \frac{\sin(\varphi_{hw}) - \tan(\varphi_{hh})}{1 + \tan(\varphi_{hh})} ,
    \label{eq:case1_bigrad}
\end{equation}
and
\begin{equation}
    a = {(T_c + R_0)\; \tan(\varphi_{hh})}.
    \label{eq:case1_smallrad}
\end{equation}

    \item {Case~2:} This geometry is represented by the red curve in the equatorial and the meridional planes as shown in Figs.~\ref{fig:torus_geo}(c) and (d), respectively, when $T_c=21.5~R_{\odot}$. The $\varphi_{hw}$ and $\varphi_{hh}$ are defined by:
\begin{equation}
    \tan(\varphi_{hw}) =  \frac{R_0 + a}{T_c},
    \label{eq:case2_hw}
\end{equation}
and
\begin{equation}
    \tan(\varphi_{hh}) =  \frac{a}{T_c + R_0}.
    \label{eq:case2_hh}
\end{equation}
Using Eqs.~(\ref{eq:case2_hw}) and (\ref{eq:case2_hh}), we derive $R_0$ and $a$:
\begin{equation}
    R_0 = T_c\; \frac{\tan(\varphi_{hw}) - \tan(\varphi_{hh})}{1 + \tan(\varphi_{hh})},
    \label{eq:case2_bigrad}
\end{equation}
and
\begin{equation}
    a = (T_c + R_0)\,\tan(\varphi_{hh}).
    \label{eq:case2_smallrad}
\end{equation}  
\end{itemize}


\begin{figure*}[ht!]
    \centering 
    \subfloat[]{\includegraphics[width=0.5\textwidth,trim={0cm 0cm 0.cm 0cm},clip=]{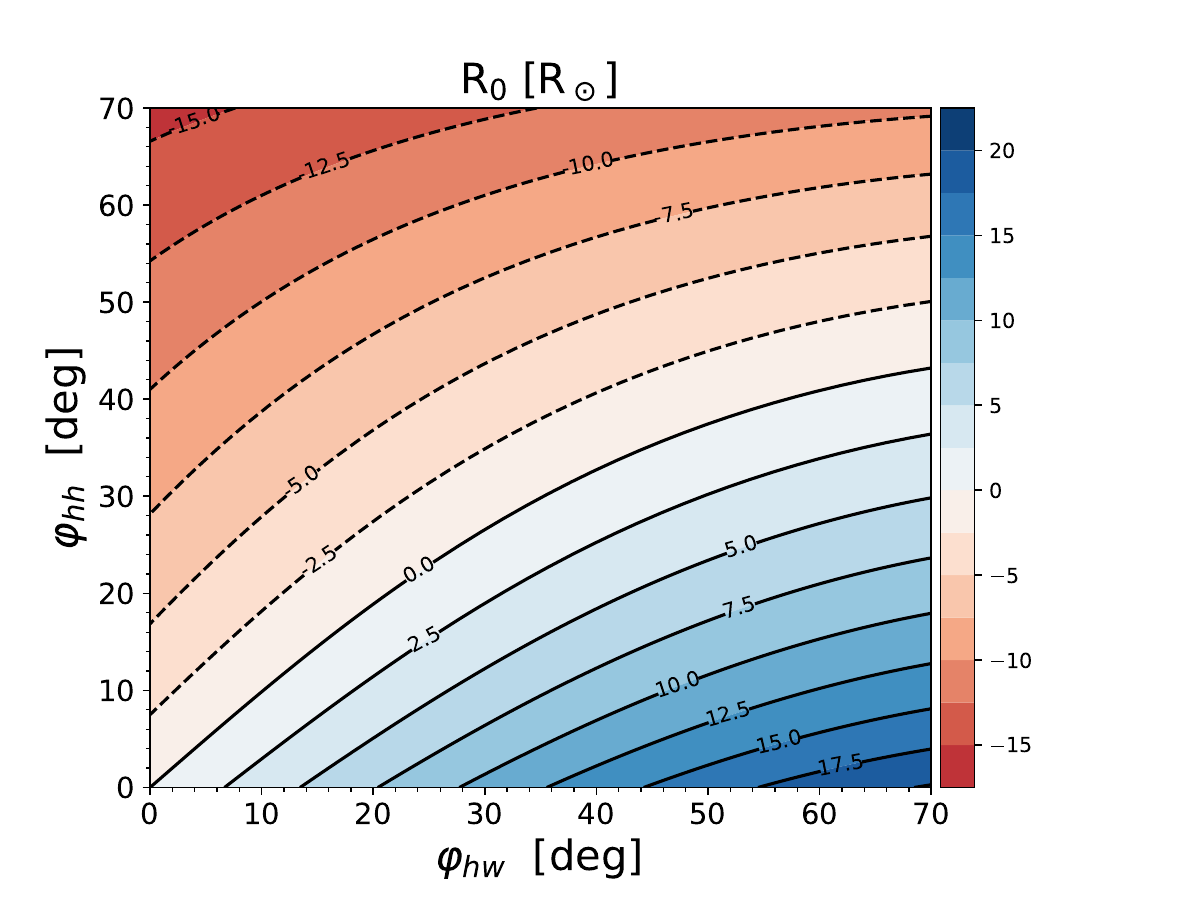}}
    \subfloat[]{\includegraphics[width=0.5\textwidth,trim={0cm 0cm 0.cm 0cm},clip=]{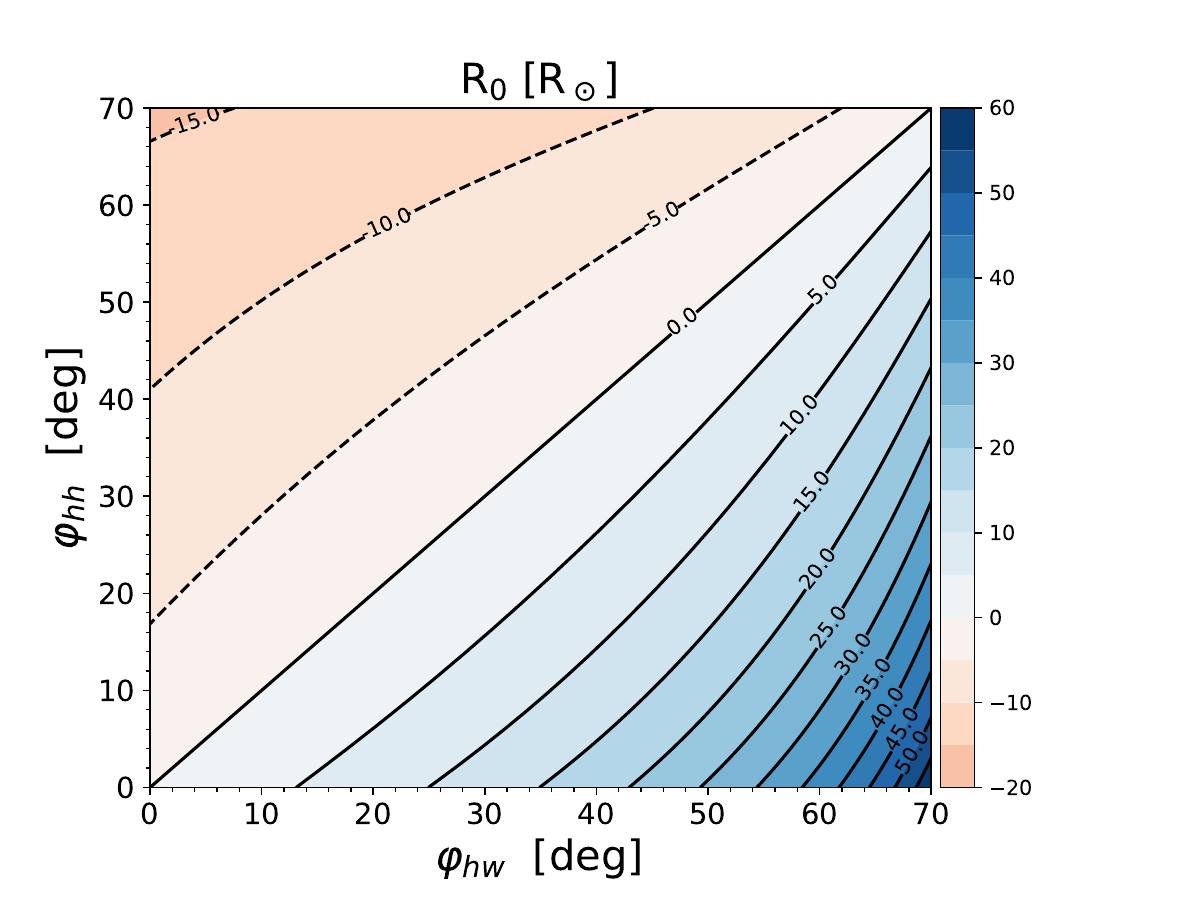}}\\
    \subfloat[]{\includegraphics[width=0.5\textwidth,trim={0cm 0cm 0.cm 0cm},clip=]{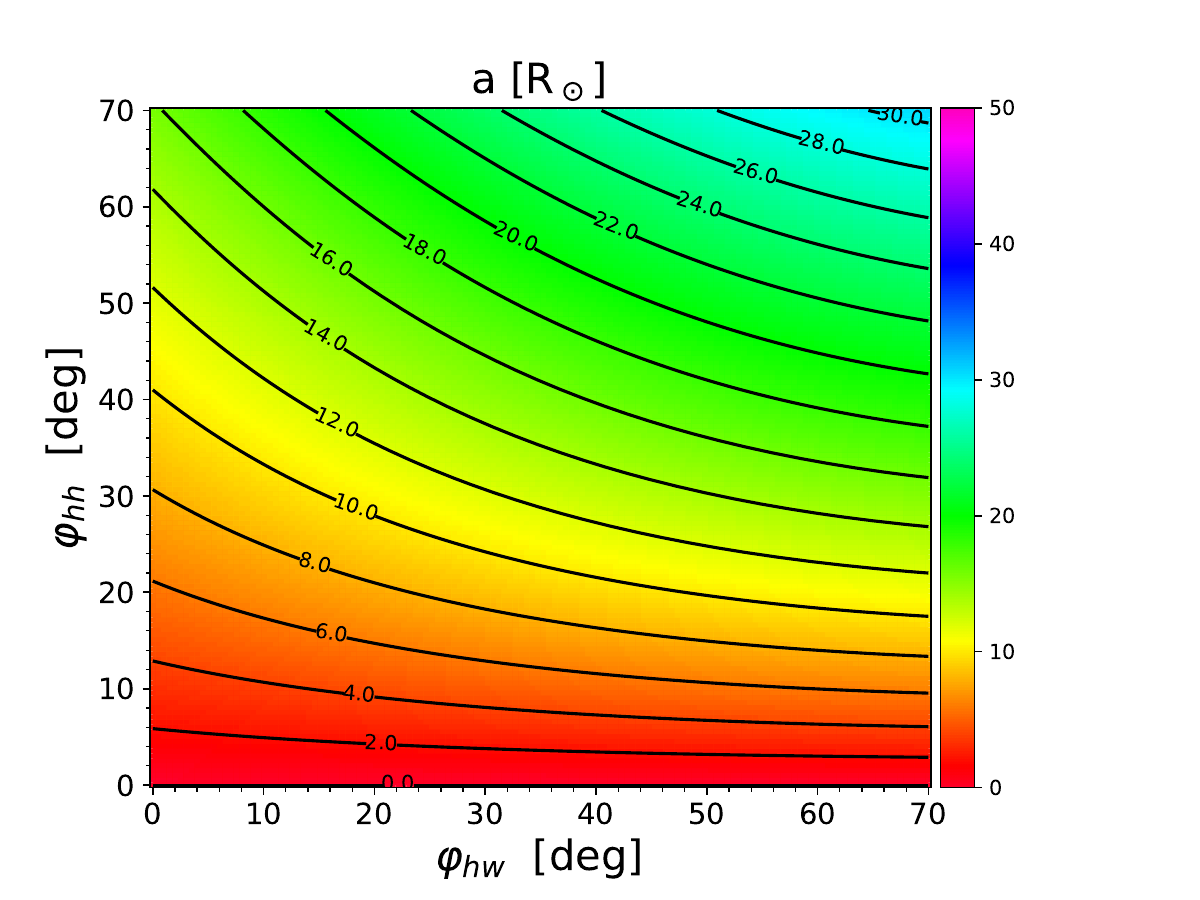}}
    \subfloat[]{\includegraphics[width=0.5\textwidth,trim={0cm 0cm 0.cm 0cm},clip=]{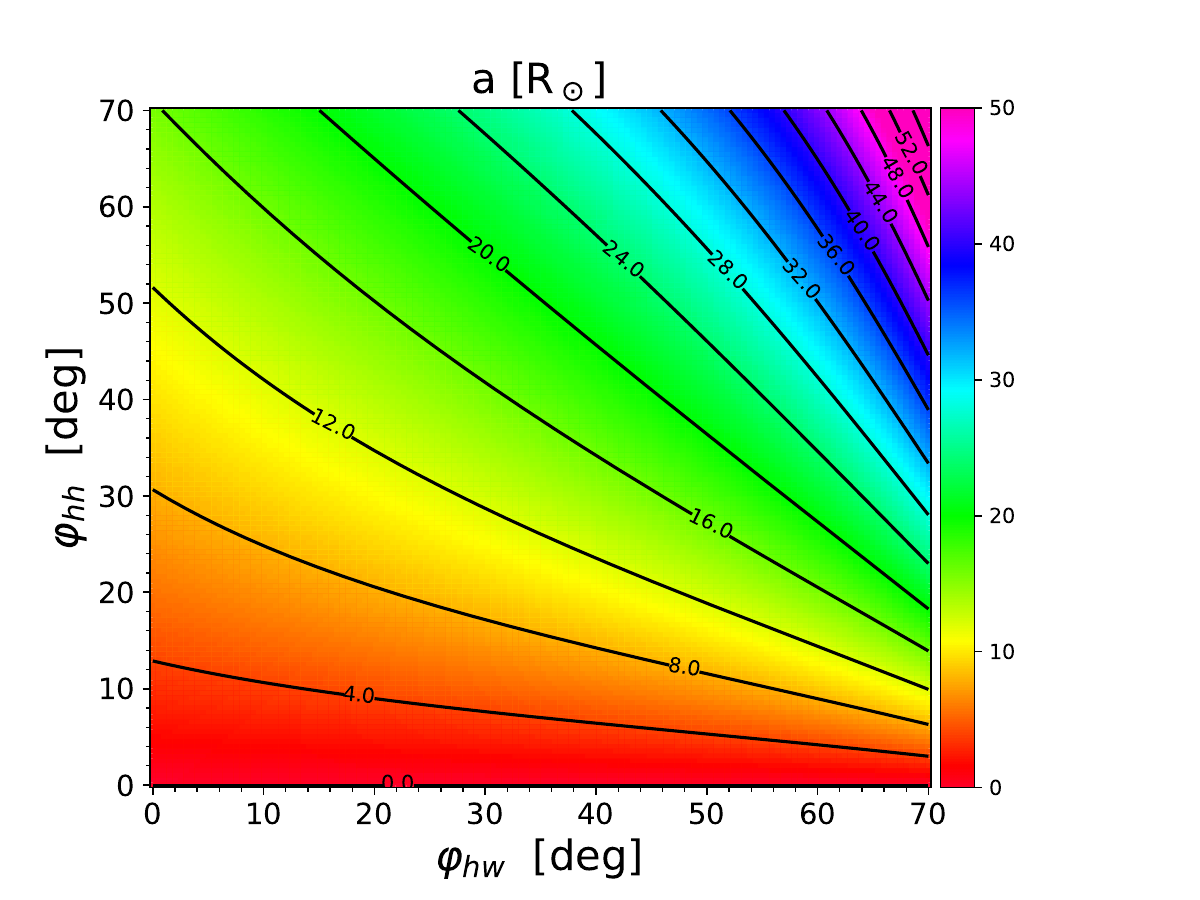}}\\
    \subfloat[]{\includegraphics[width=0.5\textwidth,trim={0cm 0cm 0.cm 0cm},clip=]{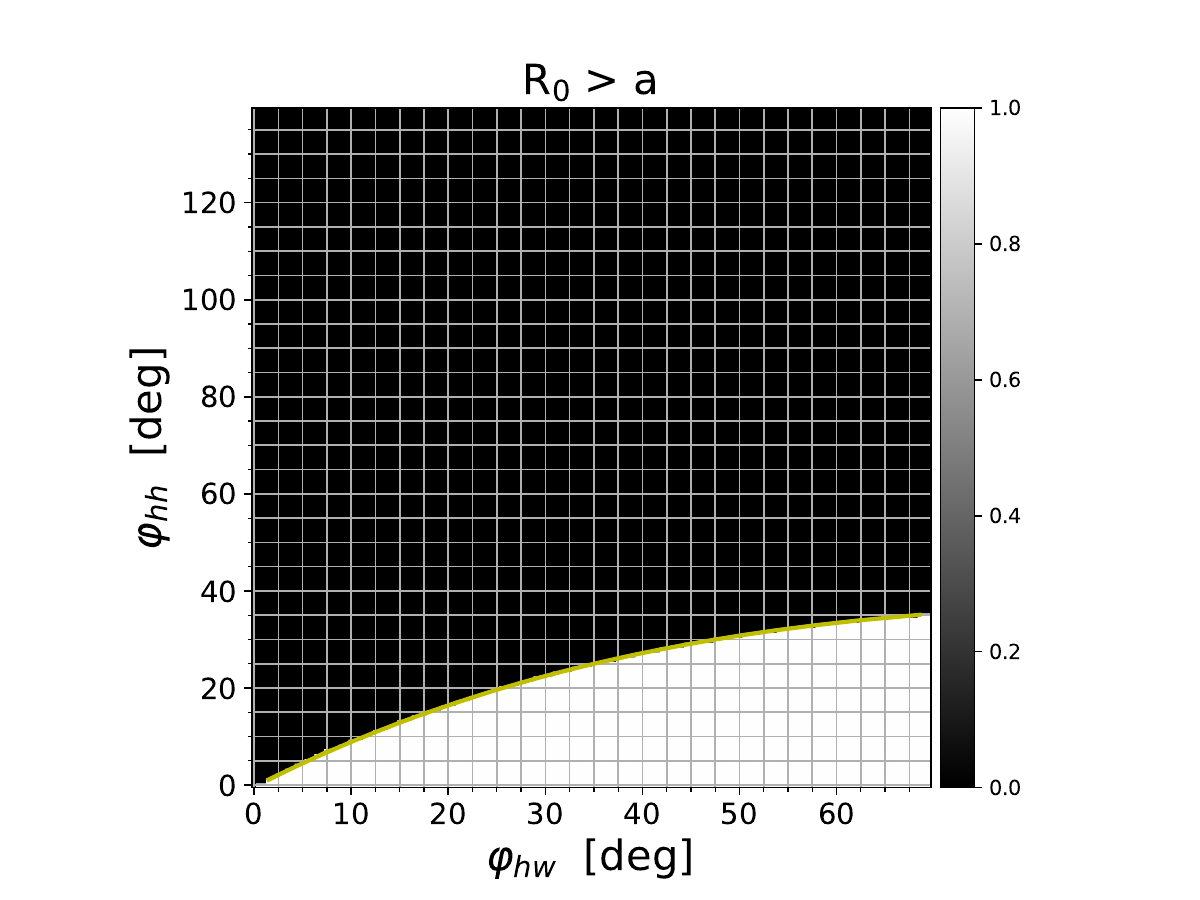}
    }
    \subfloat[]{\includegraphics[width=0.5\textwidth,trim={0cm 0cm 0.cm 0cm},clip=]{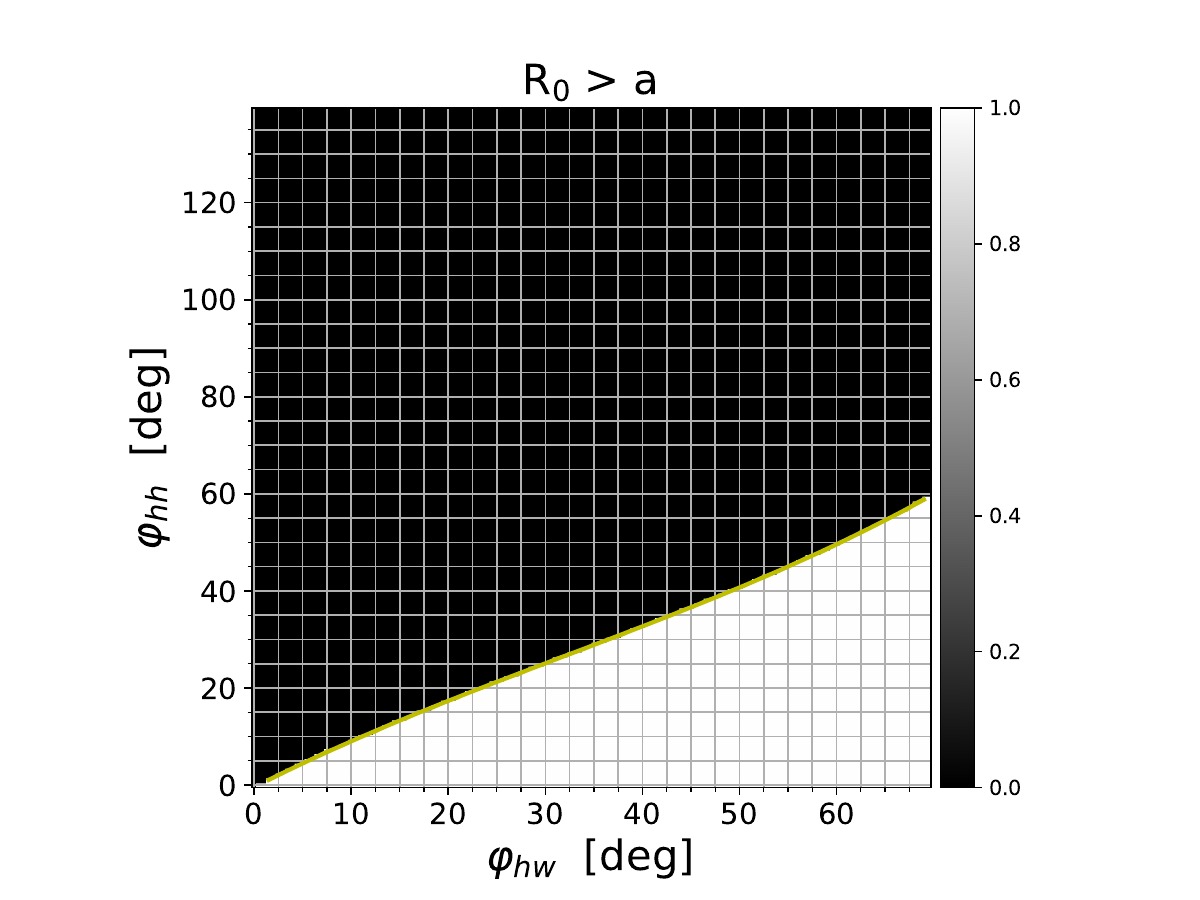}}
    \caption{Contours of major radius ($R_0$) as a function of $\varphi_{hw}$ and $\varphi_{hh}$ for (a) Case 1 and (b) Case 2 geometries. Contours of the minor radius ($a$) as a function of $\varphi_{hw}$ and $\varphi_{hw}$ for (c) Case 1 and (d). The parameter space where $R_0 \ > \ a$ is identified for (e) Case 1 and (e) Case 2, geometries, respectively.}
    \label{geo_torus_radii}
\end{figure*}

\begin{figure*}[ht!]
    \centering 
    \subfloat[Case 1: Aspect ratio parameter space]{\includegraphics[width=0.5\textwidth,trim={0cm 0cm 0.cm 0cm},clip=]{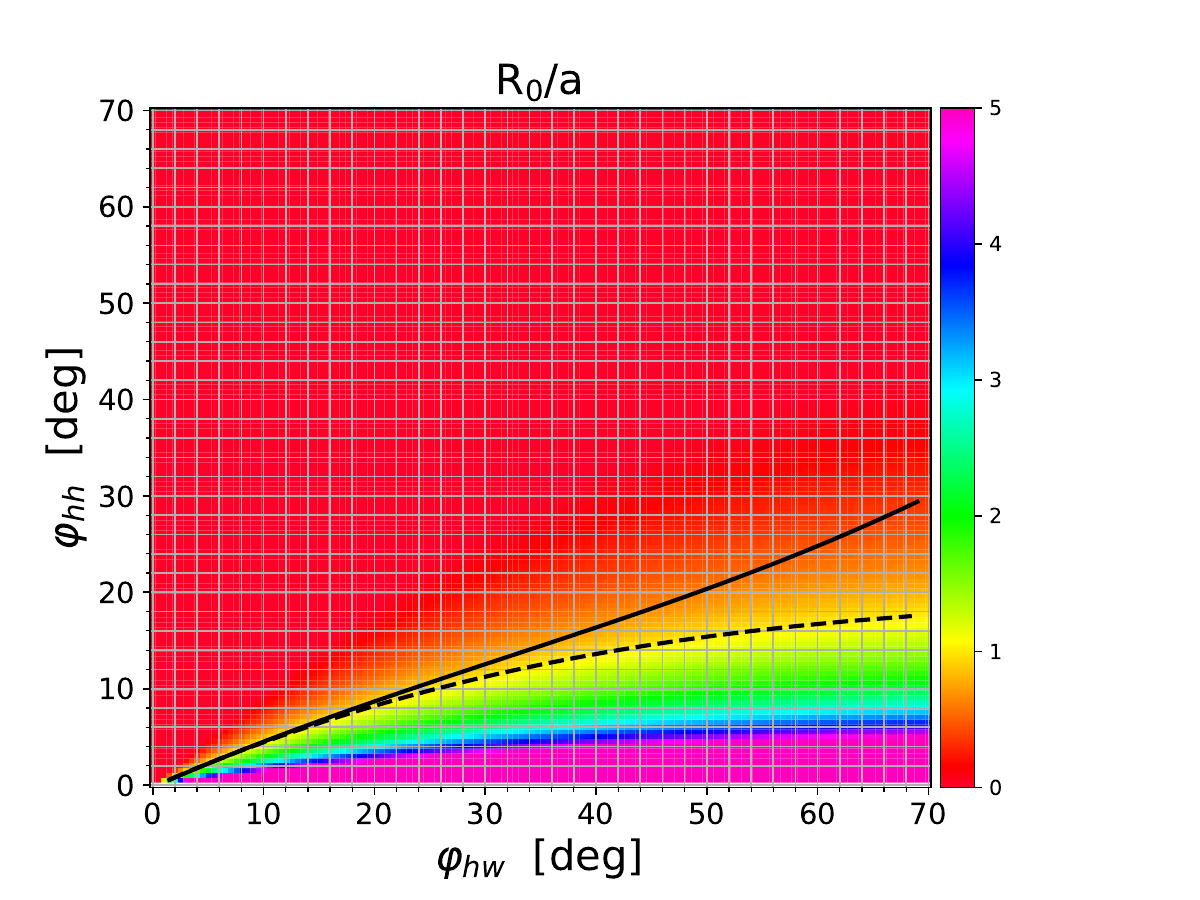}}
    \subfloat[Case 2: Aspect ratio parameter space]{\includegraphics[width=0.5\textwidth,trim={0cm 0cm 0.cm 0cm},clip=]{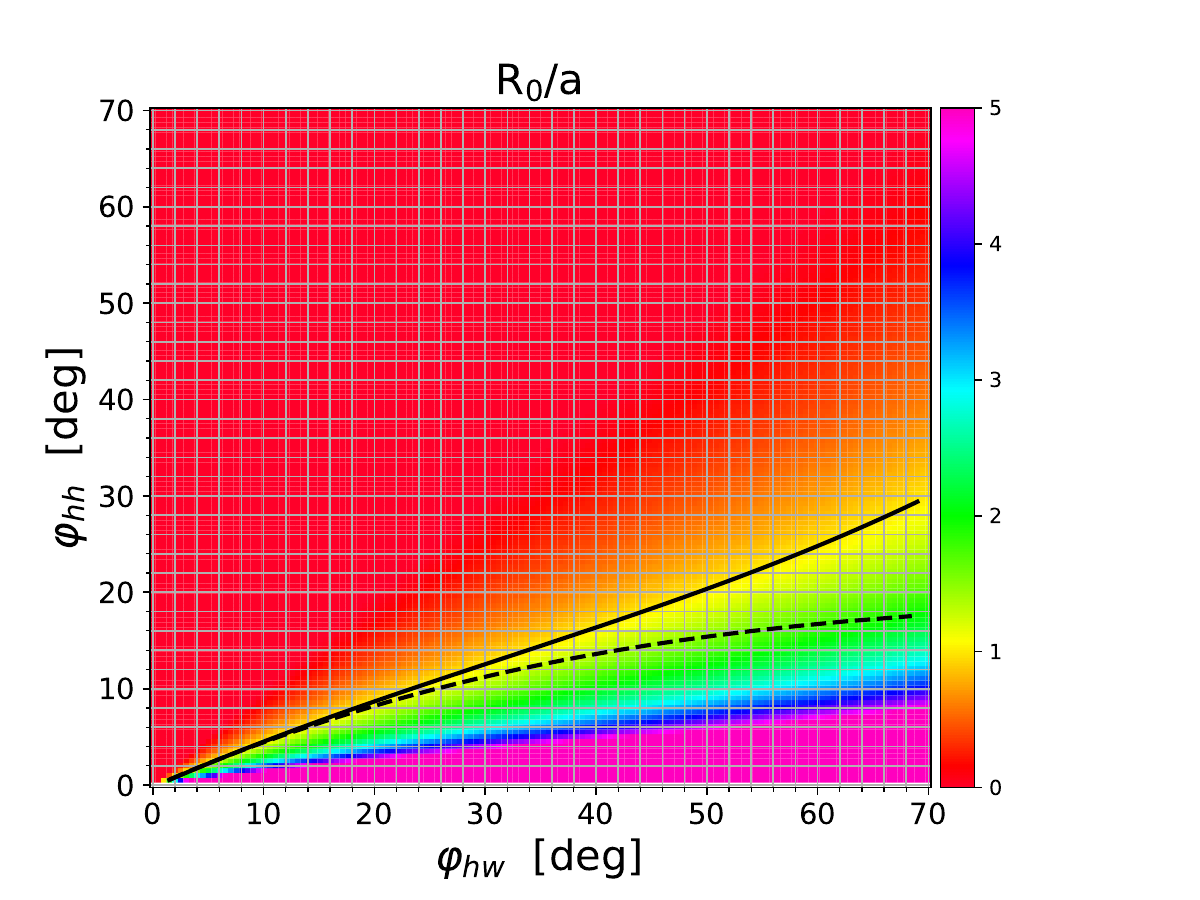}}\\
    \subfloat[Case 1: Volume parameter space]{\includegraphics[width=0.5\textwidth,trim={0cm 0cm 0.cm 0cm},clip=]{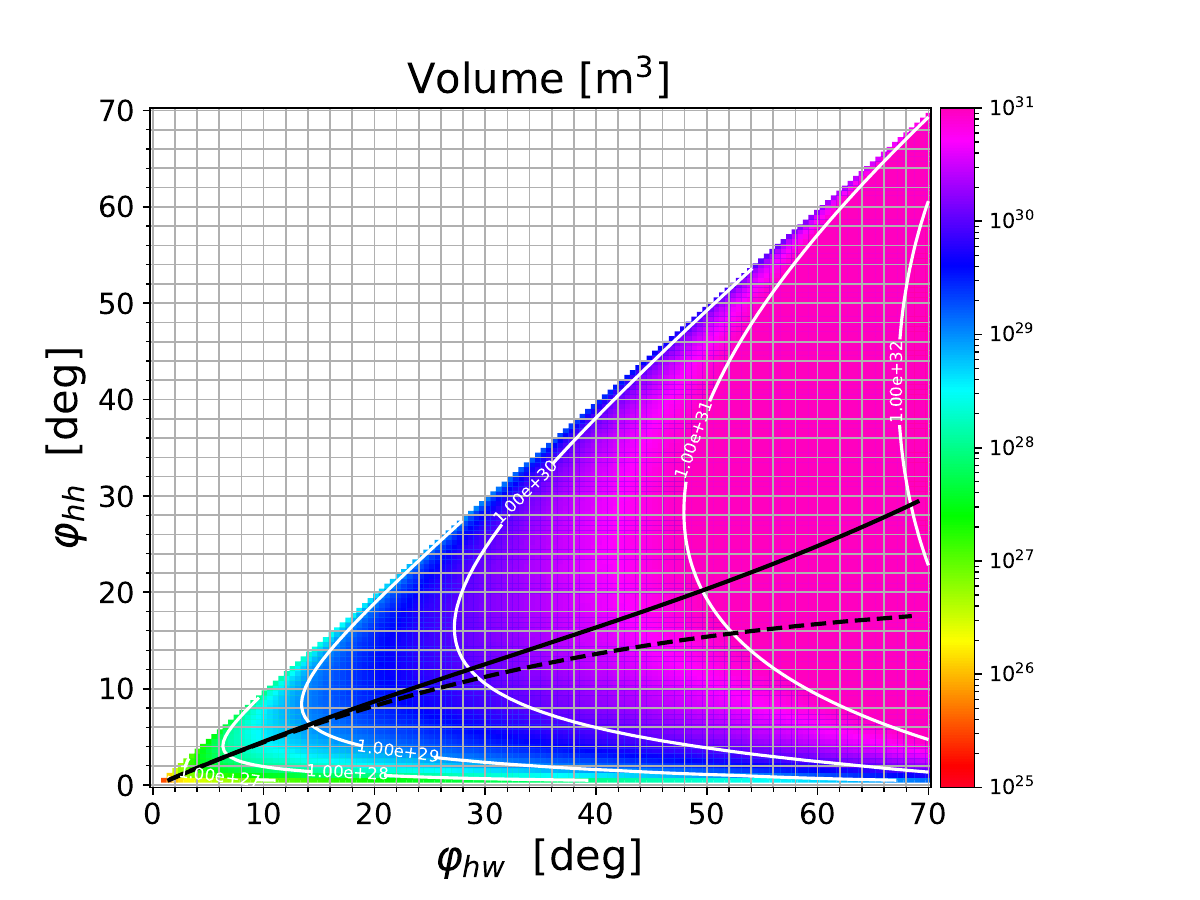}}
    \subfloat[Case 2: Volume parameter space]{\includegraphics[width=0.5\textwidth,trim={0cm 0cm 0.cm 0cm},clip=]{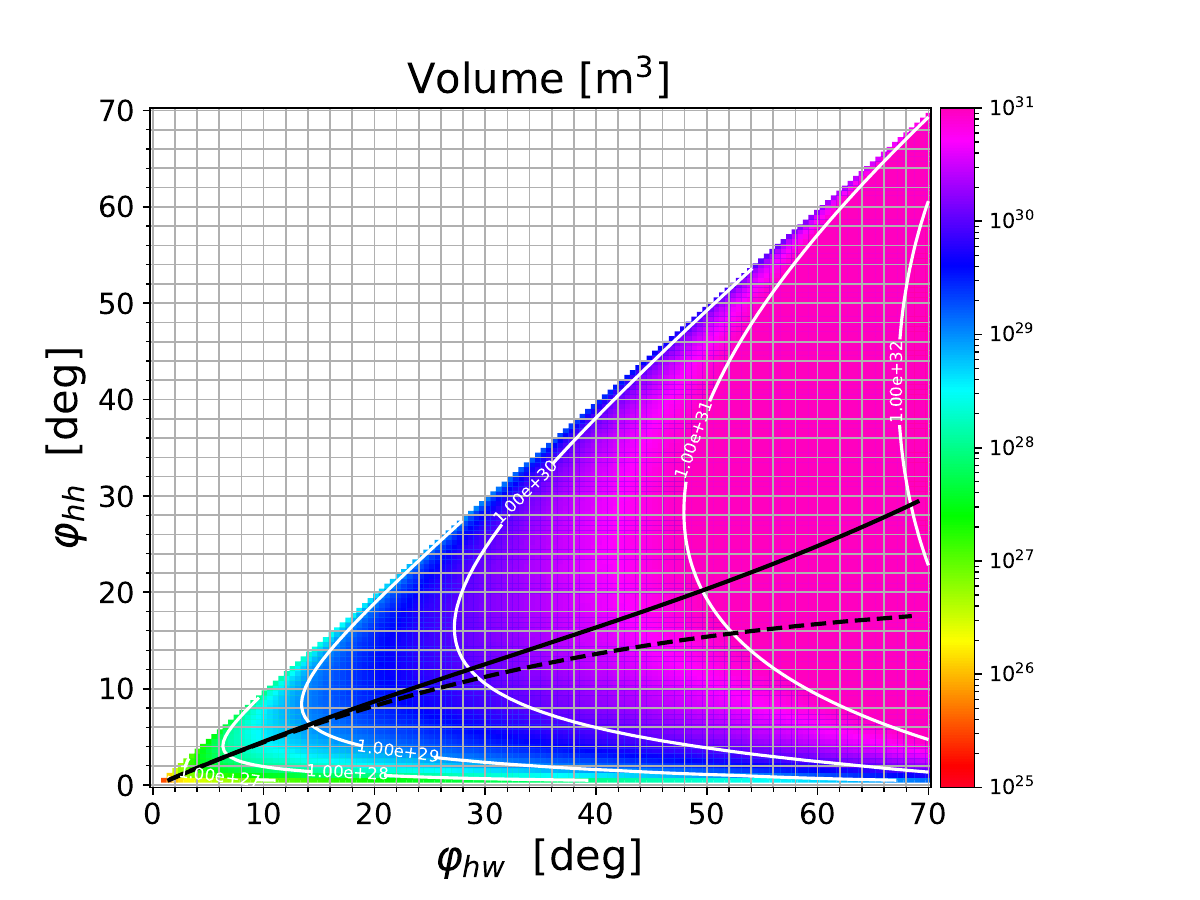}}\\
    
    \caption{Distribution of aspect ratio (a) Case 1 and (b) Case 2 geometries and the volume (c) Case 1 and (b) Case 2 geometries. The colour bar shows the magnitude of the volume. The black dashed, and the solid curves demarcate the allowed range of $\varphi_{hw}-\varphi_{hh}$ space for Case~1 and Case~2 geometries, respectively.}
    \label{fig:geo_torus_volume}
\end{figure*}

Figure~\ref{geo_torus_radii} shows the distribution of $R_0$ and $a$ of the torus as a function of $\varphi_{hw}$ and $\varphi_{hh}$ for the two geometries, and the allowed parameter space where the torus exists. 
The distribution of the $R_0$ in the $\varphi_{hw}-\varphi_{hh}$ space is shown in figs.~\ref{geo_torus_radii}(a,b) for the two geometries. $R_0$ becomes negative when $\sin(\varphi_{hw}) < \tan(\varphi_{hh})$ for the Case~1 geometry, and when $\varphi_{hw} < \varphi_{hh}$ for the Case~2 geometry. The Case~2 geometry results in a comparatively larger equatorial cross-section, resulting in a higher aspect ratio. With Case~1 we get $R_0 < 21.5$~R$_\odot$ even for wide CMEs, whereas Case~2 estimates a very large $R_0$ that cannot be injected at the EUHFORIA inner boundary. The distribution of $a$ (Figs.~\ref{geo_torus_radii}(c,d)) is similar for both geometries for the low $\varphi_{hw}-\varphi_{hh}$ space. However, for the values in the higher extreme of the parameter space, Case~2 estimates higher values of $a$ as compared to Case~1. The next step is to ensure the allowed pairs of ($a$, $R_0$) such that the torus is physical, i.e., $R_0>a$. This criterion is depicted in the boolean plots as shown in Figs~\ref{geo_torus_radii}(e) and \ref{geo_torus_radii}(f) for Case~1 and Case~2, respectively. The white space (True) in the plots corresponds to the allowed ($\varphi_{hw}$, $\varphi_{hh}$) values for both geometries. The boundary separating the allowed and non-allowed values can be fit with a univariate cubic spline (piecewise polynomial function of degree 3) plotted in yellow. This diagnostic shows that the maximum $\varphi_{hh}$ with Case~1 geometry is less than $20\degree$ even for a CME as wide as $70\degree$. Hence, with this geometry, we can construct a thin cross-section of CMEs, which could imply a potential underestimation of CME mass. Whereas, with Case~2 geometry, the range of obtained maximum $\varphi_{hh}$ is higher as compared to Case~1. We can say that the Case~2 geometry performs better in estimating the CME cross-section ($a$), although it overestimates $R_0$. Therefore, depending on the observed event, either Case~1 or Case~2 can be chosen to constrain ($a$, $R_0$) from ($\varphi_{hw}$, $\varphi_{hh}$). In general, we require the strong condition $2(a+R_0) < 21.5\;$R$_\odot$ to hold, which in turn obliges us to constrain a small $a$ for high aspect ratio cases.

{The distribution of the aspect ratio in the $\varphi_{hw}-\varphi_{hh}$ parameter space in Fig.~\ref{geo_torus_radii} is plotted in Figs~\ref{fig:geo_torus_volume}(a,b). The black dashed and solid curves are the cubic spline defining the boundary for the physical torus parameters for Case~1 and Case~2 geometries, respectively (same as in Figs.~\ref{geo_torus_radii}(e, f)). These figures suggest the range of aspect ratios that can be allowed in both geometries. Moreover, it can be observed that Case~2 favours higher aspect ratios for $\varphi_{hw} = 20\degree$ as compared to Case~1.} The volume of the Horseshoe geometry is computed as $3/4\cdot 2\pi^2R_0a^2$. Figure~\ref{fig:geo_torus_volume} illustrates the extent of the torus volume for the two geometries in $\varphi_{hw}-\varphi_{hh}$ space. The colour map covers the full range of positive volume. The order of magnitude of the volume is in the range $10^{27}-10^{31}$~m$^3$ for Case~1 and up to $10^{32}$~m$^3$ for Case~2, corresponding to the range modelled by FRi3D and spheromak models in EUHFORIA as discussed in \citet{Maharana2022}. The volume of the FRi3D model (with a flexible and extended flux rope geometry) belongs to the range $10^{29}-10^{31}\;$m$^3$. The spheromak model can also be modelled using a sine geometry and a tan geometry \citep{Scolini2019} similar to the Case~1 and Case~2 geometries defined for the torus model in our case. The range of volume is $10^{29}-10^{31}\;$m$^3$ and $10^{29}-10^{32}\;$m$^3$ for the sine and tan geometries, respectively. The Horseshoe volume profiles of Case~1 and Case~2 are similar up to $\varphi_{hw} = 20\degree$, beyond which Case~2 geometry gives a higher estimate. Depending on the volume estimated from observations of a CME, the choice of geometry and, hence, the volume can be constrained accurately. 
Using the average CME mass density of $10^{-17}\;$kg~m$^{-3}$ at $21.5$~R$_\odot$ as suggested by \citet{Temmer2021}, the torus mass for Case~1 geometry lies in the range $10^{10}-10^{14}\;$kg and for Case~2 in $10^{10}-10^{15}\;$kg. 

Statistical studies \citep{Marubashi2007,Marubashi2015} that reconstructed in situ observations of magnetic clouds at 1~au with toroidal magnetic field configurations suggest CMEs with high aspect ratios greater than 5 (the sample average aspect ratio is $\sim9$) as compared to the estimates from cylindrical configuration fitting. 
It must be noted that these reconstruction techniques estimate the global flux rope properties based on the localised fitting of a part of it. Hence, errors associated with the aspect ratio estimates are possible. We note that due to the expansion of the flux rope during propagation, the aspect ratio can increase. \citet{Vandas2002,Vandas2003} used an aspect ratio of $3-4$ for their toroidal MT CME launched into the solar wind in MHD simulations, which is not as high as the estimates obtained from the in situ reconstructions, and obtain a reasonable magnetic field at 1~au. In EUHFORIA simulations, the CME injection happens at $21.5\;$R$_\odot$, and considering the mentioned studies, it is reasonable to choose lower aspect ratios to constrain the Horseshoe model close to the Sun. 

\subsection{Speed}
Applying a 3D reconstruction model to the white light images, we fit the $\varphi_{hw}$, $\varphi_{hh}$ and the leading edge ($LE$, the bright front). 
\begin{equation}
    LE = T_c + R_0 + a\text{,}
    \label{eqn:LE}
\end{equation}
$T_c$ denotes the centre of the torus. To obtain the total speed ($v_{3D}$) at the CME apex, which is the sum of the radial speed, $v_{rad}$ (rate of change of the $T_c$), and the expansion speed, $v_{exp}$ (rate of change of the torus cross-section), we take the time derivative of the above equation and substitute Eqs.~\ref{eq:case1_bigrad} and \ref{eq:case1_smallrad} for $R_0$ and $a$,
\begin{align}
    v_{3D} = \frac{dLE}{dt} &= \frac{d T_c}{dt} + \frac{dR_0}{dt} + \frac{da}{dt} \\
    &= \Bigg(\frac{d T_c}{dt} + \frac{dR_0}{dt}\Bigg) (1 + \tan(\varphi_{hh}))\\
    &= \frac{d T_c}{dt}(1 + \sin(\varphi_{hw})).
    \label{eq:vLE}
\end{align}
The radial speed is given by 
\begin{equation}
    v_{rad} = \frac{dT_c}{dt} = \frac{v_{3D}}{1 + \sin(\varphi_{hw})}
    \label{eq:vrad}.
\end{equation}

\section{Validation events} \label{sec:validation_events}
In this work, we validate the Horseshoe model with two events. For each event, we present a brief introduction, the details of the parameters constrained by the observations, and the results of the EUHFORIA simulations. The Disturbance storm (Dst) index is computed using the empirical AK2 model of \citet{Obrien2000a,Obrien2000b} on the plasma and magnetic field properties at 1~au. The Dst model itself has some associated errors; hence, we first apply it to the observed data for each event. This serves as the reference modelled Dst. Then, we apply the Dst model to the results of the EUHFORIA simulations and assess the performance of EUHFORIA results in predicting Dst as compared to the reference modelled Dst.

\subsection{Event~1: 12 July 2012} \label{sec:event1}
The aim is to validate the Horseshoe model in predicting the arrival time and magnetic field profile caused due to the impact of CME at 1~au. The first studied event is an isolated, non-interacting, Earth-directed CME. This textbook event of 12 July 2012, triggered by an X1.4 flare from NOAA Active Region (AR) 11520 located at S17E06, is one of the popular geo-effective events \citep{Hu2016,Webb2017,Gopalswamy2018}. The interplanetary propagation of this CME has been previously modelled in the framework of EUHFORIA, using the magnetised CME models -- spheromak \citep{Scolini2019} and the FRi3D \citep{Maharana2022}. A shock driven by the fast CME (with an average projected speed of $890\;$km~s$^{-1}$) arrived at Earth on 14 July 2012 at 17:39~UT. The magnetic cloud signatures were recorded between 15 July 2012 at 06:00~UT and 17 July 2012 at 05:00~UT, with a temporally long negative $B_z$ signature (minimum $B_z=-18\;$nT). The prolonged southward $B_z$ resulted in an intense storm with the main phase starting around 10.00~UT on 15 July 2012 and maintained a $\mathrm{Dst}< -100\;$nT until around 9.00~UT on 16 July 2012 (minimum $\mathrm{Dst}=-122\;$nT on 15 July). The observed in situ signatures are plotted in Fig.~\ref{fig:event1_euhforia} and \ref{fig:event1_euhforia_all} along with the simulation results.

\subsubsection{Observationally constrained parameters for the EUHFORIA simulation}
The 3D reconstruction of the CME, performed using the FRi3D reconstruction tool \citep{Isavnin2016}, is provided in Appendix~\ref{app:event1}. Applying the geometrical transformations given by Eqns.~\ref{eq:case1_bigrad} and \ref{eq:case1_smallrad} to the fitted $\varphi_{hw}=48\degree$ and $\varphi_{hh}=11\degree$, the $a$ and $R_0$ are obtained as 6~R$_\odot$ and 10~R$_\odot$, respectively. \citet{Temmer2021} estimate the mass and density of the CME in Event~1 as $1.84\cdot10^{13}\;$kg and $1.75\cdot10^{-17}\;$kg~m$^{-3}$, respectively, resulting in a volume $3.2\cdot10^{30}\;$m$^{-3}$. Using the constrained $R_0$, $a$, and the uniform $10^{-17}\;$kg~m$^{-3}$ density, the volume of the CME modelled by the Horseshoe model is $1.8\cdot10^{13}\;$kg which is in agreement with the observed estimate. The reconstruction provides us with the $v_{3D}$, which is converted to the $v_{rad}$ using Eq.~\ref{eq:vrad}. From the fitting, the geometrical inclination (unsigned tilt) is 51$\degree$, consistent with the PIL orientation at the source region. The magnetic field properties of the CME could be extracted due to the clear association between the flare and the CME. The chirality is right-handed and the polarity is west to east. The direction of the axial magnetic field by incorporating the polarity and chirality is 129$\degree$ for the Horseshoe model. The signed $\varphi_{p}$ is $4.32\cdot 10^{21}\;$Mx with an error of $0.66\cdot 10^{21}\;$Mx ($15\%$). The $B_0$ is computed using the observed $\phi_{p}$ value in the inverted Eq.~\ref{eq:phi_p_final}. All the parameters used for the EUHFORIA simulation using the Horseshoe model are reported in Table~\ref{tab:event1_euh_params}.

\subsubsection{EUHFORIA simulations}
\begin{figure*}[ht!]
    \centering 
    {\includegraphics[width=0.75\textwidth,trim={0cm 0cm 0cm 0cm},clip=]{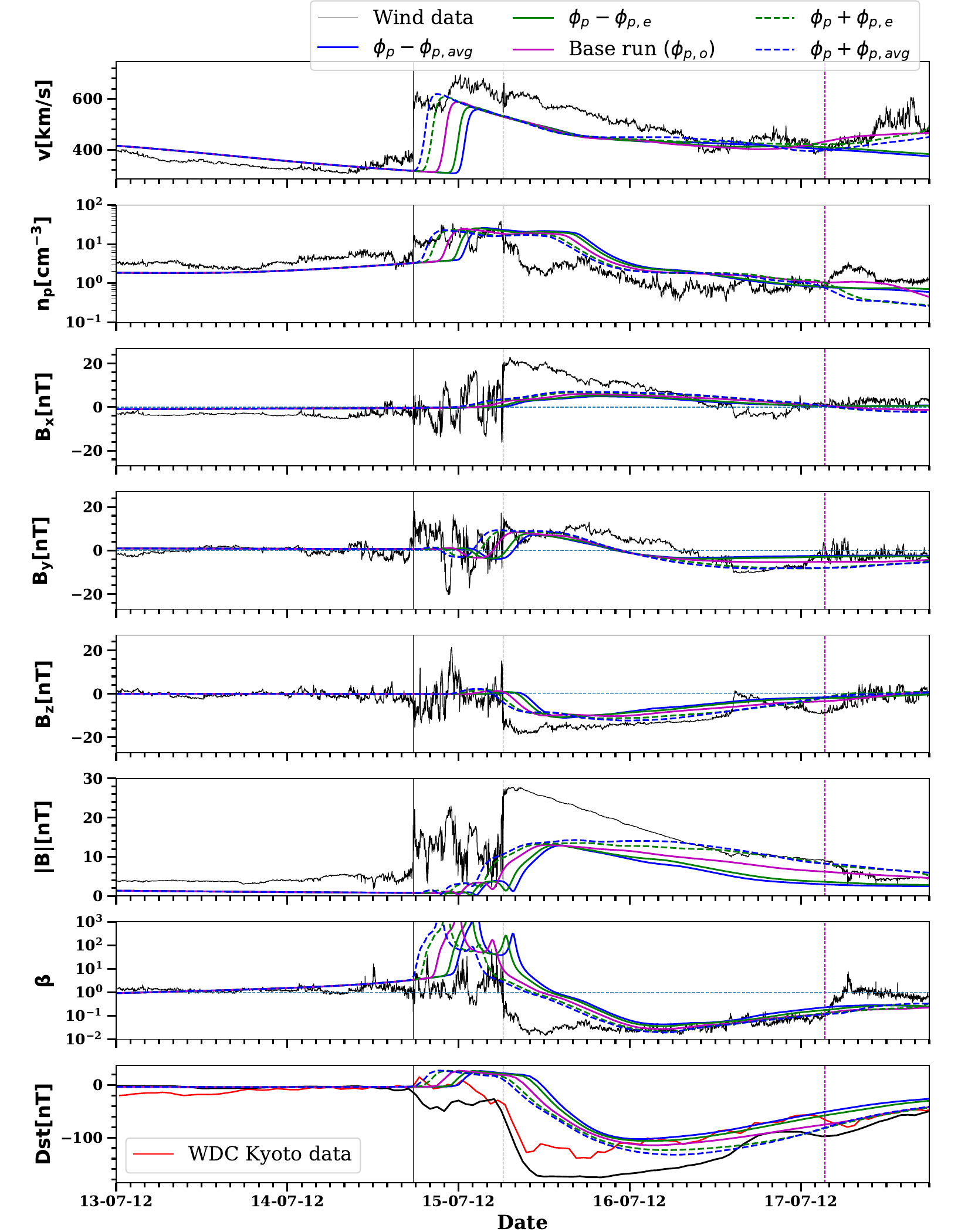}} 
    \caption{Results of Horseshoe ensemble simulations for Event~1 with varying $\phi_p$ obtained using EUHFORIA. Panels from top to bottom: speed ($v$), proton number density ($n_p$), $B_x$, $B_y$, $B_z$, magnetic field strength ($|B|$), plasma beta ($\beta$), and Dst. The observations from the WIND spacecraft are plotted in black in all the plasma and magnetic field properties panels. In the Dst panel, the observed Dst from WDC, Kyoto is plotted in red, the reference modelled Dst using the observed data in black, and the Dst using EUHFORIA simulations in other colours. The solid black vertical line marks the shock arrival time, and the dashed magenta lines show the start and end of the magnetic cloud as reported in the WIND ICME catalogue (\url{https://wind.nasa.gov/ICME_catalog/ICME_catalog_viewer.php}).}
    \label{fig:event1_euhforia}
\end{figure*}

We created an ensemble of simulations based on the varying $B_0$ obtained from $\phi_{p,o}$, $\phi_{p,o}\pm\Delta\phi_{p,e}$ and $\phi_{p,o}\pm\Delta\phi_{p,avg}$. For Event~1, $\phi_{p,o}\pm\Delta\phi_{p,e}$ and $\phi_{p,o}\pm\Delta\phi_{p,avg}$ are associated with $RE=15\%$ and $RE_{avg}=23\%$, respectively. The results of the ensemble run, in comparison with observations (plotted with the black curve), are presented in Fig.~\ref{fig:event1_euhforia}. The first two panels show the speed ($v$) and proton number density ($n_p$), which are qualitatively modelled similarly for all the ensembles with a difference in the arrival time of the shock, characterised by the sharp peak in $v$. Upon interaction with the solar wind, a non-zero Lorentz force ($\mathbf{j} \times \mathbf{B}$) can develop due to local misalignment between the currents and magnetic fields within the evolving flux rope. The higher the magnetic flux, the higher the Lorentz force, leading to enhanced CME acceleration and expansion and earlier arrival time at Earth \citep{Subramanian2009}. Panels 3, 4, 5 and 6 {of Fig.~\ref{fig:event1_euhforia}} show the $B_x$, $B_y$, $B_z$, and the total magnetic field ($|B|$). The magnetic field components show similar behaviour as the observations, i.e., they match qualitatively. The magnetic field profiles show not only the difference in arrival time of the ensemble runs but also the expansion of the magnetic cloud. The distinction between the sheath and the magnetic cloud regions can be made using the plasma beta ($\beta$, shown in panel 7), i.e., the ratio between the plasma pressure and the magnetic pressure. The sheath region is characterised by $\beta \gg 1$, and the magnetic cloud boundary begins when the $\beta$ starts falling towards $1$, and it stretches through the $\beta<1$ region. Finally, panel~8 shows the predicted Dst, computed with the empirical model \citep{Obrien2000a,Obrien2000b} using the EUHFORIA output. The observed Dst provided by the World Data Center for Geomagnetism, Kyoto (\url{http://wdc.kugi.kyoto-u.ac.jp/dstdir/}) is plotted in red, and the reference modelled Dst (computed using observations) is plotted in black. The colourful lines represent the results of the ensemble runs, as mentioned in the legend.

\begin{figure*}[ht!]
    \centering {\includegraphics[width=0.75\textwidth,trim={0cm 0cm 0cm 0cm},clip=]{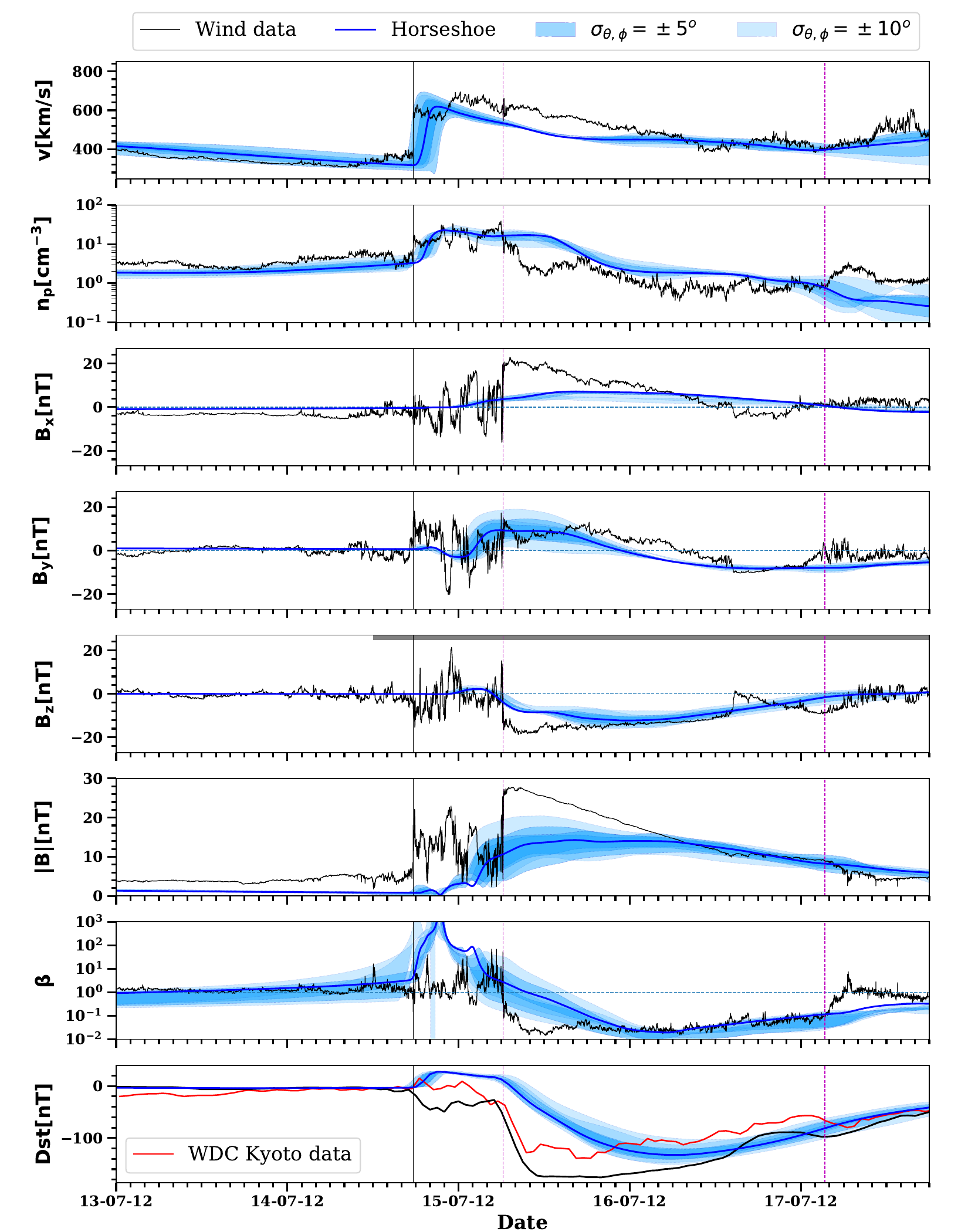}} 
    \caption{Results of the best Horseshoe ensemble simulation of Event~1 obtained using EUHFORIA at Earth (solid blue line) and at virtual satellites in the $5-10\degree$ latitudinal and longitudinal offset around Earth (shaded region). All other plot details are similar to Fig.~\ref{fig:event1_euhforia}. The grey bar at the top of the $B_z$ panel depicts the timespan of the dynamic time warping analysis in Section~\ref{sec:results_discussion}.}
    \label{fig:event1_euhforia_all}
\end{figure*}

The simulation with $\phi_{p,o} + \Delta\phi_{p,avg}$ performed the best in predicting the arrival time 2012-07-14 at about 18:13~UT, which is only 34 minutes later than the observed shock arrival, 
and with also accurately predicted magnetic field profile. The most accurately predicted simulation time profiles of the Horseshoe model are plotted in Fig.~\ref{fig:event1_euhforia_all}. Virtual spacecraft are placed at an offset of latitude and longitude around Earth ($\sigma_{\theta,\phi}$) with the step of $5\degree$, to capture the variability of the plasma and magnetic field properties in the vicinity of Earth. The profiles at the location of these virtual spacecraft are plotted in the shades of dark and light blue for $\sigma_{\theta,\phi}=\pm5\degree$ and $\sigma_{\theta,\phi}=\pm10\degree$, respectively. The percentage difference in minimum $B_z$ is computed as:
\begin{equation}
    \Delta min(B_z) = \frac{min(B_{z,euh}) - min(B_{z,obs})}{min(B_{z,obs})} \times 100,
    \label{eq:relative_Bz}
\end{equation} 
\noindent where $B_{z,euh}$ and $B_{z,obs}$ are the EUHFORIA-simulated and in situ observed time profiles, respectively. 

\begin{figure*}
    \centering
    \subfloat[]{\includegraphics[width=0.45\textwidth,trim={0cm 0cm 0cm 0cm},clip=]{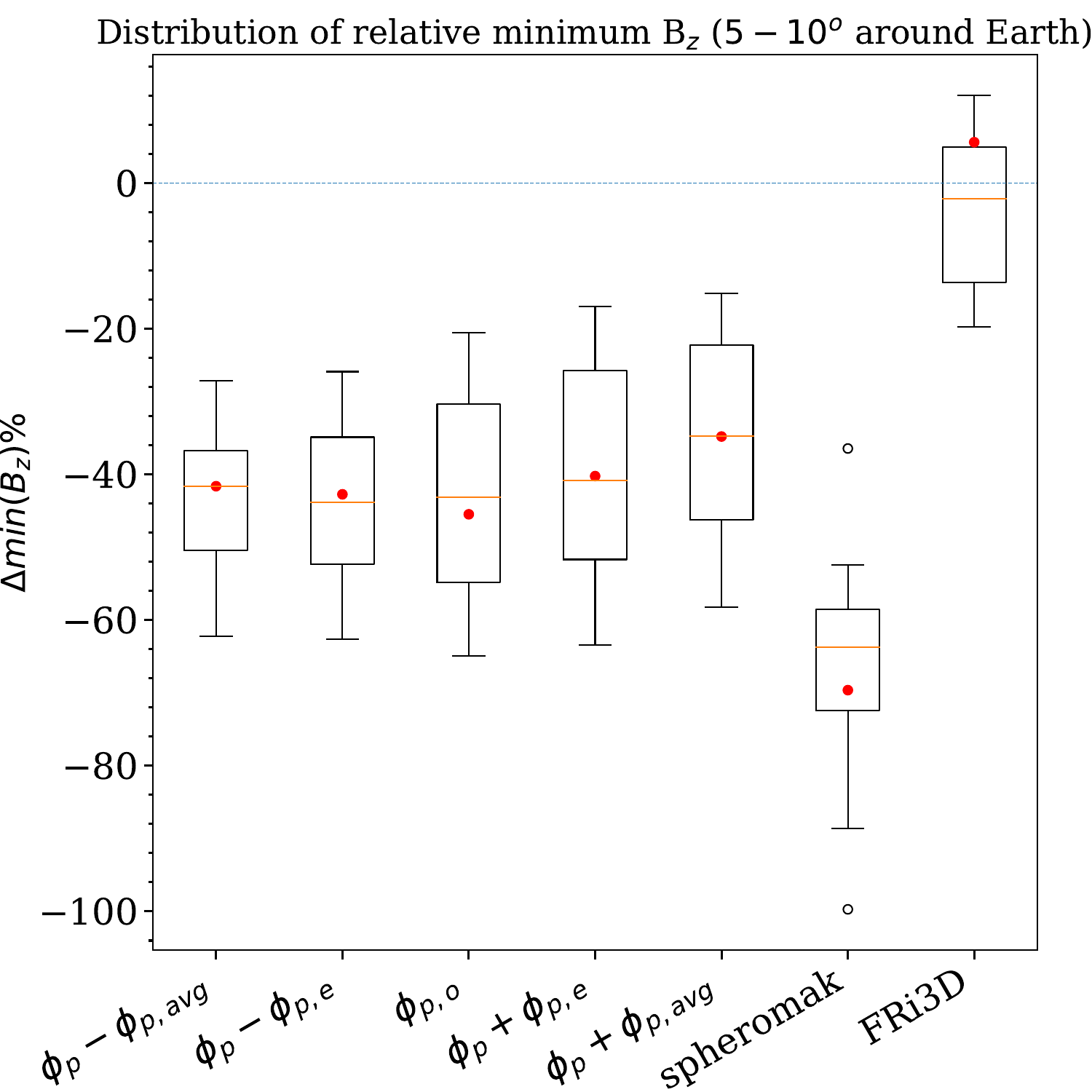}}
    \subfloat[]{\includegraphics[width=0.45\textwidth,trim={0cm 0cm 0cm 0cm},clip=]{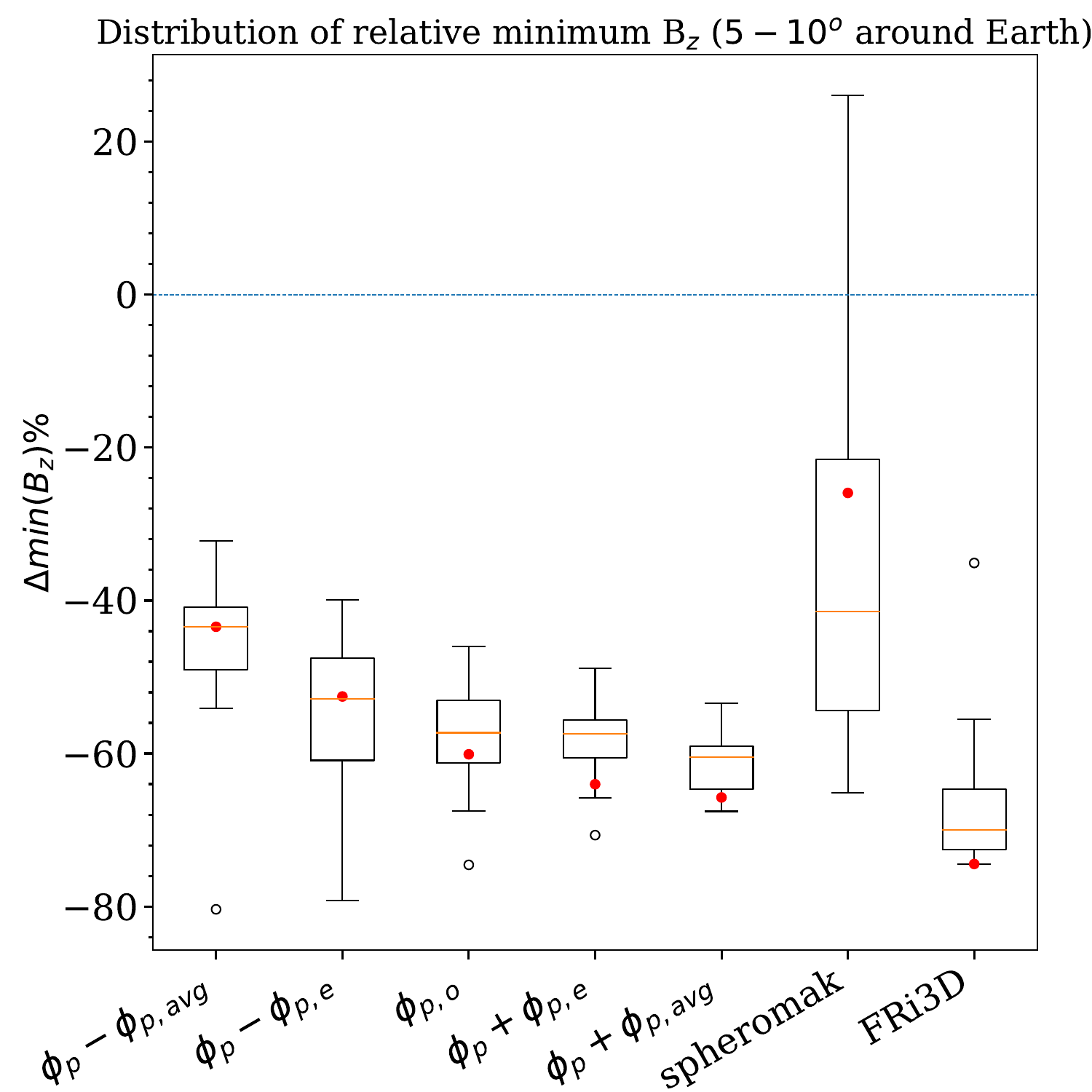}}
    \caption{A boxplot of the spread of the relative minimum $B_z$ ($\Delta min(B_z)$) values at Earth and the virtual spacecraft at an offset of latitude and longitude of 5-10$\degree$ - (a) Event 1, (b) Event 2. The red dot and the orange line in the boxes represent the $\Delta min(B_z)$ at the location of Earth and the median $\Delta min(B_z)$ of the distribution, respectively. The dashed blue line indicates $\Delta min(B_z)=0$ as a reference.}
    \label{fig:events_boxplot}
\end{figure*}

The best Horseshoe simulation results in $\Delta min(B_z) = -35\%$, i.e., the simulated minimum $B_z$ is underestimated by 35$\%$ compared to observations. In the vicinity of Earth (at all the virtual spacecraft), $\Delta min(B_z)$ lies in the range [$-60\%,-15\%$]. For the comparison of the results from different ensemble runs of Event~1, the $\Delta min(B_z)$ at Earth and the virtual spacecraft is plotted for each ensemble run in Fig.~\ref{fig:events_boxplot}(a). The red dot and the orange line in boxes represent the $\Delta min(B_z)$ at the location of Earth and the median $\Delta min(B_z)$ of the distribution, respectively. These values lie within the interquartile range of the data for all the ensemble runs. The distribution of the absolute median $\Delta min(B_z)$ reduces with the increase in $B_0$, which is consistent with the result that increasing $\phi_p$ leads to more negative $B_z$ values (closer to the observations) in the case of Event~1. The absolute median $\Delta min(B_z)$ obtained with the poloidal fluxes $\phi_p-\Delta\phi_{p,avg}$ and $\phi_p-\Delta\phi_{p,e}$ are lower than that of $\phi_{p,o}$. This is an exception to the trend that the absolute $\Delta min(B_z)$ reduces with an increase in $\phi_{p}$. However, the whiskers (extent of the farthest data point from the box) of those two cases still lie within the range of whiskers of the $\phi_{p,o}$ case. This implies that the predicted $B_z$ values at different virtual spacecraft were more localised around the predicted $B_z$ at Earth for those two cases and within the uncertainty range of $\phi_{p,o}$, and hence consistent with the trend mentioned above.

\begin{figure*}[ht!]
    \centering {\includegraphics[width=0.75\textwidth,trim={0cm 0cm 0cm 0cm},clip=]{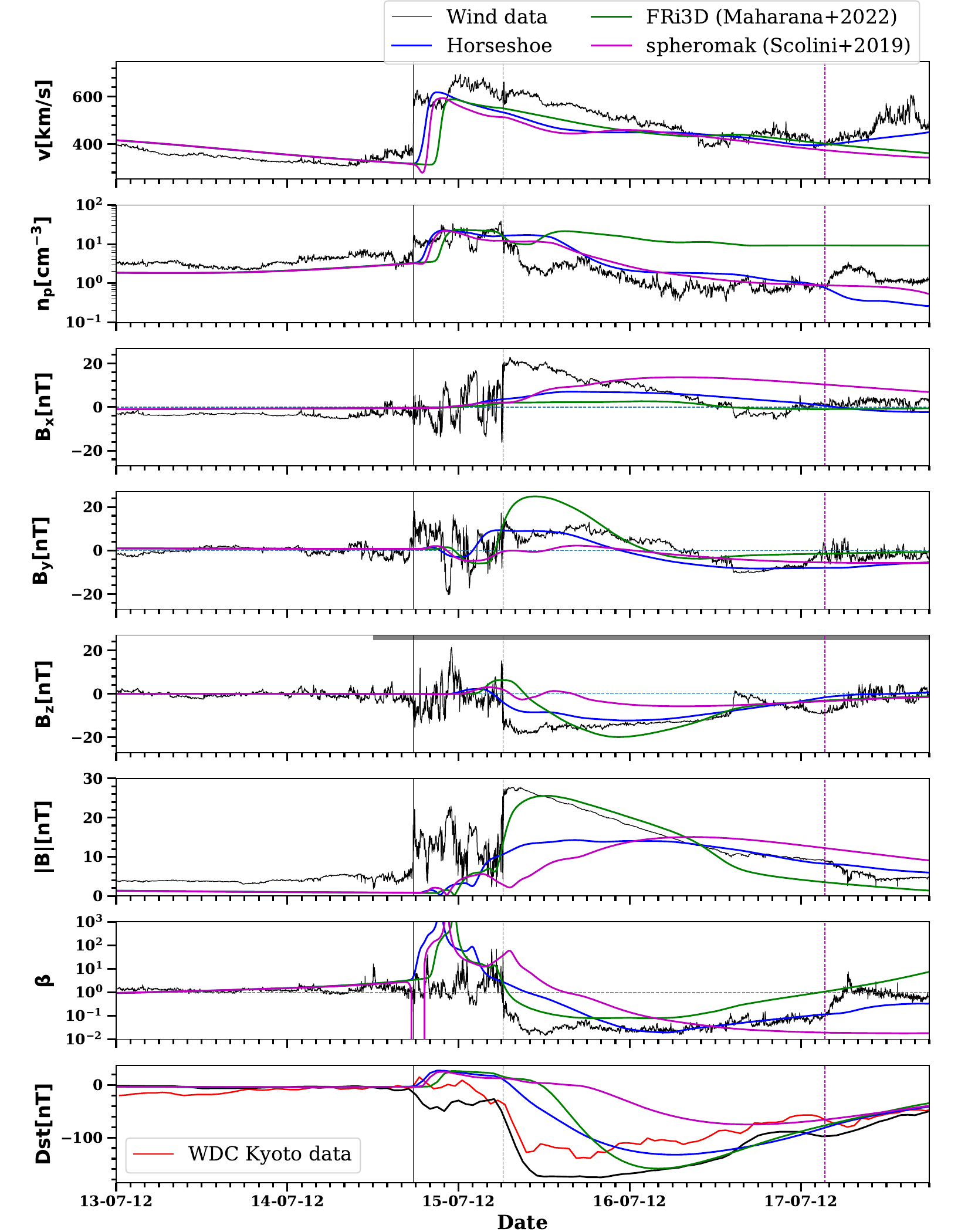}}
    \caption{The comparison of the predicted CME profiles modelled with spheromak, FRi3D and Horseshoe models for Event~1. The description of plot panels is the same as Fig.~\ref{fig:event1_euhforia}. The grey bar at the top of the $B_z$ panel depicts the timespan of the dynamic time warping analysis in Section~\ref{sec:results_discussion}.}
    \label{fig:comp_models_euhforia_event1}
\end{figure*}

To assess the performance of the Horseshoe model, we refer to the results of Event~1 obtained using the spheromak model \citep{Scolini2019} and the FRi3D model \citep{Maharana2022}, as illustrated in Fig.~\ref{fig:comp_models_euhforia_event1}. The quantitative analysis of the performance of the CME models in predicting the full $B_z$ profile at Earth will be discussed in Section~\ref{sec:results_discussion}. We introduce Fig.~\ref{fig:comp_models_euhforia_event1} here to set the context for the minimum $B_z$ analysis. All the Horseshoe ensemble runs perform better than the spheromak model as shown in Fig.~\ref{fig:events_boxplot}(a) in predicting the minimum $B_z$. 
Although the FRi3D model has an absolute median $\Delta min(B_z)$ closer to zero, the positive median $\Delta min(B_z)$ implies that the minimum $B_z$ is overestimated as compared to observations in that case. The best simulation with the Horseshoe model, i.e., the $\phi_p+\Delta\phi_{p,avg}$ case, has overlaps between its first quartile and the third quartile of the spheromak model and between its third quartile and the first quartile of the FRi3D model. This implies that the minimum $B_z$ prediction capability of the Horseshoe model is intermediate between the spheromak and FRi3D models. 
The relevant information regarding the $\Delta min(B_z)$ from the boxplot analysis for Event~1 is provided in Table~\ref{tab:boxplot}.

The last panel of Fig.~\ref{fig:event1_euhforia} shows the results of Dst modelled for the ensemble runs, compared to the measured Dst. The reference modelled $min(\mathrm{Dst}) = -176\;$nT) is overestimated (Dst$_{obs} = -139\;$nT). The $min(\mathrm{Dst})$ modelled by the $\phi_p+\Delta\phi_{p,avg}$ ensemble run ($-133\;$nT) is the best prediction and is underestimated by $24\%$ with respect to the reference prediction. As a comparison, the FRi3D model predicts the $min(B_z)=-159\;$nT and the spheromak model predicts $-75\;$nT, which are underestimated by $9\%$ and $57\%$, respectively, with respect to the reference prediction (Fig.~\ref{fig:comp_models_euhforia_event1}). It can be inferred that the Horseshoe model is an improvement over the spheromak model in predicting the minimum $B_z$ and Dst and that it has the potential to, with further optimisation, match the FRi3D model. 
\begin{table*}
\centering
\begin{tabular}{ p{3.0cm}||p{1.8cm} p{1.8cm} p{1.8cm}}
 \hline
 \hline
 \multicolumn{4}{c}{\textbf{Input parameters}} \\
 \hline
 CME model   &  spheromak & FRi3D & Horseshoe\\
 \hline
   \multicolumn{4}{c}{Geometrical} \\
 \hline
 Insertion time   & 2012-07-12 19:24 & 2012-07-12 19:02 & 2012-07-12 19:28 \\
 Radial speed   & $763$~km~s$^{-1}$ &  $664$~km~s$^{-1}$ & $683$~km~s$^{-1}$\\
 Latitude    & $-8\degree$ & $-8\degree$ & $-6\degree$\\
 Longitude   & $-4\degree$ & $-4\degree$ & $0 \degree$\\
 Half-width  & - & $38\degree$ & $48\degree$\\
 Half-height & - & $36.8\degree$ & $11\degree$\\
 Radius      & $16.8$~R$_\odot$ & - & -\\
 Minor Radius      & - & - & $6$~R$_\odot$\\
 Major Radius      & - & - & $10$~R$_\odot$\\
 Toroidal height & - & $12.29$~R$_\odot$ & - \\
 \hline
    \multicolumn{4}{c}{Deformation} \\
 \hline
 Flattening  & - & $0.3$ & - \\
 Pancaking   & - & $0.44$ & - \\
 Skew        & - & $0$ & - \\
 \hline
 \multicolumn{4}{c}{{Magnetic field}} \\
 \hline
 Chirality    & $+1$ & $-1^*$ & $+1$\\ 
 Polarity    & - & $-1$ & -\\
 Tilt        & $-135 \degree$ & $45 \degree$ & $129 \degree$\\
 Toroidal magnetic flux & $1 \cdot 10^{14}\;$Wb & - & -\\
 Total magnetic flux & $2.4 \cdot 10^{14}\;$Wb & $0.5 \cdot 10^{14}\;$Wb & - \\
 Axial magnetic field strength & - & - & $2.1\cdot10^{-6}\;$nT\\
 Twist       & - & $1.0$ & - \\ 
 \hline
  \multicolumn{4}{c}{{Plasma parameters}} \\
 \hline
 Mass density     & $10^{-18}\;$kg~m$^{-3}$ & $ 10^{-17}\;$kg~m$^{-3}$ & $10^{-17}\;$kg~m$^{-3}$\\
 Temperature & $0.8 \cdot 10^6\;$K & $0.8 \cdot 10^6\;$K & $0.8 \cdot 10^6\;$K\\

 \hline
 \hline
\end{tabular}
\caption{CME input parameters used in EUHFORIA simulations of Event~1 (12 July 2012) employing the spheromak, FRi3D and Horseshoe models. $^*$FRi3D chirality is implemented with an opposite convention, i.e., -1 for right-handedness and +1 for left-handedness.}
\label{tab:event1_euh_params}
\end{table*}

\begin{table*}
\begin{center}
\begin{tabular}{l  l  c  c  c  c  c  c  c}
\hline
\hline
Event & $\Delta min(B_z)$ (in $\%$) & $\phi_p - \phi_{p,avg}$ & $\phi_p - \phi_{p,e}$ & $\phi_p$ & $\phi_p - \phi_{p,e}$ & $\phi_p + \phi_{p,avg}$ & spheromak & FRi3D\\
\hline
\hline
Event~1 & Maximum &  -41 & -47 & -53 & -56 & -59 & -21 & -65 \\
 & Median & -43 & -53 & -57 & -57  & -60 & -41 & -70 \\
 & Minimum&  -54 & -79 & -67 & -66 & -68 & -65 & -74 \\
 & Earth & -43 & -52 & -60 &  -64 & -66 & -26 & -74 \\

\hline
\hline

Event~2 & Maximum &  -37 & -35 & -30 & -26 & -22 & -58 &  5\\
 & Median & -42  & -44 & -43 & -41 & -35 & -64 & -2 \\
 & Minimum &  -62 & -63 & -65 & -63 & -58 & -89 & -19 \\
 & Earth &  -42 & -43 & -45 & -40 & -35 & -70 & 6 \\

\hline
\hline
\end{tabular}
\end{center}
\caption{The maximum, median, and minimum values of the $\Delta min(B_z)$ distribution (in $\%$) over different virtual spacecraft including at the location of Earth, and the $\Delta min(B_z)$ at Earth are tabulated for the ensemble runs of the Horseshoe model, spheromak model and FRi3D model. This data is provided for both Event~1 and Event~2. Here, the absolute maximum value of the distribution in all the cases is closer to zero, i.e., they correspond to the virtual spacecraft with the best fit of the modelled $min(B_z)$ to observations.}
\label{tab:boxplot}
\end{table*}

\subsection{Event~2: 10 September 2014} \label{sec:event2}
This event involves two successive CMEs erupting between 8-10 September 2014. On 8 September 2014, around 23:12~UT, the first CME (hereafter, CME1) was associated with the M4.6 flare originating from the NOAA AR 12158. The second CME (hereafter, CME2) erupted on 10 September 2014 at 17:21~UT from the same AR 12158, and it was associated with the X1.6 flare. Both CMEs were detected by C2 and C3 instruments onboard LASCO and COR2B instruments onboard STEREO-B. Previous studies have addressed the details of CME eruption, propagation, and geo-effectiveness of the CME2 in detail \citep{Cho2017,Webb2017,An2019,Maharana2023}. The arrival of CME1 was not considered in these studies as it was just a flank hit at Earth and was not reported in ICME catalogues \citep[e.g.,][]{Richardson2010,Nieves2018}. However, CME2 was a candidate of the ISEST VarSITI campaign\footnote{\url{http://solar.gmu.edu/heliophysics/index.php/ISEST}} and was used to perform the exercise of real-time forecasting. Due to its head-on impact with a strong negative $B_z$, the CME2 was forecasted to cause severely disturbed geomagnetic conditions. However, it arrived at Earth with a positive $B_z$ in its magnetic cloud and a negative $B_z$ in the sheath ahead, causing only moderate storm (Dst$\sim–88\;$nT) conditions. Later, a thorough analysis of the remote observations of the source region \citep{Vemareddy2016a,Dudik2016,Zhao2016} and the in situ measurements of the interplanetary CME \citep{Marubashi2017,Kilpua2021,An2019} suggested a possible northward deflection and significant rotation of the CME in the low corona which was not captured in observations. A close-to-flank passage of the flux rope at Earth, hence, led to erroneous space weather predictions. The comprehensive story of the erroneous space weather prediction associated with this event (CME1+CME2) is provided in \citet{Maharana2023}. These authors analysed the orientation of CME2 at the source region, close to 0.1~au and at 1~au and hypothesised that the rotation must have happened in the low corona (see Fig.~8 and associated text). The authors corroborated the claim with simulations and reproduced the negative $B_z$ in the sheath created due to the interaction between CME1 and CME2. 

The shock driven by CME1 (S1) was observed on 11 September at 22:50~UT based on the IPShock catalogue\footnote{IPShock catalogue – \url{http://ipshocks.fi/}} \citep{Kilpua2015}. The start of the magnetic ejecta associated with CME1 (ME1) was identified on 12 September at 8:45~UT. It continued until the shock associated with CME2 (SE2) arrived at L1 on 12 September at 15:17~UT as per the WIND ICME catalogue\footnote{WIND ICME catalogue – \url{https://wind.nasa.gov/ICME_catalog/ICME_catalog_viewer.php}}. The magnetic ejecta associated with CME2 (ME2) passes through L1 between 12 September at 21:36 UT \citep[corrected by ][]{Maharana2023} and 14 September at 11:38~UT, as reported in the WIND ICME catalogue. 

In \citet{Maharana2023}, this event was modelled using the combination of the spheromak and FRi3D models for CME1 and CME2, respectively. This is because combining two spheromaks resulted in erroneous $B_z$ profiles at Earth, and using the FRi3D model for two successive CMEs was numerically challenging. 
For the consistent analysis of the performance of different flux rope models in EUHFORIA, we performed a simulation where CME1 was modelled with spheromak, and CME2 was modelled with the horseshoe model. We keep the parameters of CME1 the same as in \citep{Maharana2023} and focus on optimising the Horseshoe model parameters of CME2 through a parametric study by varying the $\phi_p$ in the error range. In addition, we provide results of the first attempt to model two successive CMEs with the Horseshoe model. The Horseshoe parameters for CME1 will be further optimised in future studies as there is much scope for improvement. 

\subsubsection{Observationally constrained parameters for EUHFORIA simulation}
Similar to Event~1, we performed the 3D reconstruction of CME2 with the FRi3D reconstruction tool to fit a toroidal geometry. The details can be found in Appendix~\ref{app:event2}. Our fitting of CME2 for constraining the Horseshoe model is consistent with the parameters obtained by \citet{Maharana2023}, who derived the parameters for constraining the FRi3D model. Overall, the CME is fitted to be thinner (low $\varphi_{hh}$), keeping in mind the toroidal geometry for the Horseshoe model, as compared to the reconstructed $\varphi_{hh}$ for the spheromak and FRi3D models. We obtain $R_0 = 10\;R_\odot$ and $a=6\;R_\odot$ applying the geometrical transformations given by Eqns.~\ref{eq:case1_bigrad} and \ref{eq:case1_smallrad} on $\varphi_{hw}=52\degree$ and $\varphi_{hh}=12\degree$. The $v_{rad}$ is obtained from the reconstructed $v_{3D}$ using Eq.~\ref{eq:vrad}.
The $\phi_{p,o}$ values are taken from the \citet{Kazachenko2023} catalogue. The unsigned $\varphi_p$ for CME2 is $1.22\cdot 10^{22}\;$Mx with an error $1.21\cdot 10^{21}\;$Mx ($10\%$). Considering the left-handed chirality and the E-W polarity of the CME2 inferred by \citet{Maharana2023} and the geometrical orientation obtained from the reconstruction, we use a tilt of $-35\degree$ for the Horseshoe model. The parameters of CME1 are the same as when modelled by spheromak in \citet{Maharana2023}. 
We note that the CME1 was shifted in longitude towards the Sun-Earth line to model the interaction with CME2 in EUHFORIA. Since the spheromak has no legs, the shift was necessary to simulate the flank encounter \citep{Maharana2023}. Further, we also constrained CME1 for the Horseshoe model. Applying the similar methodology as for CME2, we obtain $R_0 = 7\;R_\odot$ and $a=7\;R_\odot$. CME1 has an unsigned poloidal flux $\varphi_p=8.4\cdot 10^{21}\;$Mx with an error $1.58\cdot 10^{21}\;$Mx, and a left-handed chirality. The parameters of the Horseshoe model used for CME2 are listed in Table~\ref{tab:event2_euh_params} along with the spheromak and FRi3D parameters for comparison. 

\subsubsection{EUHFORIA simulations}
\begin{figure*}[ht!]
    \centering {\includegraphics[width=0.75\textwidth,trim={0cm 0cm 0cm 0cm},clip=]{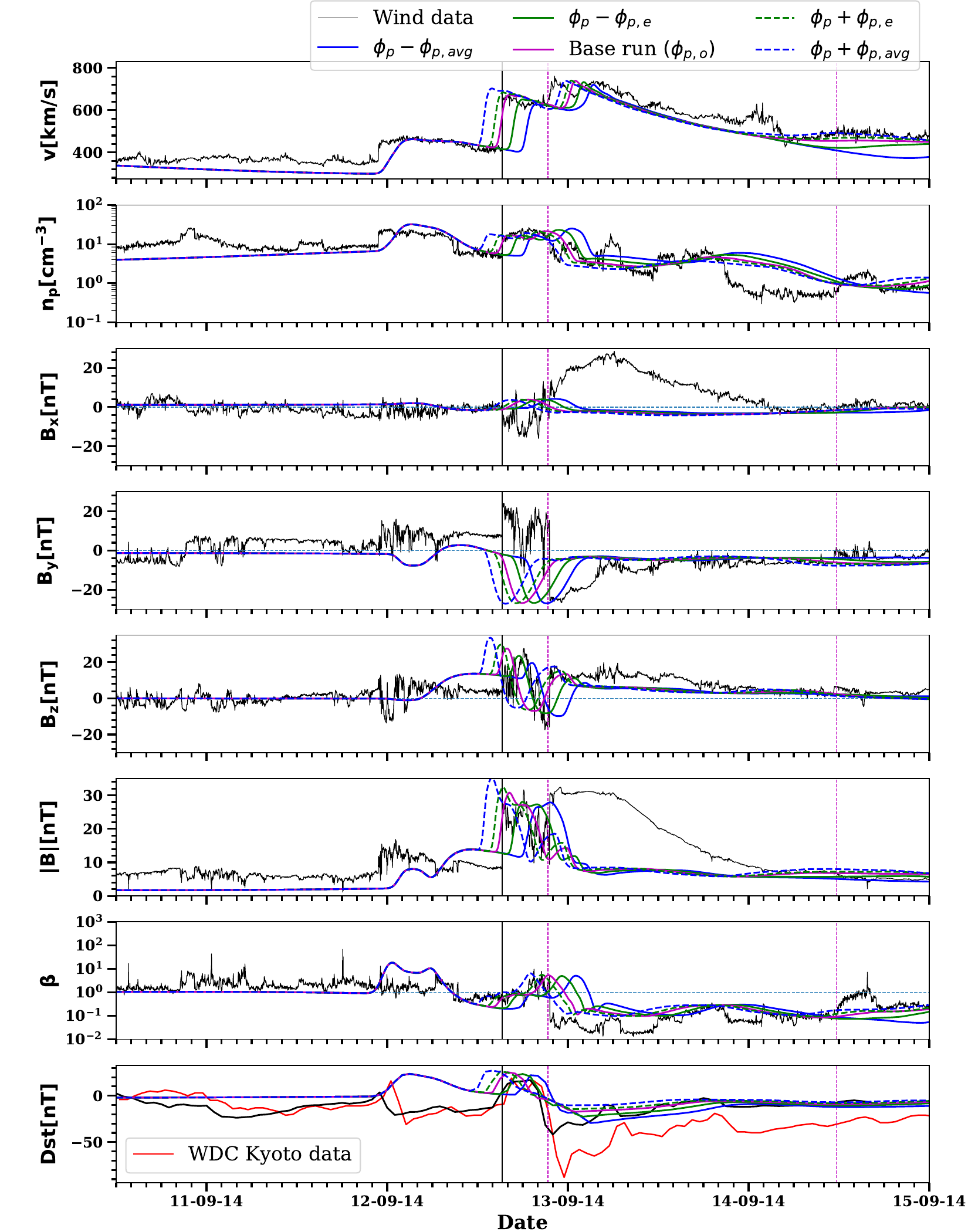}}
    \caption{Results of Horseshoe ensemble simulations for Event~2 with varying $\phi_p$ obtained using EUHFORIA. 
    The detailed plot description is the same as for Fig.~\ref{fig:event1_euhforia}.}
    \label{fig:event2_euhforia}
\end{figure*}

\begin{figure*}[ht!]
    \centering {\includegraphics[width=0.75\textwidth,trim={0cm 0cm 0cm 0cm},clip=]{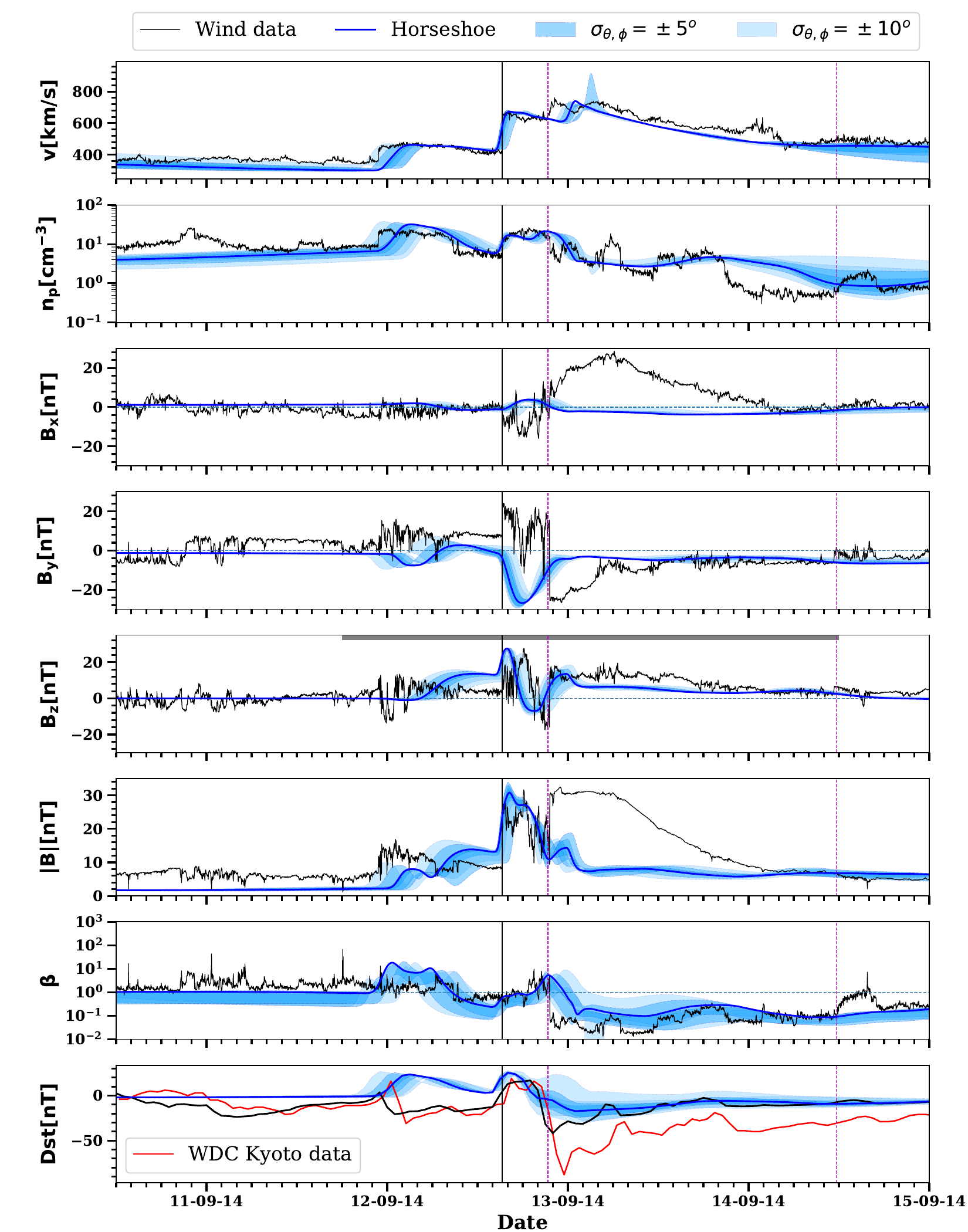}} 
    \caption{Results of the best Horseshoe ensemble simulation of Event~1 obtained using EUHFORIA.
    The detailed plot description is the same as for Fig.~\ref{fig:event1_euhforia_all}. The grey bar at the top of the $B_z$ panel depicts the timespan of the dynamic time warping analysis in Section~\ref{sec:results_discussion}.}
    \label{fig:event2_euhforia_all}
\end{figure*}

In this section, we discuss the results of the ensemble modelling of CME2 based on changing the $B_0$ parameter and keeping the properties of CME1 constant, as shown in Figure~\ref{fig:event2_euhforia}. The arrival time and $n_p$ of the ensembles qualitatively match the observations. CME arrives earlier for the higher poloidal flux, $\phi_p$, values due to the higher Lorentz force, a trend similar to as seen for Event~1. All ensemble modelling results show an underestimation in the $B_x$ and $B_y$ profiles of ME2. The $B_z$ inside ME2 is positive and increases with the increase in $\phi_p$. There is an overestimation of $B_y$ in the sheath ($\beta\gg1$ region before the magnetic cloud). On the other hand, the coherent positive and negative features of the $B_z$ component in the sheath region are well reproduced by the model. The negative value of the $B_z$ component, during this event, is formed in the sheath of CME2 due to the interaction between CME1 and CME2 when the CME2 compresses the negative $B_z$ part of ME1 in the trailing part of CME1. We noticed that the higher the difference in the speed of the CMEs, the more negative the minimum $B_z$. When the CME2 is slower (low $\phi_p$), it creates a continuous compression of CME1 over a longer period before reaching Earth. Hence, if we want to model the most negative minimum $B_z$, then we predict a delayed arrival of CME2. So, there is a trade-off between obtaining the best arrival time or the minimum $B_z$ for this case based on varying just $\phi_p$. In future work, a more comprehensive parameter study must be performed by varying the $\varphi_{hh}$ and $\varphi_{hw}$ to check the contribution of the extent and mass of the CME2 in the formation of the compressed features in the sheath ahead of it. For the current study, we choose, as the best case, the base run with a poloidal flux $\phi_{p,0}$ as it models the most accurate arrival time and reasonably well the negative $B_z$ in the sheath. 
The time series plot of the best case is shown in Fig.~\ref{fig:event2_euhforia_all} with the spatial variability of the plasma and magnetic field at $\sigma=\pm5\degree$ and $\sigma=\pm10\degree$ around Earth. 

The boxplots of the $\Delta min(B_z)$ (viz.\ Eqn.~\ref{eq:relative_Bz}) for the Horseshoe ensemble runs, along with the simulations of CME2 performed with the spheromak and the FRi3D models \citep{Maharana2023} are presented in Fig.~\ref{fig:events_boxplot}(b). A comparison of the performance of the three magnetised CME models in modelling Event~2 is provided in Section~\ref{sec:results_discussion}. The trends are opposite to the one obtained for Event~1, as the lowest $\phi_p$ results in the least $\Delta min(B_z)$ as explained above. The range of $\Delta min(B_z)$ of the spheromak model is extended over [$-65\%$, $25\%$], but it does not mean that the model captures the observed minimum negative $B_z$. It is because of the wrongly predicted $B_z$ component using the spheromak model, possibly due to spheromak rotation in the heliosphere \citep{Asvestari2022}. All the ensembles of the Horseshoe model must be rather compared with the FRi3D simulation as the $B_z$ profile is more correctly modelled by it. For the best run, $\Delta min(B_z)$ lies in the range [$-68\%$, $-46\%$] at the position of all the virtual spacecraft and is $-60\%$ at Earth, meaning that $min(B_z)$ is underestimated everywhere. The boxplots of the Horseshoe ensembles overlap with that of the spheromak and the FRi3D models and imply that the Horseshoe model performs better than both models in predicting the $min(B_z)$ in the case of Event~2 (Fig.~\ref{fig:events_boxplot}(b)). The reference modelled $min(\mathrm{Dst}) = -47\;$nT) is underestimated ($min(\mathrm{Dst}_{obs}) = -88\;$nT). The $min(\mathrm{Dst})$ modelled with the best Horseshoe ensemble run ($\phi_{p,0}$ case) is $-17\;$nT which is an underestimation by $63\%$ as compared to the reference prediction. The Horseshoe result is close to that of the FRi3D prediction of $-18\;$nT (underestimated by $61\%$ compared to the reference model). The relevant information about the $\Delta min(B_z)$ analysis using boxplots for Event~2 is summarised in Table~\ref{tab:boxplot}. \\

\begin{table*}
\centering
\begin{tabular}{ p{3.0cm}||p{1.8cm} | p{1.8cm} p{1.8cm} p{1.8cm}}
 \hline
 \hline
 \multicolumn{5}{c}{\textbf{Input parameters}} \\
 \hline
 CME & CME1 &  & CME2 & \\
 \hline
 CME model & spheromak & spheromak & FRi3D & Horseshoe\\
  \hline
  \multicolumn{5}{c}{Geometrical} \\
 \hline
 Insertion time   & 2014-09-09 20:14~UT  & 2014-09-10 20:39~UT & 2014-09-10 20:14~UT & 2014-09-10 20:04~UT\\
 Radial speed   & $450$~km~s$^{-1}$ & $719$~km~s$^{-1}$ &  $500$~km~s$^{-1}$ & $607$~km~s$^{-1}$\\
 Latitude    & $22\degree$ & $0\degree$ & $24\degree$ & $30\degree$\\
 Longitude   & $-14\degree$ & $23\degree$ & $15\degree$ & $7\degree$\\
 Half-width  & - & - & $50\degree$ & $52\degree$\\
 Half-height & - & - & $30\degree$ & $12\degree$\\
 Radius      & $21$~R$_\odot$ & $20$~R$_\odot$ & - & -\\
 Minor Radius      & - & - & - & $6$~R$_\odot$\\
 Major Radius      & - & - & - & $10$~R$_\odot$\\
 Toroidal height & - & - & $13.6$~R$_{\odot}$ & \\
\hline
 \multicolumn{5}{c}{{Magnetic field}} \\
 \hline
 Chirality   & $-1$ & $+1$$^*$ & $-1$\\ 

 Polarity    & - & - & $+1$ & - \\
 Tilt        & $-135\degree$ & $-45\degree$ $^{**}$ & $45\degree$ & $-35\degree$\\
 Toroidal magnetic flux & $0.5\cdot 10^{14}\;$Wb & $1\cdot 10^{14}\;$Wb & - & \\
 Total magnetic flux & - & - & $5\cdot10^{13}\;$Wb & \\
 Axial magnetic field strength & - & - & - & $2.3\cdot10^{-6}$~T\\
 Twist       & - & - & $1.5$ & - \\
 \hline
 \multicolumn{5}{c}{{Deformation}} \\
 \hline
 Flattening  & - & - & $0.5$ & -\\
 Pancaking   & - & - & $0.5$ & -\\

\hline
 \multicolumn{5}{c}{{Plasma parameters}} \\
\hline
 Mass density     & $10^{-18}\;$kg~m$^{-3}$ & $10^{-18}\;$kg~m$^{-3}$ & $10^{-17}\;$kg~m$^{-3}$ & $10^{-17}\;$kg~m$^{-3}$\\
 Temperature & $0.8 \cdot 10^6\;$K & $0.8 \cdot 10^6\;$K & $0.8 \cdot 10^6\;$K & $0.8 \cdot 10^6\;$K \\
 \hline
 \hline
\end{tabular}
\caption{CME parameters used in EUHFORIA simulations of Event~2 employing the spheromak model for CME1, and spheromak, FRi3D and Horseshoe models for CME2.}
\label{tab:event2_euh_params}
\end{table*}

\begin{figure*}[ht!]
    \centering {\includegraphics[width=0.75\textwidth,trim={0cm 0cm 0cm 0cm},clip=]{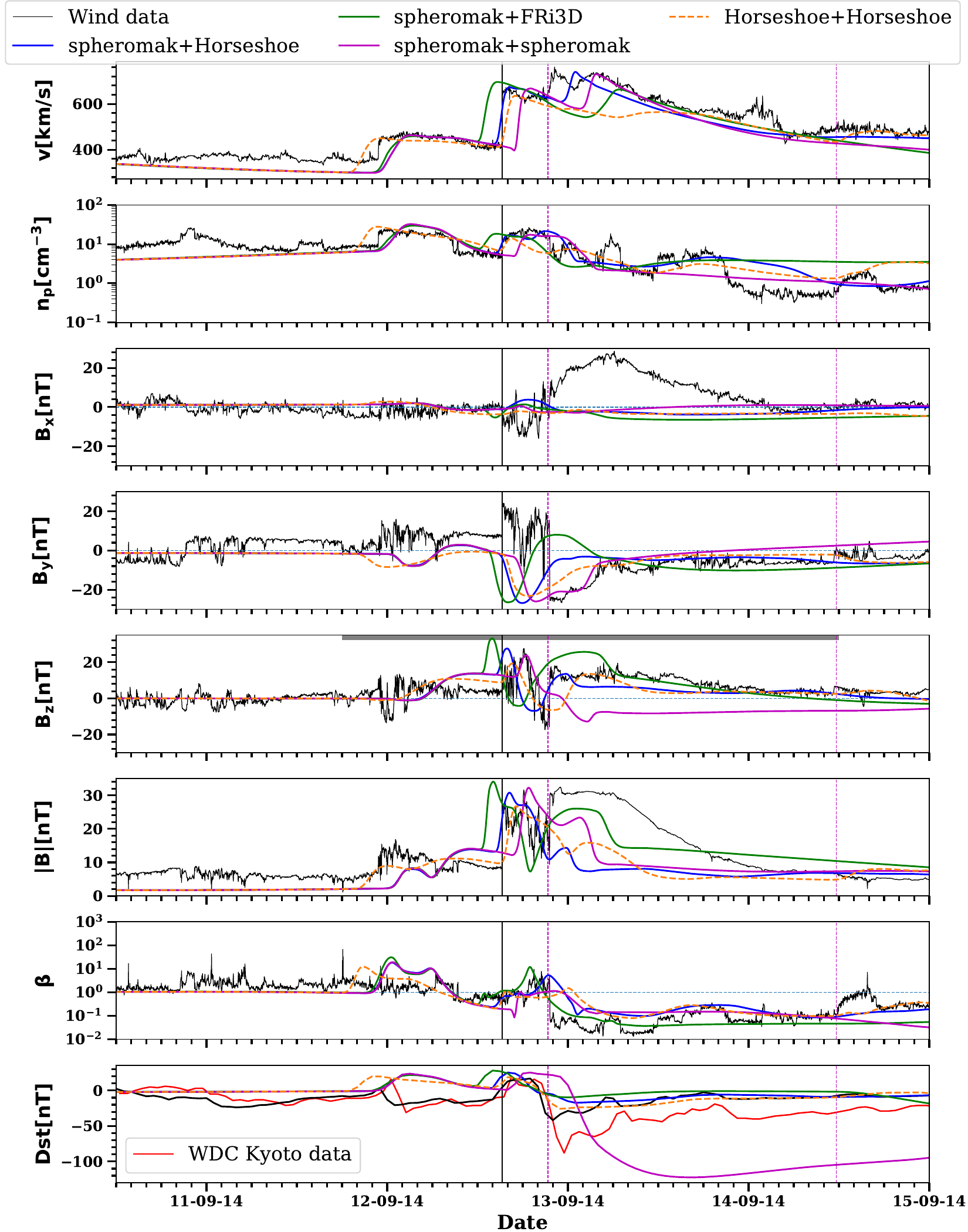}} 
    \caption{The comparison of the predicted CME profiles modelled with spheromak, FRi3D and Horseshoe models for Event~2 (CME1 modelled with spheromak). An additional simulation with both CME1 and CME2 modelled with the Horseshoe model is provided (dashed orange line). The description of plot panels is the same as Fig.~\ref{fig:event1_euhforia}. The grey bar at the top of the $B_z$ panel depicts the timespan of the dynamic time warping analysis in Section~\ref{sec:results_discussion}.}
    \label{fig:comp_models_euhforia_event2}
\end{figure*}

\noindent {\bf Horseshoe+Horseshoe simulation: } 
With Event~2 we validated the usage of the Horseshoe model in combination with other magnetised CME models. We go a step further to model both CMEs with the Horseshoe model. The location of the CME source region (inferred from the 3D reconstruction) is kept the same as in the spheromak model. The geometrical and the $B_0$ values are constrained for the Horseshoe model as per methodologies described in Section~\ref{sec:constrain_params}. The results are plotted in Fig.~\ref{fig:comp_models_euhforia_event2} (orange dashed line). The CME1 shock arrival is predicted by the Horseshoe+Horseshoe simulation is $2$~hours before the observed arrival time. Whereas, in the case of the spheromak+Horseshoe simulation, the arrival time was delayed by $\sim2$~hours. The Horseshoe+Horseshoe simulation predicts the Dst (see the last panel of Fig.~\ref{fig:comp_models_euhforia_event2}) better than the spheromak+Horseshoe case. These results highlight the possibility of using the Horseshoe model for the modelling of the successive CMEs. This mitigates the limitation of the FRi3D model to numerically inject two successive CMEs and improves the prediction of magnetic field components as compared to the spheromak model. The Horseshoe simulations will be further optimised in future studies, and more work will be carried out to model successive CMEs.

\section{Analysis and discussion} 
\label{sec:results_discussion}

In this section, the evaluation of the performance of the three magnetised CME models (spheromak, FRi3D, and Horseshoe) is done based on three criteria. First, we assess the ability of the models to predict the arrival time of the CME shock. Further, we evaluate the accuracy of the models based on their $B_z$ prediction capabilities and, finally, based on the speed of their numerical computations. Metrics are a way to quantitatively assess which model provides the best results for improving space weather forecasting. We begin with identifying metrics for evaluation based on criterion~2. Metrics like the mean absolute error (MAE) or the root mean square error (RMSE) provide the averaged uncertainties for the whole time series. Whereas advanced metrics like Dynamic Time Warping \citep[DTW;][]{Keogh2001,Gorecki2013,Laperre2020,Samara2022} determine how similar two-time series are by performing optimal temporal alignments between the common features \citep{Müller2007}. 

To accurately predict the geomagnetic indices, it is crucial to obtain not only the minimum negative $B_z$ caused by a CME impact but also the commencement and the duration of the negative $B_z$ values. Hence, we employ the DTW technique to assess the accuracy of the EUHFORIA-modelled $B_z$ time profiles compared to the observations. The DTW technique is a distance measure similar to the Euclidean distance. It estimates the similarities between two time series containing similar patterns but differing in time. This technique has been previously applied for assessing the EUHFORIA-modelled solar wind speed profiles with observations by \citet{Samara2022}, and the Dst computed using EUHFORIA modelled data \citet{maharana2024}. We employ the open-source code developed by \citet{Samara2022} \footnote{\url{https://github.com/SamaraEvangelia/DTW\_
ForSolarWindEvaluation}} for our analysis.

We first apply DTW between the observed $B_z$ data from WIND and the $B_z$ modelled by employing different CME models in EUHFORIA. We ensure the following for the correct application of the algorithm: (a) the first and the last points of one sequence are matched with the first and last points of the other; (b) the mapping is monotonically increasing in time; and (c) there is no data gap, i.e., every point in the two sequences is matched with at least one point in the other. To limit the ``pathological alignment problem'' that creates singularities, i.e., when one point of a sequence is matched with multiple points in the other, we apply a time `window' for the alignment. Windowing restricts the mapping of the points to a certain time window and restricts the number of singularities. We also smooth the observed 1-minute cadence data at L1 (containing high-frequency fluctuations) optimally to match the trends in the smooth-modelled data better. The fluctuations serve as local minima and maxima and influence the DTW results by creating more singularities and increasing the DTW cost. However, the limit of smoothing should be carefully determined not to miss the important features in the data. To apply the algorithm, first, the DTW cost matrix is computed based on the following equation: 
\begin{equation}
\centering
    D(i,j) = \delta(s_{i},q_{i})+\hbox{min}\{D(i-1, j-1), D(i-1, j), D(i, j-1)\},
\label{eq:DTW}
\end{equation}
\noindent where $D(i,j)$ is the cumulative DTW cost or distance, and $\delta(s_{i},q_{i}) = |s_{i} - q_{i}|$ corresponds to the Euclidean distance between the point $s_{i}$ from one time series and the point $q_{i}$ from the other time series. The first element of the array $D(0,0)$ is equal to $\delta(s_{0},q_{0})$. The last element of this cost matrix, the DTW score, is presumed to be a quantification of alignment between the two sequences. To evaluate the performance of different sequences with respect to a single time series, we calculate the sequence similarity factor (SSF) for each DTW analysis. SSF quantifies how good the modelled result is compared to an ideal (observations) and a non-ideal ($B_z = 0$, reference) prediction scenario. It is the ratio between the DTW score of the observed and modelled $B_z$ time series and the DTW score between the observed and reference scenario time series. Namely, it is defined as:
\begin{linenomath}
\begin{equation}
    \text{SSF} = \frac{DTW_{score}(O, M)}{DTW_{score}(O, N)}, \ \text{SSF} \in [0, \infty),
\end{equation}
\end{linenomath}
\noindent where $O$, $M$, and $N$ represent the observed, modelled, and non-ideal cases, respectively. We evaluate the CME models by their SSF values for each validation event. In addition, we present, for each event, the plots of the DTW alignment maps, the histograms representing the distribution of the time differences ($\Delta t$) and amplitude differences ($\Delta B_z$) between the observed and the modelled sequences ($\Delta~=~$Observed - Modelled).

To assess the performance of the CME models based on criterion~3, we compare the computational time of the simulations. As the goal is to improve the models for operational space weather purposes, optimising the computational time is necessary. For consistency, all the simulations were performed using EUHFORIA (ver 2.0) on the wICE cluster of the Vlaams Supercomputer Centrum ({\url{http://www.vscentrum.be}}) utilising two nodes with 72 cores per node (144 parallel processes).

\subsection*{Event~1}
\begin{figure*}[ht!]
    \centering 
    \subfloat[]{\includegraphics[width=0.5\textwidth,trim={0cm 0cm 0cm 0cm},clip=]{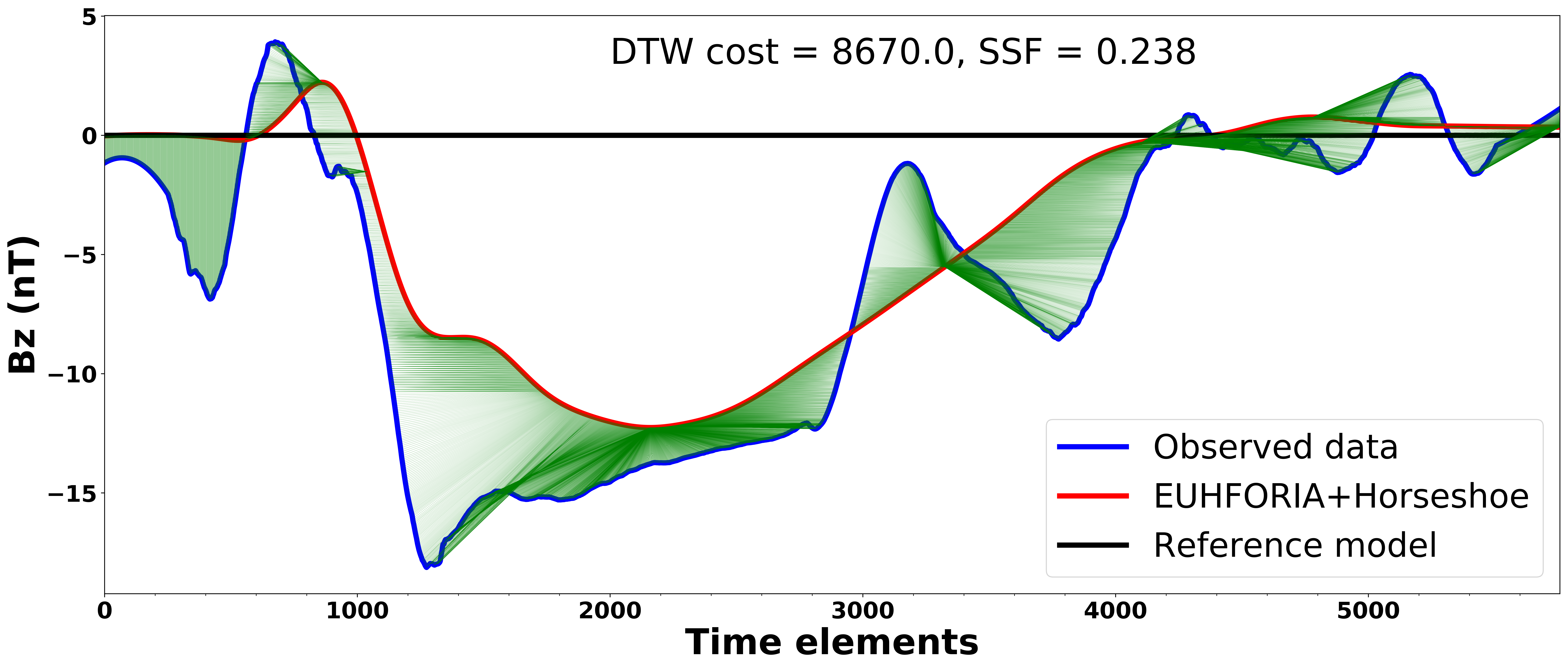}}
    \subfloat[]{\includegraphics[width=0.25\textwidth,trim={0cm 0cm 0cm 0cm},clip=]{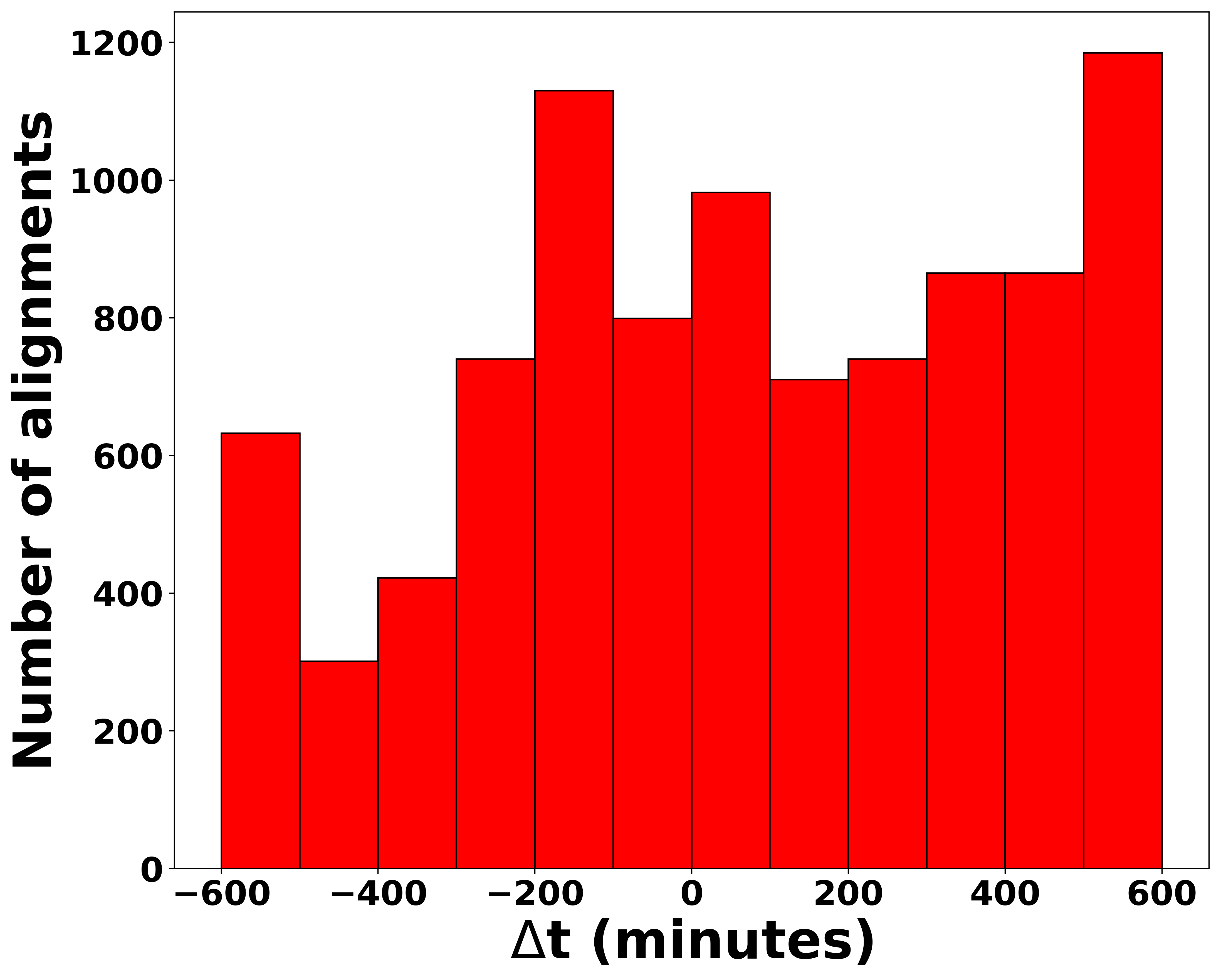}} 
    \subfloat[]{\includegraphics[width=0.25\textwidth,trim={0cm 0cm 0cm 0cm},clip=]{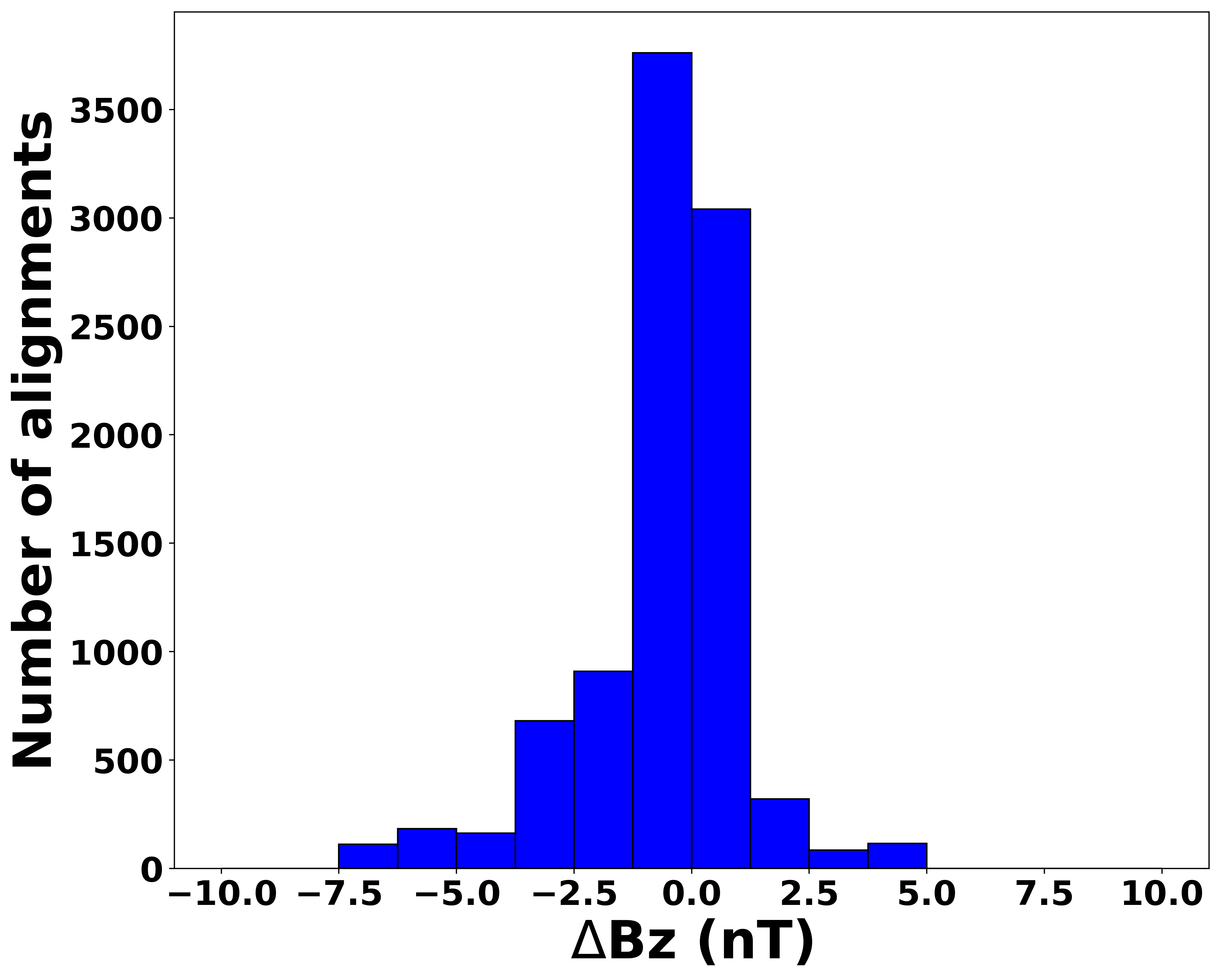}}\\
    \subfloat[]{\includegraphics[width=0.5\textwidth,trim={0cm 0cm 0cm 0cm},clip=]{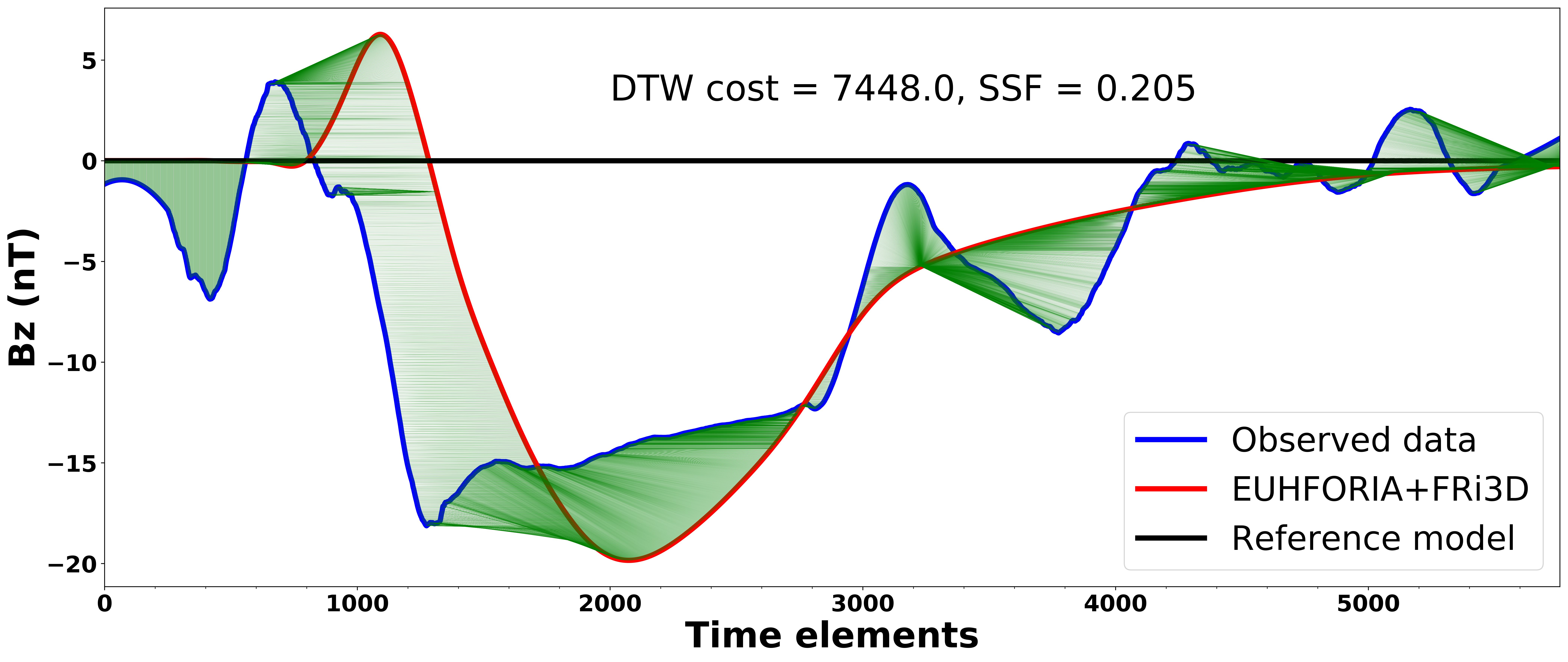}}
    \subfloat[]{\includegraphics[width=0.25\textwidth,trim={0cm 0cm 0cm 0cm},clip=]{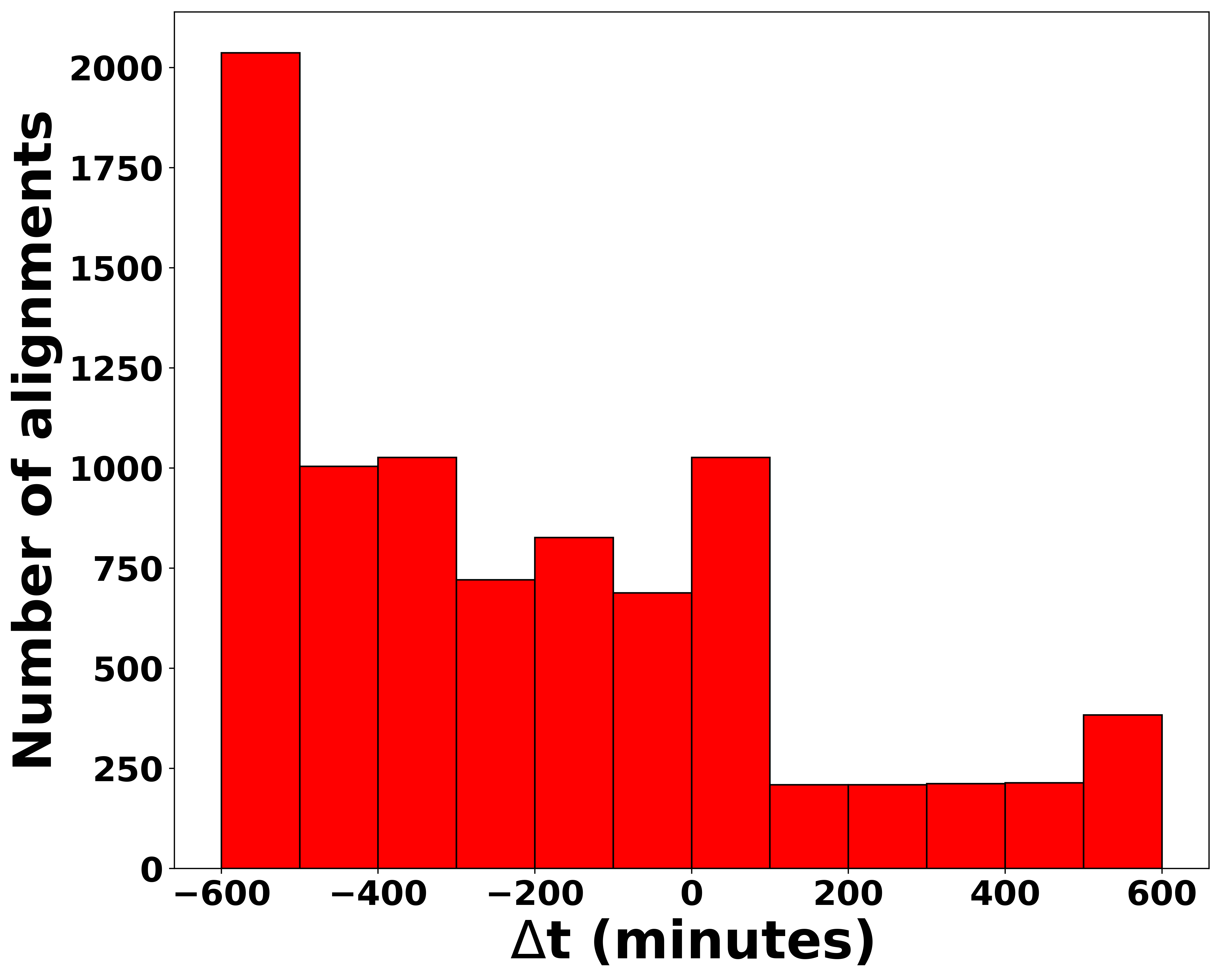}} 
    \subfloat[]{\includegraphics[width=0.25\textwidth,trim={0cm 0cm 0cm 0cm},clip=]{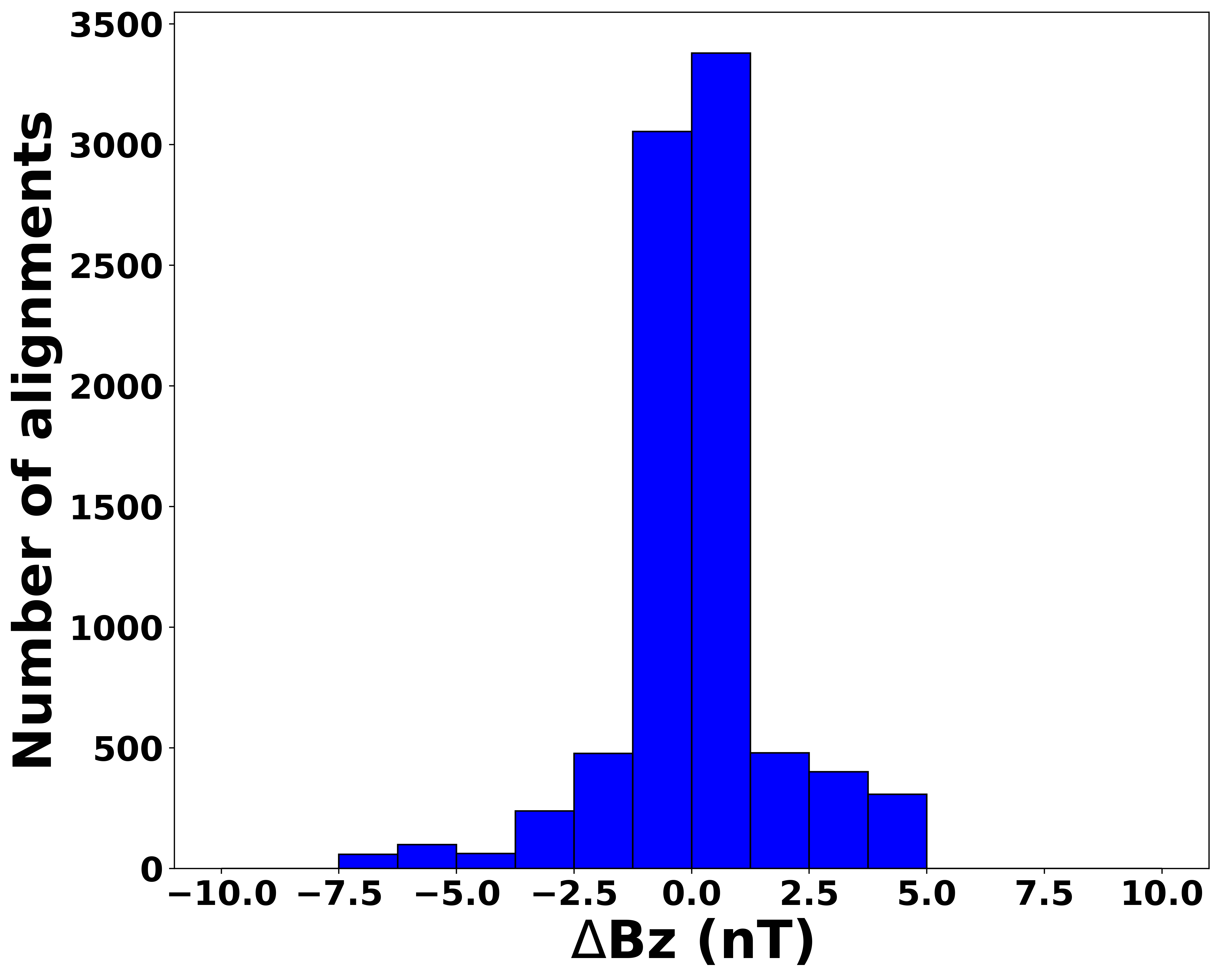}}\\ 

    \subfloat[]{\includegraphics[width=0.5\textwidth,trim={0cm 0cm 0cm 0cm},clip=]{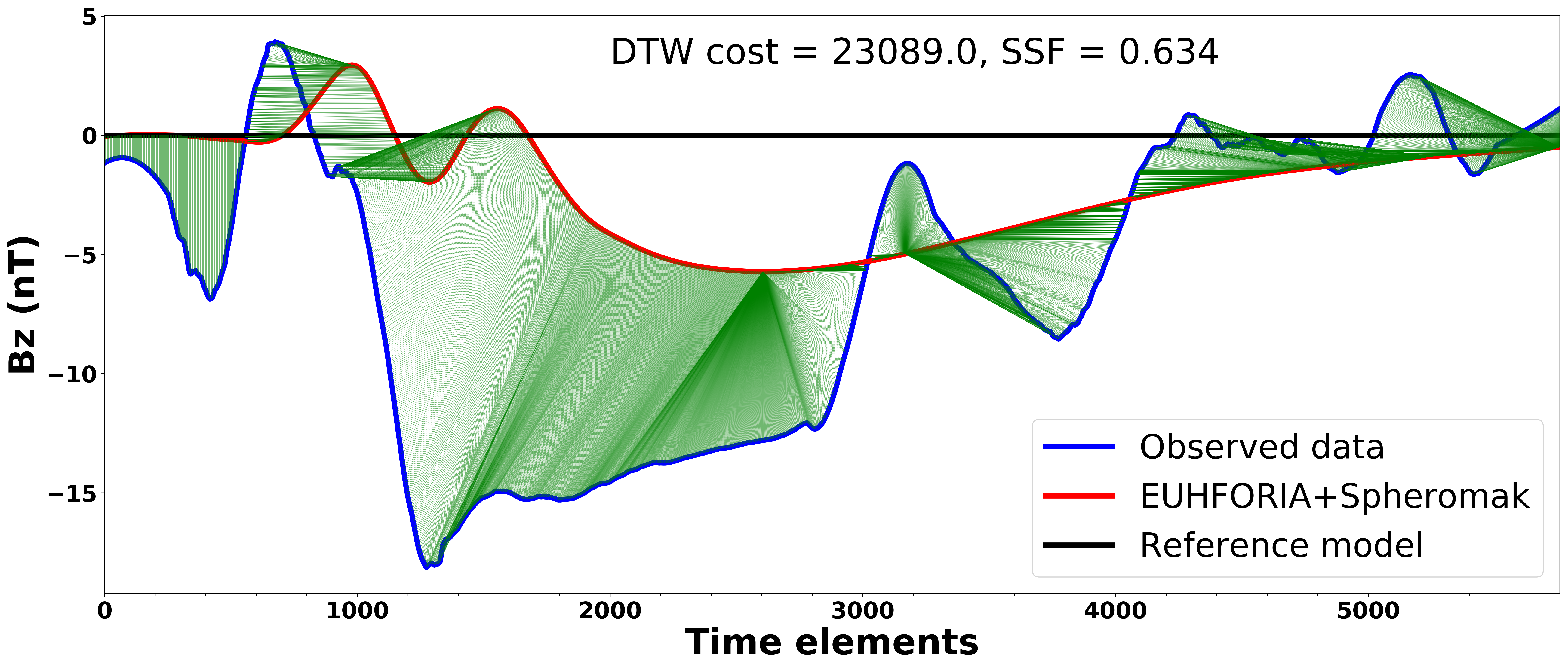}}
    \subfloat[]{\includegraphics[width=0.25\textwidth,trim={0cm 0cm 0cm 0cm},clip=]{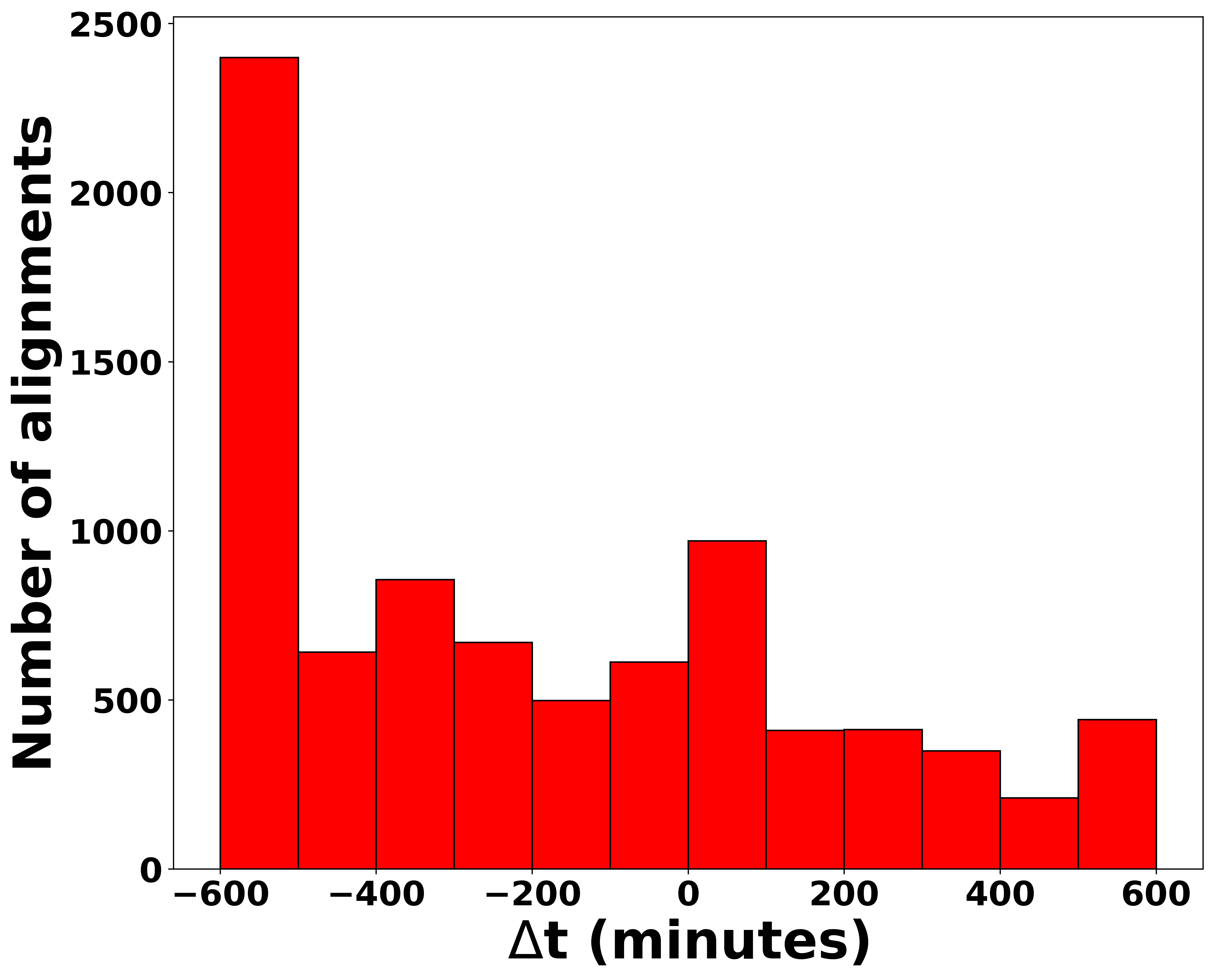}} 
    \subfloat[]{\includegraphics[width=0.25\textwidth,trim={0cm 0cm 0cm 0cm},clip=]{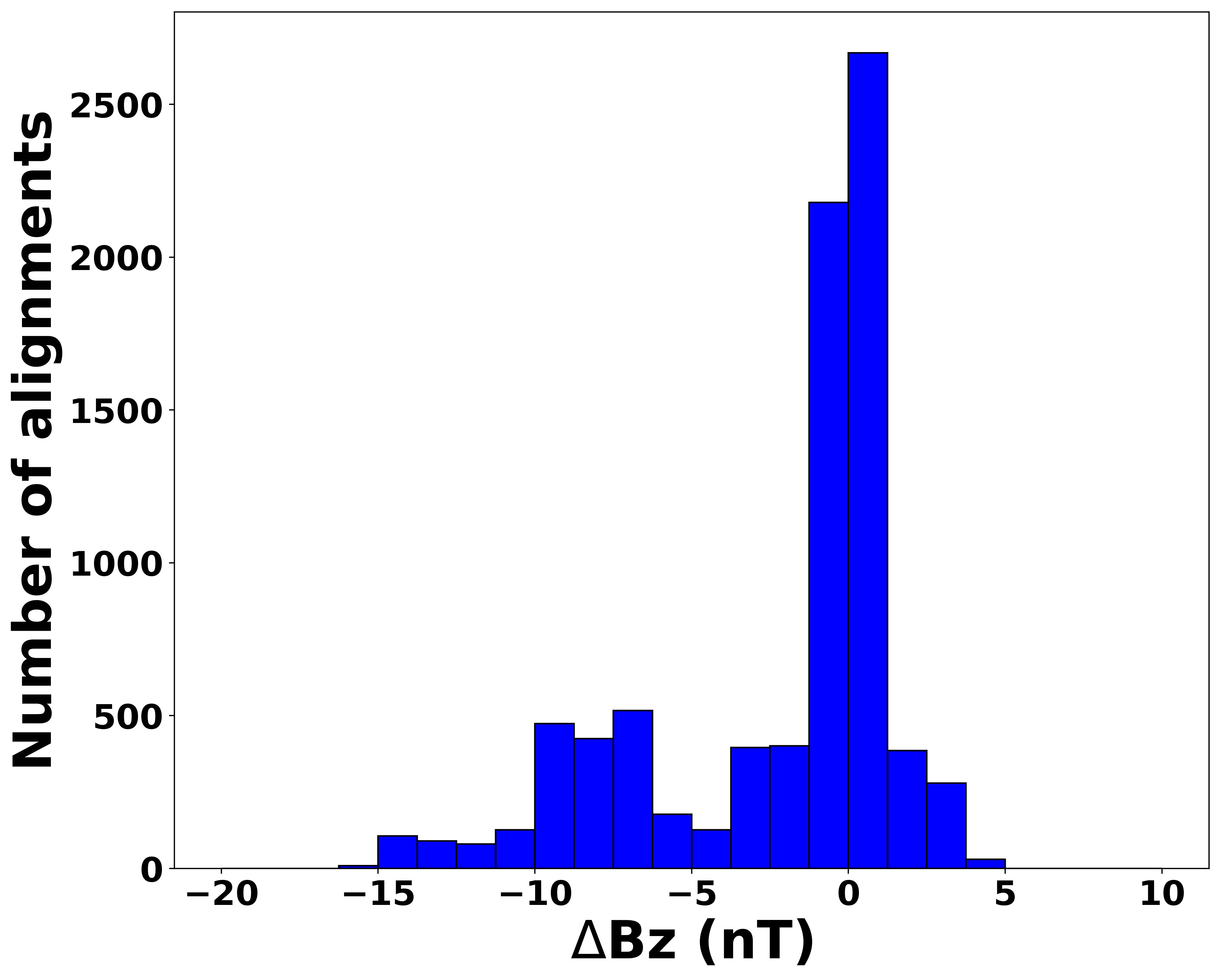}}
    \caption{DTW analysis of Event~1 for all CME models. 
    Rows 1 (a, b and c), 2 (d, e and f) and 3 (g, h and i) show the results for the Horseshoe model, FRi3D model and the spheromak model, respectively. Columns 1, 2 and 3 depict the DTW alignment {between the observed (blue) and modelled (red) time series}, histograms of time differences between the aligned points, and the histograms of the $B_z$ differences between the aligned points, respectively.}
    \label{fig:event1_dtw}
\end{figure*}

The in situ plasma characteristics, magnetic field, and the Dst predictions at Earth are presented in Fig.~\ref{fig:comp_models_euhforia_event1}. Out of three considered models, the Horseshoe model predicts the most accurate shock arrival time with $34$~minutes of delay compared to observed shock arrival. Unlike in the case of the FRi3D model, the $n_p$ profile obtained with the Horseshoe model decreases to the ambient solar wind density values after the CME passage as in the observations.
The DTW alignments between the observed $B_z$ and the corresponding modelled profiles using the Horseshoe, FRi3D and the spheromak models for Event~1 are illustrated in Fig.~\ref{fig:event1_dtw} (panels a, d and g), respectively. DTW is applied for the time period between 06:00~UT on 14 July 2012 and 12:00~UT on 18 July 2012, which covers the $B_z$ profile of interest {(the shaded grey region in the $B_z$ panel of Fig.~\ref{fig:comp_models_euhforia_event1})}. We have applied a window of $600\;$minutes for this event by visual inspection as most of the features could be matched within that time frame. Smoothing of the observed data is done over $500\;$minutes. Panels b, e and h of Fig.~\ref{fig:event1_dtw} show the histograms of the $\Delta t$ for Horseshoe, FRi3D, and spheromak models, respectively. The $\Delta t$ is mostly negative for the FRi3D and the spheromak models, which means that they predict the features later than observed. The Horseshoe model has a quite flat distribution of the $\Delta t$, with most of the alignments in the positive spectrum. This implies that the features observed by the model occur earlier. The $\Delta B_z$ histograms for the Horseshoe, FRi3D, and spheromak models are shown in Fig.~\ref{fig:event1_dtw} (panels c, f and i), respectively. Most of the $\Delta B_z$ alignments lie between $\pm2.5\;$nT for all three models. However, there are multiple alignments with $\Delta B_z$ in the range $[-10,-5]\;$nT in the case of the spheromak model, implying the underestimation of the modelled $B_z$. 

The SSF (viz.\ Table~\ref{tab:dtw}) for the Horseshoe model is lower than in the case of the Spheromak model, meaning that the Horseshoe model better reconstructs the overall $B_z$ profile. At the same time, the SSF for the Horseshoe model is only slightly higher than FRi3D, with the temporal alignments quite homogeneously distributed for the Horseshoe model. The $min(B_z)$ might not be as negative as predicted by FRi3D. However, the time alignment of different features in the $B_z$ is better for the Horseshoe CME.
The Horseshoe model performs the best as per criterion~1 as it gives the best estimate of the arrival time. Considering criterion~2, the SSF of Horseshoe is quite close to the FRi3D model, making its predictions more reliable than the spheromak model. Using the same number of {processing} cores and resolution, the computational time for the simulations with the spheromak model is 20 minutes, for the Horseshoe model is $3\;$hours $24\;$minutes, and for the FRi3D model is more than three times longer amounting to $9\;$hours and $2\;$minutes. Evaluating the models based on criterion~3, the Horseshoe can be placed as intermediate between the spheromak and the FRi3D models. So, by combining evaluations based on both criteria, the Horseshoe CME model presents promising capabilities for reliable $B_z$ predictions.

\subsection*{Event~2}
The in situ plasma characteristics, magnetic field, and the Dst predictions at Earth are provided in Fig.~\ref{fig:comp_models_euhforia_event2}. For consistency, we compare those simulations where CME1 is modelled with spheromak for all, and CME2 is modelled with Horseshoe, FRi3D, and spheromak models. With the Horseshoe model, the predicted arrival time of S2 (shock associated with CME2) matches the observations. FRi3D predicted the arrival time $\sim2$~hours in advance, and with spheromak, the estimated arrival time is delayed by $\sim2$~hours, with respect to the observed arrival times. 

\begin{figure*}[ht!]
    \centering 
    \subfloat[]{\includegraphics[width=0.5\textwidth,trim={0cm 0cm 0cm 0cm},clip=]{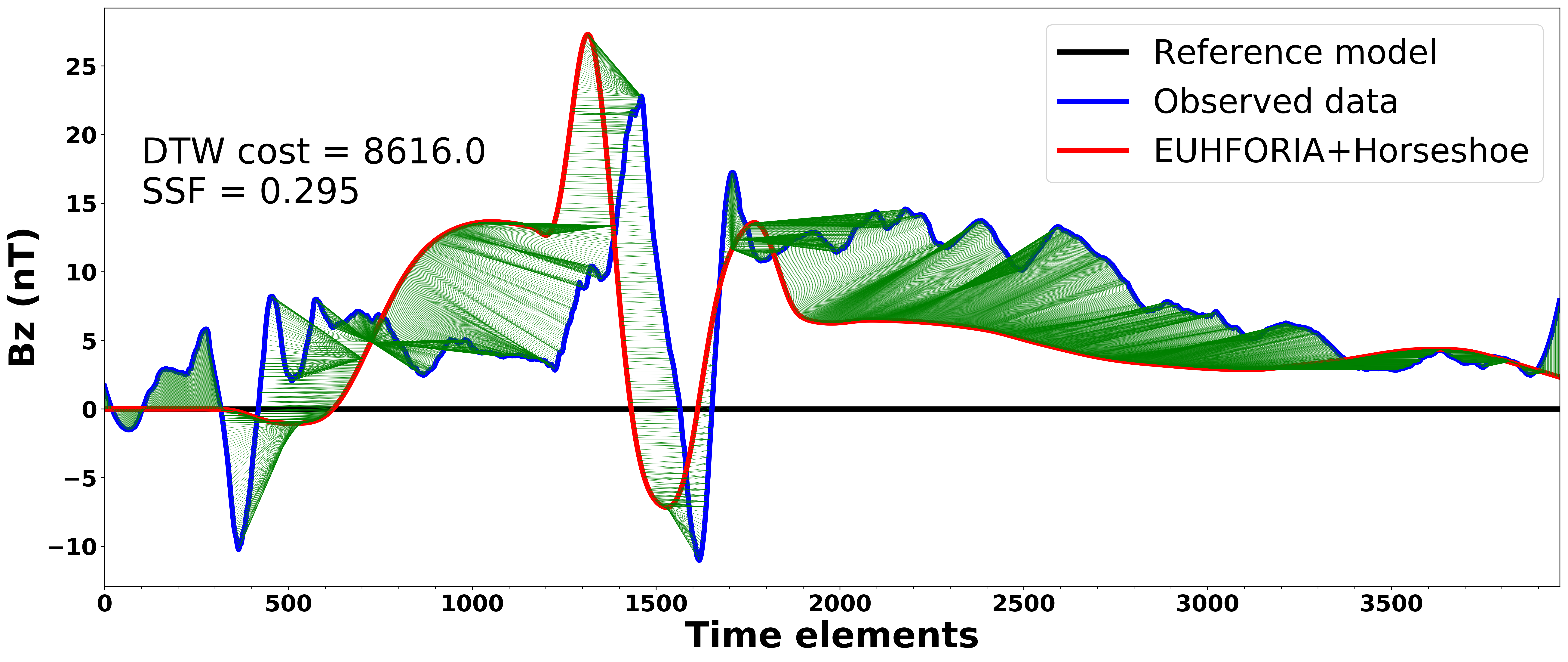}}
    \subfloat[]{\includegraphics[width=0.25\textwidth,trim={0cm 0cm 0cm 0cm},clip=]{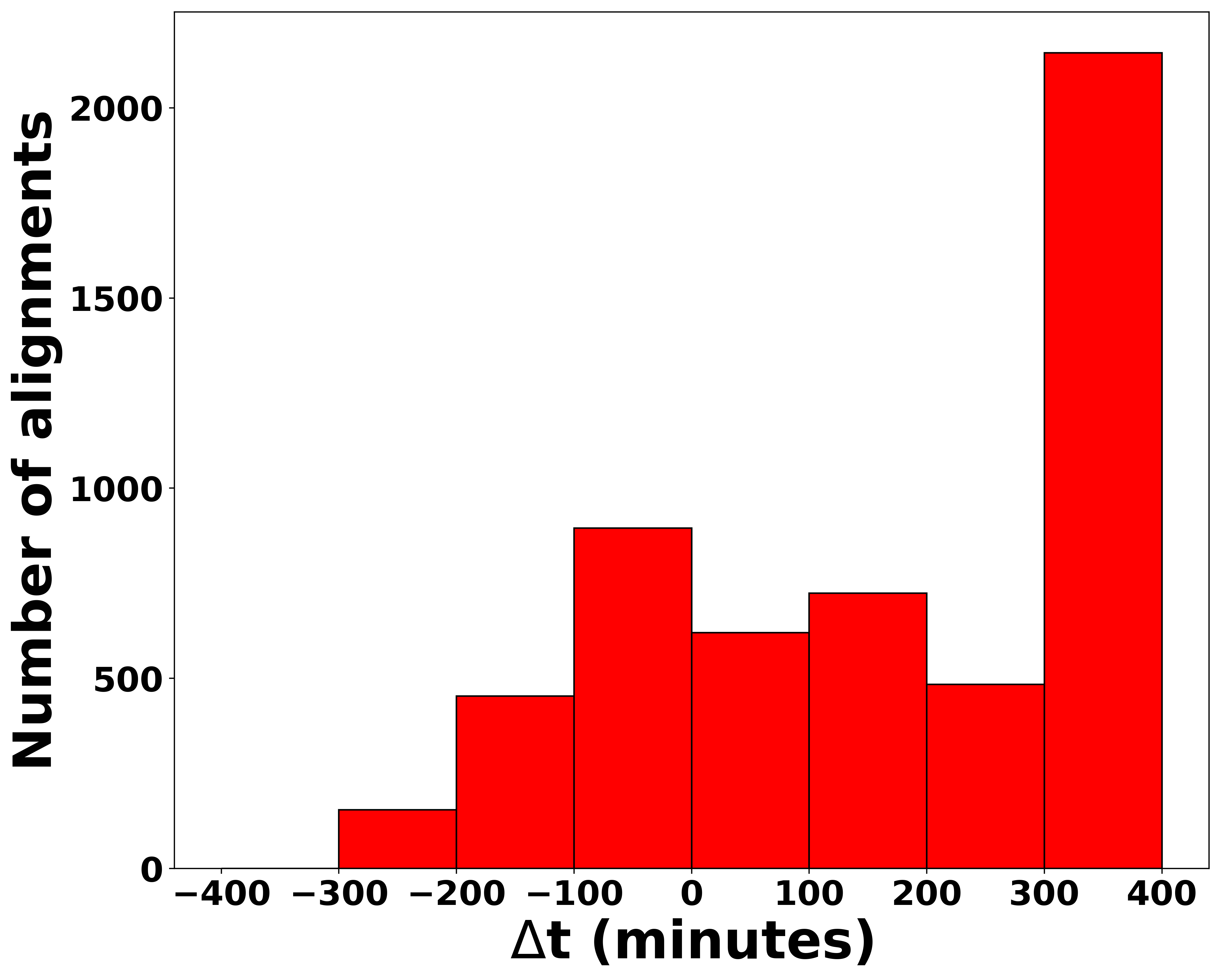}} 
    \subfloat[]{\includegraphics[width=0.25\textwidth,trim={0cm 0cm 0cm 0cm},clip=]{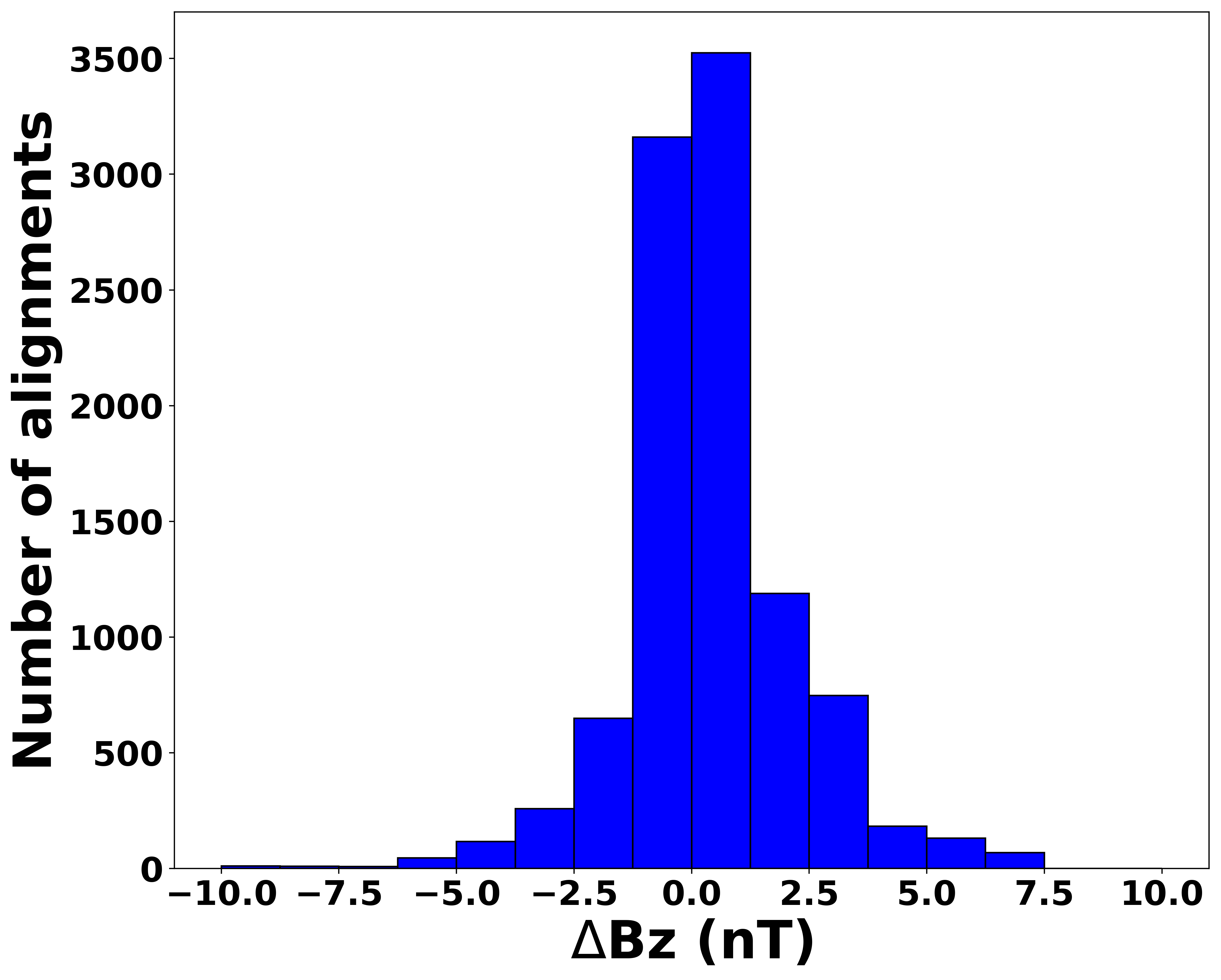}}\\
    
    \subfloat[]{\includegraphics[width=0.5\textwidth,trim={0cm 0cm 0cm 0cm},clip=]{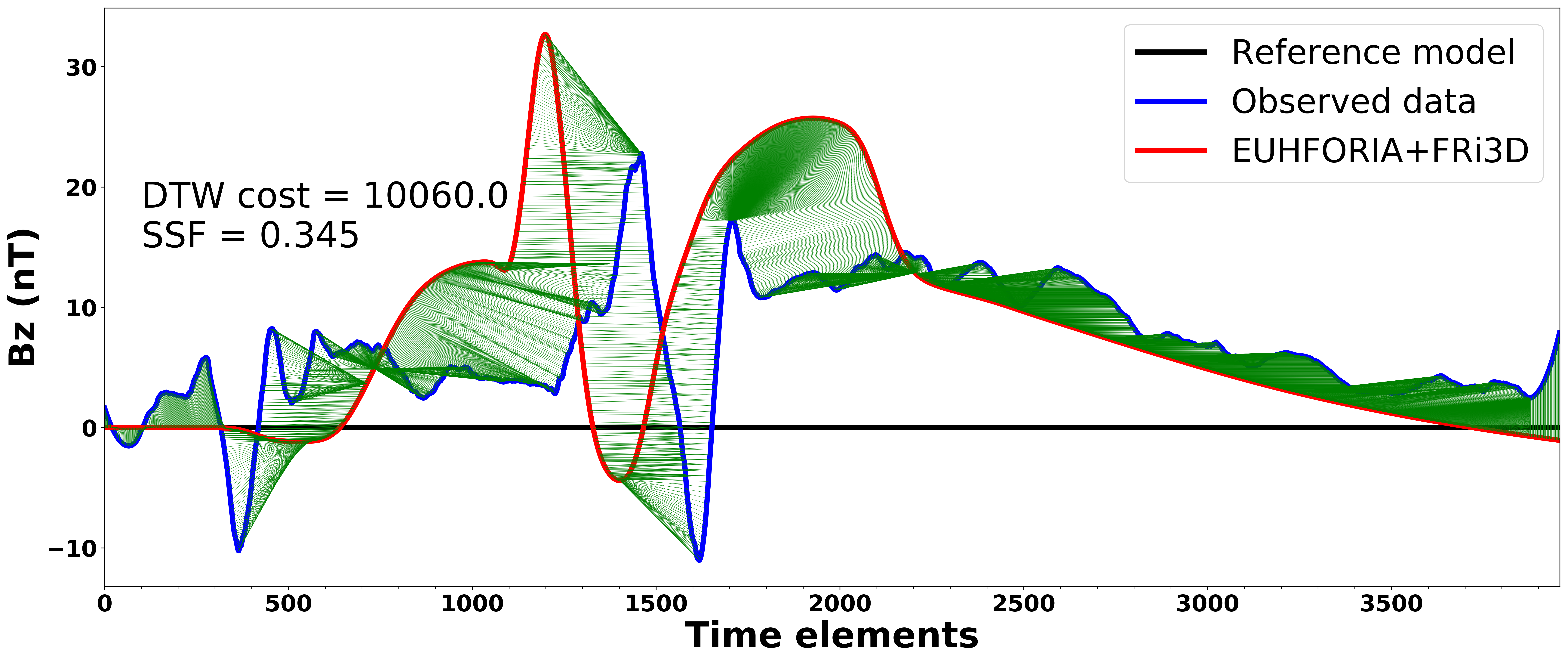}}
    \subfloat[]{\includegraphics[width=0.25\textwidth,trim={0cm 0cm 0cm 0cm},clip=]{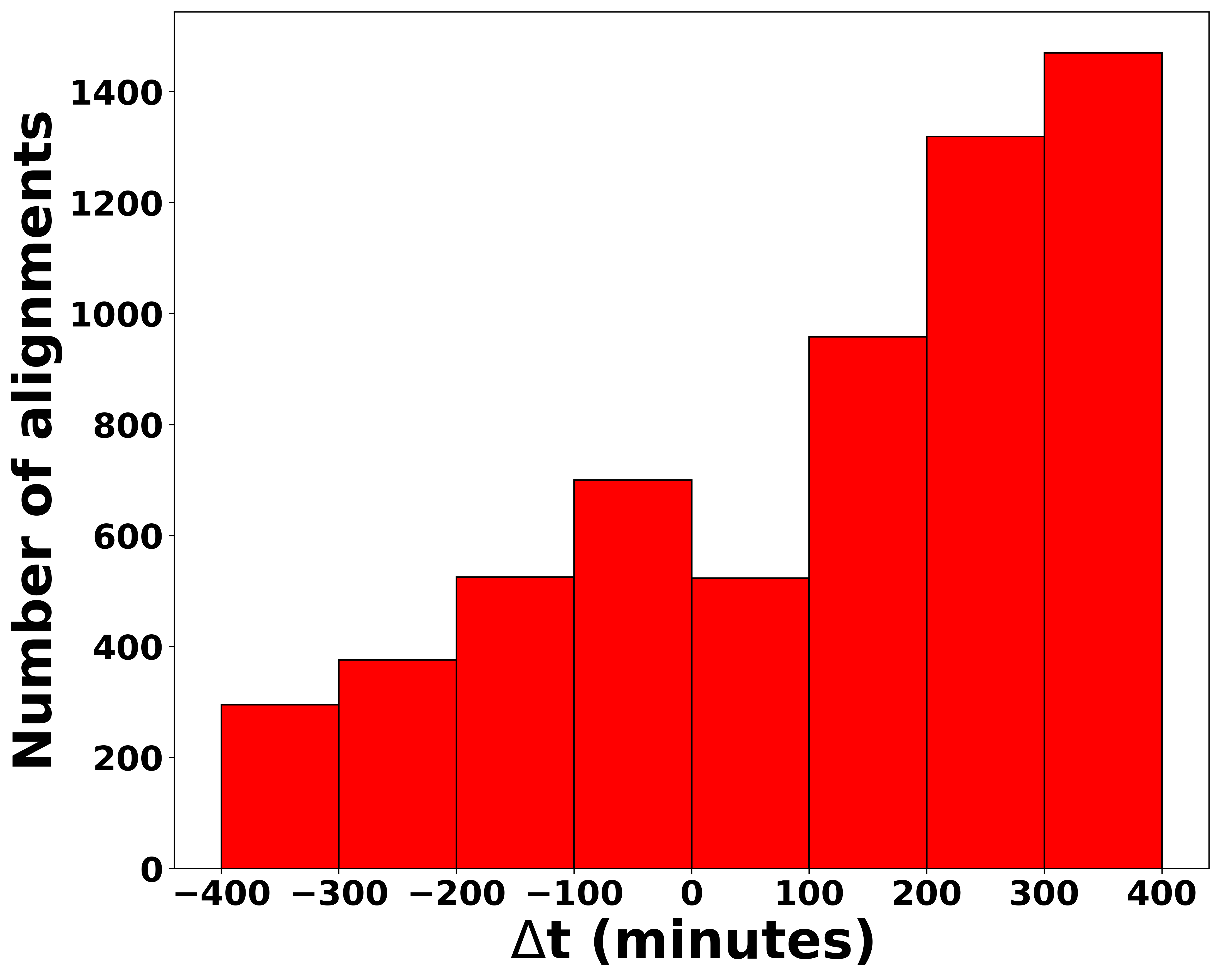}} 
    \subfloat[]{\includegraphics[width=0.25\textwidth,trim={0cm 0cm 0cm 0cm},clip=]{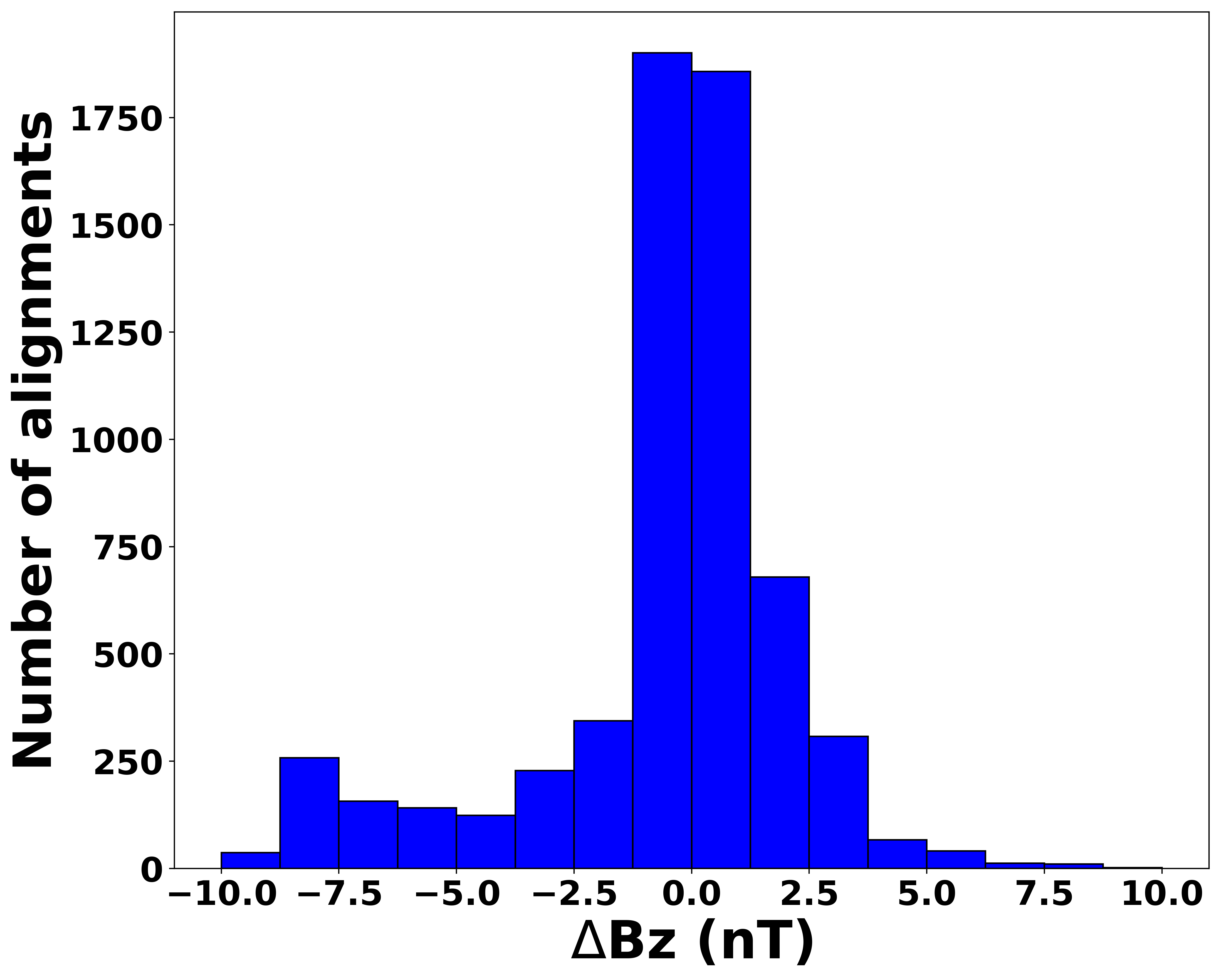}}\\ 
    
    \subfloat[]{\includegraphics[width=0.5\textwidth,trim={0cm 0cm 0cm 0cm},clip=]{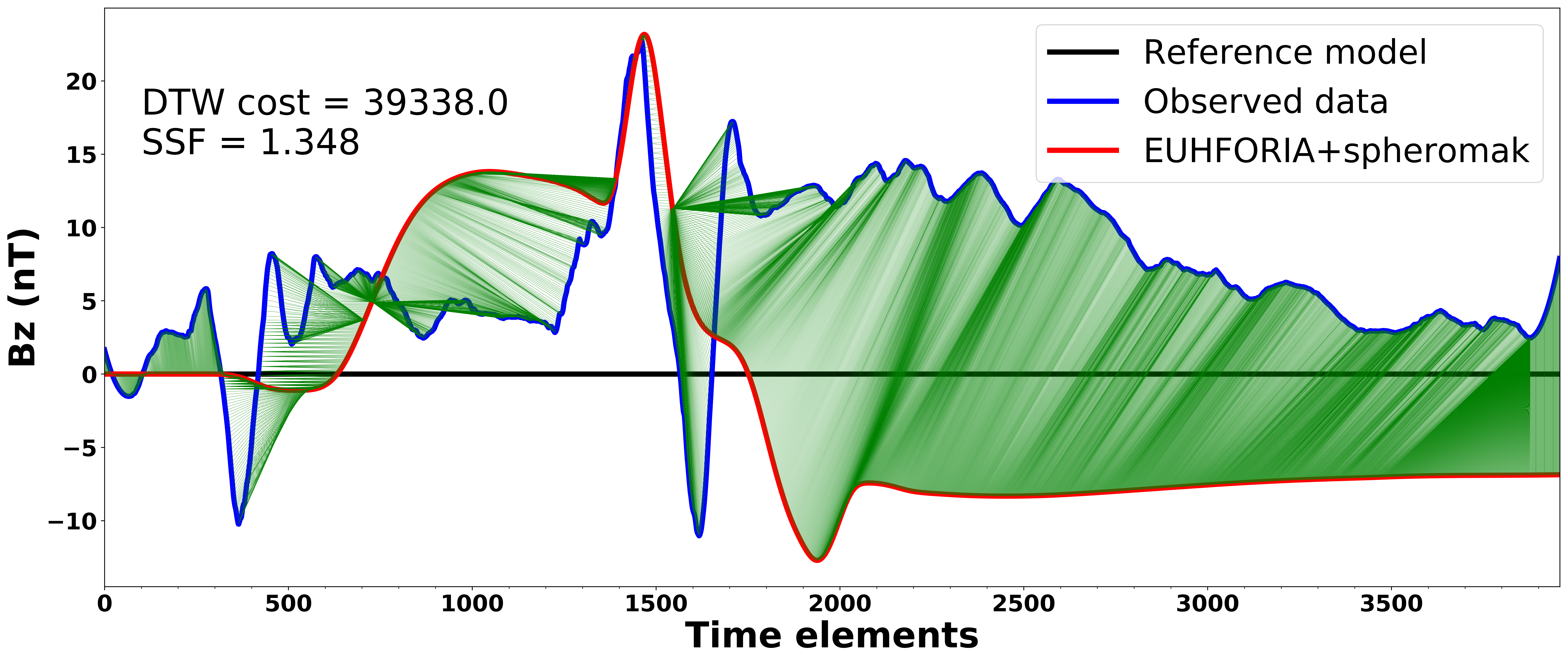}}
    \subfloat[]{\includegraphics[width=0.25\textwidth,trim={0cm 0cm 0cm 0cm},clip=]{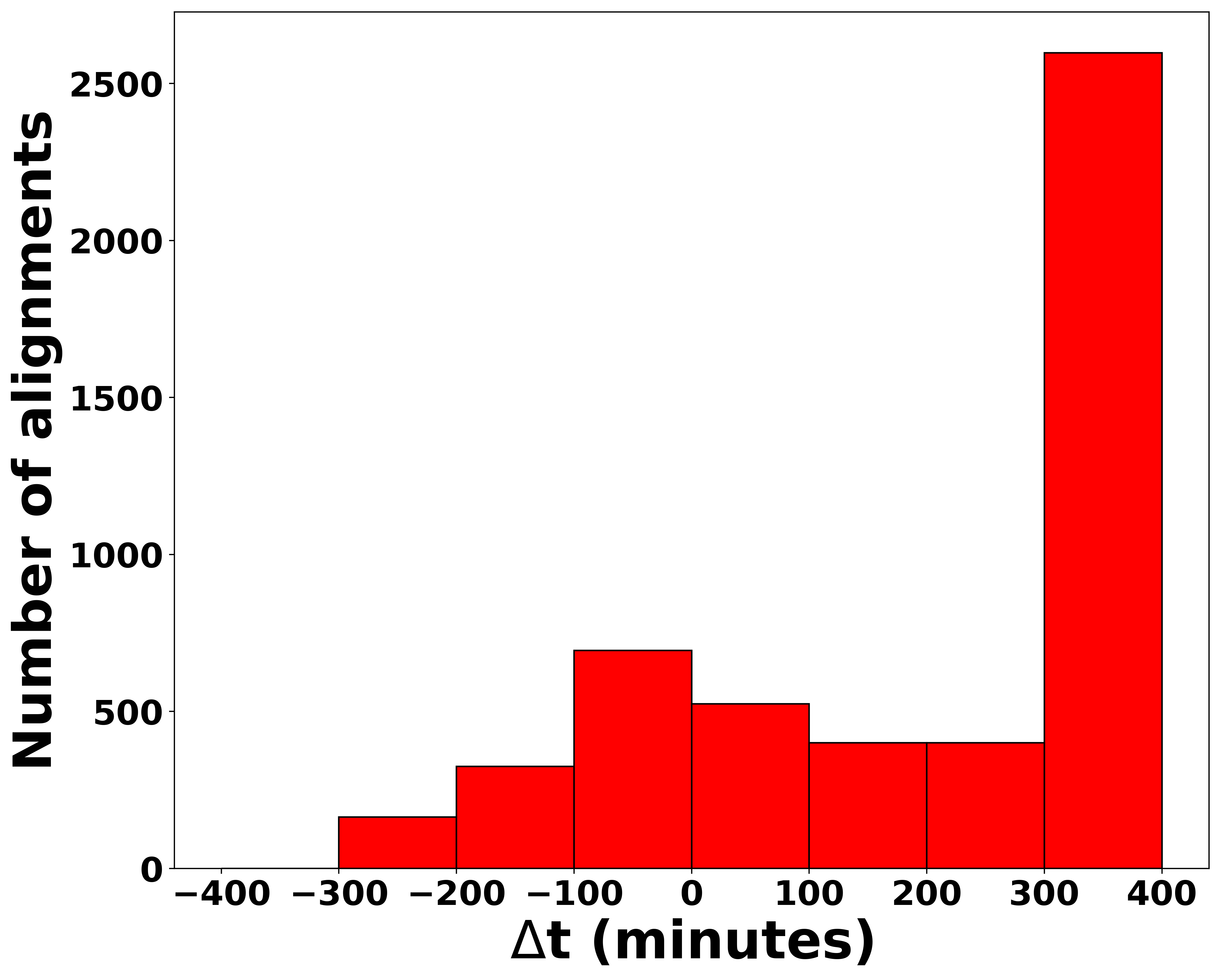}} 
    \subfloat[]{\includegraphics[width=0.25\textwidth,trim={0cm 0cm 0cm 0cm},clip=]{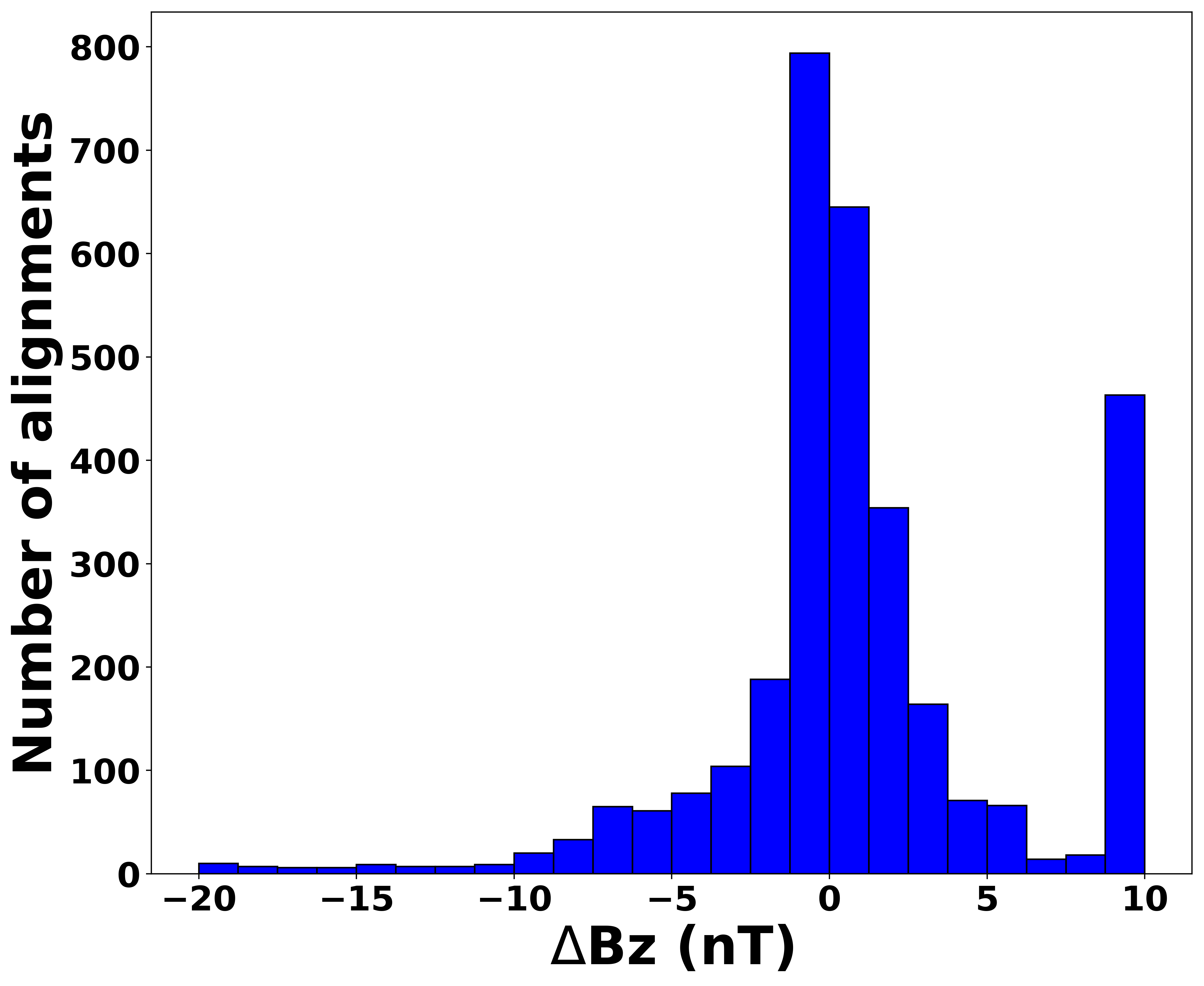}}
    \caption{DTW analysis of Event~2 for all CME models. Rows 1, 2 and 3 show the results for the Horseshoe model, FRi3D model and the spheromak model, respectively. Columns 1, 2 and 3 depict the DTW alignment {between the observed (blue) and modelled (red) time series}, histograms of time differences between the aligned points, and the histograms of the $B_z$ differences between the aligned points, respectively.}
    \label{fig:event2_dtw}
\end{figure*}

DTW is applied for the time period between 18:00~UT on 11 September 2014 and 12:00~UT on 14 September 2014 {(the shaded grey region in the $B_z$ panel of Fig.~\ref{fig:comp_models_euhforia_event2})}. This is a complicated case for applying DTW because of the observed temporally fast fluctuations in the sheath that are close to the features we want to capture. Smoothing of those features in observed time series, to the extent required to be mapped with the modelled sequences, results in the reduction of the magnitudes of the peaks of the positive and negative $B_z$ in the sheath region. 
Choosing a smaller smoothing window increases the number of singularities in the alignment. Hence, we optimally smooth over $150\;$minutes to preserve the magnitude of the negative $B_z$ in the sheath region. That allows us to compare the minimum negative $B_z$ predicted by the CME models in EUHFORIA. 
Panels (b, e and h) of Fig.~\ref{fig:event2_dtw} show the distribution of $\Delta t$ in the alignments of the Horseshoe, FRi3D, and spheromak models, respectively, for Event~2. The maximum number of alignments for both Horseshoe and Spheromak models lie within [$300$, $400$]~minutes ($\Delta t$), which implies that the modelled profiles predict the majority of features earlier than observed. For the FRi3D model, although the maximum alignments fall in [$300$, $400$]~minutes, there is an increasing trend in the distribution of $\Delta t$ towards $400$~minutes (i.e., the number of early predictions are more distributed over various $\Delta t$ as compared to the other CME models). The histograms of $\Delta B_z$ for the Horseshoe, FRi3D, and the spheromak models are shown in Fig.~\ref{fig:event2_dtw} (panels c, f and i), respectively. Most of the alignments lie in the $\Delta B_z$ range [$-2.5$, $2.5$]~nT in the case of both the Horseshoe and the FRi3D models. For spheromak, the alignments are between [$-5$, $5$]~nT of $\Delta B_z$, and also, a large number of alignments are close to $10\;$nT. This is a consequence of the erroneous prediction of negative $B_z$ in ME2 by the spheromak model as opposed to the observed positive $B_z$.

The Horseshoe model performs the best as per criterion~1, i.e., it gives the best estimate of the arrival time. The SSF for the Horseshoe-modelled profile is better than in the case of both FRi3D and spheromak models and, hence, is the best candidate as per criterion~2. The Horseshoe simulation best reproduces the positive-to-negative switch in the sheath $B_z$. However, FRi3D matches the $B_z$ strength in the magnetic cloud better, except for the initial overestimation of the positive $B_z$ after the passage of the sheath. The computational times of the simulations with the spheromak, Horseshoe and FRi3D models are 32 minutes, 8 hours, 8 minutes, and 15 hours 22 minutes, respectively. For the FRi3D model, the majority of the total computational time is spent on the calculation of the magnetic field at the EUHFORIA inner boundary (i.e., mask computation). In the case of the horseshoe model, however, due to its complete analytical form, the time spent on the mask computation is reduced. The spheromak model still consumes the least time in computing the mask due to its purely force-free nature during injection into the heliospheric domain. The force-free nature is not completely satisfied when changing the geometry of the full torus into Horseshoe. {This results in some unrealistic high-speed parcels near the inner boundary, emerging from the cells where the legs of the Horseshoe CME are connected. Hence, pronounced gradients are formed in the speed profile in the computational domain close to the inner boundary. To maintain the stability as per Courant-Friedrichs-Lewy (CFL) condition, the allowed numerical timestep, \texttt{dt}, becomes much lower for the Horseshoe model. That is why more iterations are necessary to reach the end of the simulations, which extends the overall computational duration. Such problems are reported in previous works, e.g., \citet{regnault2023,linan2024}}. 
{\bf Therefore, despite the analytic formulation of the Horseshoe magnetic field, the computational time is still high because of how we disconnect the CME. As the high-speed artefacts occur closer to the boundary, they do not affect the predictions at Earth. However, we acknowledge that this is a crucial problem in the case of successive CME injection or for predicting space weather impacts at locations closer to the inner boundary. Hence, we would propose detailed future work to mitigate this issue.} 

The Horseshoe model performs intermediately amongst the models based on criterion~3. It meets both our criteria of speed and accuracy reasonably as compared to the other CME models in the framework of EUHFORIA. This highlights the potential of the Horseshoe model towards a reliable and efficient operational space weather forecasting model.\\

\begin{table}
\begin{center}
\begin{tabular}{l  c  c}
\hline
\hline
SSF & Event~1 & Event 2\\
\hline
\hline
spheromak & 0.634 & 1.348\\

FRi3D  & 0.205 & 0.345\\

Horseshoe & 0.238 & 0.295\\
\hline
\hline
\end{tabular}
\end{center}
\caption{Sequence similarity factor (SSF) for Event~1 and Event~2 modelled using the spheromak, FRi3D and Horseshoe models.}
\label{tab:dtw}
\end{table} 

\subsubsection*{General remarks}
\citet{Isavnin2016} point out that the FRi3D model overestimates the magnetic flux budget of CMEs due to the underestimation of the magnetic field line twist near the flux rope boundary. The fact that FRi3D performs better in matching the total magnetic field is because of the overestimation of $B_y$ and $B_z$ components ($B_x$ is not properly modelled for the events in this study). The Horseshoe model, with the current method of constraining magnetic field properties, provides a reasonable magnitude of $B_y$ and $B_z$ but still fails to model the $B_x$ component. We note that $B_x$ is not modelled accurately by any of the models we studied. The reasons for not capturing $B_x$ could be the erroneous reconstruction of the CME geometry and location of launch that does not make the impact of CME through its flank. The other reason could be because of the circular cross-section of the models. Or the CME was deflected during its propagation between $0.1-1.0\;$au, which our simulations do not capture self-consistently. 


\section{Summary and outlook}\label{sec:conclusion}
The following are the key takeaways of this study:
\begin{enumerate}
    \item We implemented in EUHFORIA the Horseshoe CME model. Our Horseshoe model has a modified Miller-Turner magnetic configuration and a modified torus-like geometry that better mimics the CME leg structures. We pointed out the differences in the magnetic field profiles predicted by the full torus and the Horseshoe models at 1~au. Based on the results, we were able to recommend the use of the Horseshoe model in space weather forecasting for its better computational performance and modelling accuracy.
    \item The methodologies to constrain the geometrical, plasma, and magnetic field parameters of the Horseshoe model from observations were designed, and their efficiency was demonstrated with two examples.
    \item We validated the Horseshoe model with two events -- first, a single non-interacting CME event of 12 July 2012 (Event~1); and second, a CME-CME interaction event of 8-10 September 2014 (Event~2). Variability in the predicted plasma and magnetic field time profiles due to the errors in observational constraining of the axial magnetic field strength has been addressed through ensemble modelling. In addition, the changes in the predicted $min(B_z)$ compared to the observations at spatial locations around Earth were also analysed. The Horseshoe model predicts the CME shock arrival time to Earth and the minimum negative $B_z$ component comparable to that of the FRi3D model (a realistic flux rope model) in both events. The computational time of the Horseshoe model is intermediate between the spheromak and the FRi3D models.
    \item The first attempt to test the capability of the Horseshoe model to inject two successive CMEs into the EUHFORIA was successfully shown with Event~2. This mitigates the limitation of the FRi3D model and provides reasonable results, producing an upgrade over the spheromak model. Further validation of the Horseshoe model in modelling multiple CMEs must be done. 
    \item The modelled $B_z$ component of the magnetic field, obtained with the Horseshoe model, was compared with the ones obtained with the FRi3D and spheromak models in the framework of EUHFORIA using the Dynamic Time Warping (DTW) technique.
\end{enumerate}
Overall, the Horseshoe model is promising in predicting the CME arrival time and magnetic field properties in an operational setup. We note that it is numerically more stable to launch initially force-free magnetic field configurations (spheromak) than the non-force-free ones. The full toroidal mMT topology is divergence-free and approximately force-free initially. Although not the most accurate in reproducing the magnetic field profiles in the heliosphere, it is numerically more stable and computationally less expensive. Whereas incomplete injection of the force-free magnetic field through the inner boundary gives rise to residual $\mathbf{j}\times\mathbf{B}$ forces that create excess $\nabla\cdot\mathbf{B}$. As the Constrained Transport methodology in EUHFORIA is not efficient in cleaning the $\nabla\cdot\mathbf{B}$, it is difficult to sustain the stability of CME models with the legs which continuously inject additional magnetic field. 
One of the next steps is to implement the Horseshoe model in Icarus \citep[][]{Verbeke2022,Baratashvili2022} which is an advanced version of EUHFORIA implemented in the framework of MPI-AMRVAC \citep[][]{Keppens2003,Xia2018} and uses a parabolic divergence cleaning method which is more efficient than the Constrained Transport in EUHFORIA. Another future work will be to contract the back part of the torus and inject it fully through the inner boundary to conserve the force-free fields. {The commonly used constant-alpha (force-free) magnetic field models, used to fit interplanetary CMEs, exhibit an increasing twist outward towards the flux rope boundary (e.g., the mMT topology as mentioned in Section~\ref{subsec:CME_Model}). However, most of the observed interplanetary CMEs have a uniform twist distribution \citep{hu2015}. Hence, \citet{vandas2019} prescribe a flat twist profile to a cylindrical flux rope, which renders the flux rope non-force-free. On large scales, the assumption of a force-free magnetic field is generally invalid \citep{Kilpua2011}. Hence, non-force-free magnetic field topology with a uniform twist profile can be used for interplanetary CME models, although their numerical implementation might pose challenges. In future works, such uniform twist distribution must be developed for toroidal flux ropes to explore more accurate modelling of CME evolution. } 

%
%

\section*{Acknowledgement}
The authors acknowledge support from the projects C14/19/089 (C1 project Internal Funds KU Leuven), G0B5823N and G002523N (WEAVE) (FWO-Vlaanderen), 4000134474 (SIDC Data Exploitation, ESA Prodex), Belspo project B2/191/P1/SWiM,  the FEDtWin project PERIHELION and the financial support received to attend the third BINA workshop under the Belgo-Indian Network for Astronomy and astrophysics (BINA) project approved by the International Division, Department of Science and Technology (DST, Govt. of India; DST/INT/BELG/P-09/2017) and the Belgian Federal Science Policy Office (BELSPO, Govt. of Belgium; BL/33/IN12). We thank Dr.\ Christine Verbeke for insightful discussions about CME model implementation in EUHFORIA, and Dr. \ Brigitte Schmieder for their constant support and suggestions for the improvement of this model.
\bibliographystyle{aa} 
\bibliography{biblio.bib}

\appendix
\section{Geometrical reconstruction of the events}
In this section, we provide the details of the geometrical reconstruction of the CMEs from the white light images. As this procedure can be subjective and, hence, introduces uncertainties in CME arrival time predictions \citep{Verbeke2023}, we provide our rationale behind the features we fit. We use the white-light image tool of Flux Rope in 3D model \citep[FRi3D,][]{Isavnin2016}. 

\subsection{Event 1: July 12, 2012}\label{app:event1}
\begin{figure}[ht!]
    \centering
    {\includegraphics[width=0.5\textwidth,trim={0.5cm 0cm 0cm 0cm},clip=]{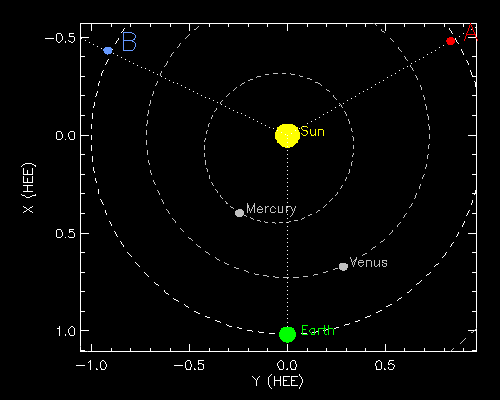}}
    \caption{The position of the STEREO-A and STEREO-B spacecraft during the CME eruption on 12 July 2012. The image was created using the freely available `Where is STEREO' tool (\url{https://stereo-ssc.nascom.nasa.gov/cgi-bin/make_where_gif}).}
    \label{fig:20120712_recon_position}
\end{figure}

\begin{figure}[ht!]
    \centering 
    {\includegraphics[width=0.5\textwidth,trim={4.5cm 3.cm 4.5cm 3.cm},clip=]{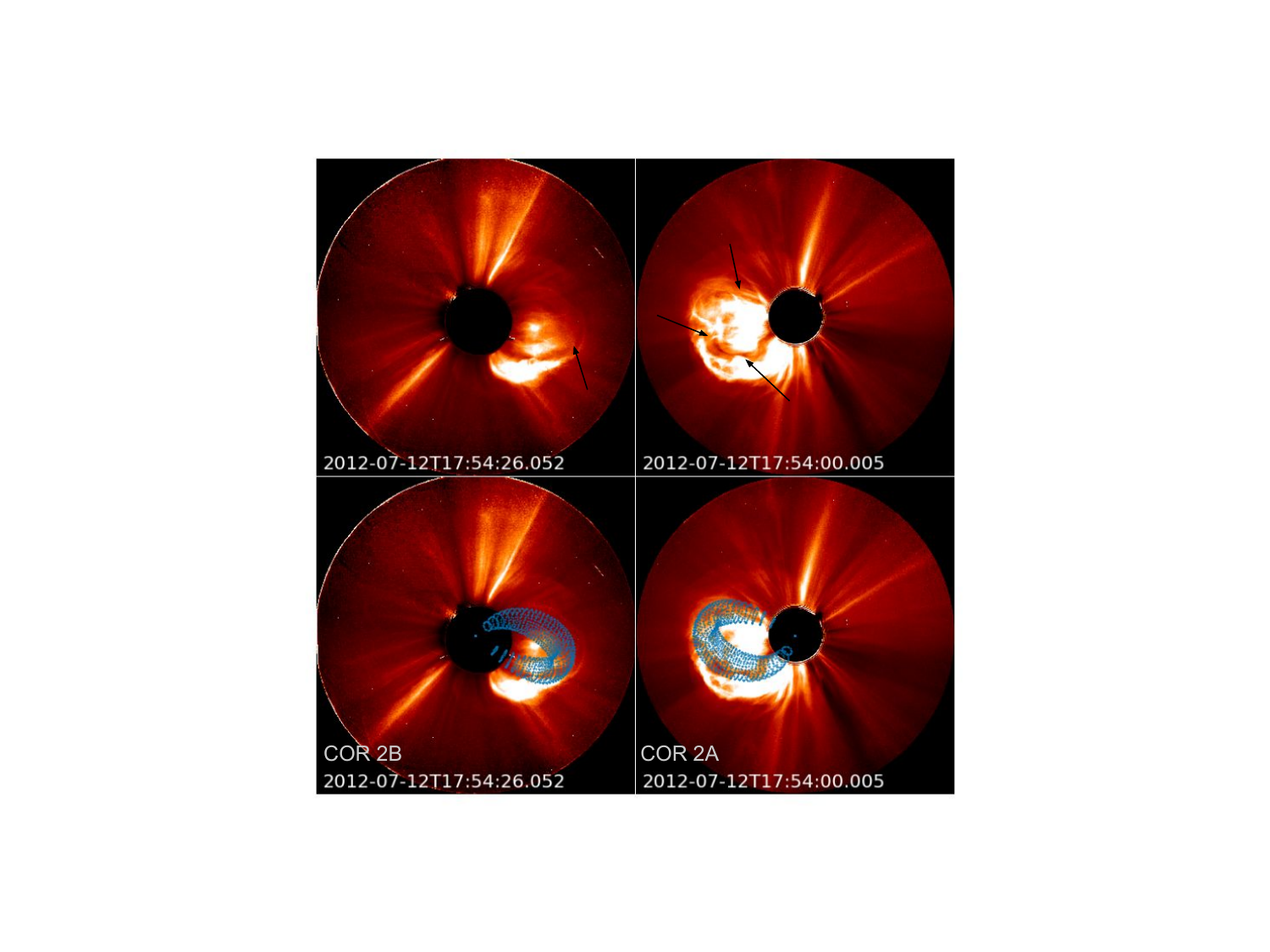}}
    \caption{The white light images as observed from COR2-A and COR2-B and the same images overlaid with FRi3D reconstructed wire frame in the bottom panels. The arrows show different projected white light features used as proxies to reconstruct the 3D structure as discussed in the text.}
    \label{fig:20120712_recon_fri3d}
\end{figure}

Although this event has been reconstructed before in previous studies \citep{Scolini2019,Maharana2022}, we repeat the procedure to match the constraints of a toroidal geometry to validate the Horseshoe model. The spacecraft STEREO-A (STA) and STEREO-B (STB) observed the CME as a limb event in their field of view (FOV). They were separated by $125\degree$ in longitude and $10\degree$ in latitude (Fig.~\ref{fig:20120712_recon_position}). COR2A and CORB are the coronagraphs onboard STA and STB, respectively. LASCO spacecraft observed the CME as a halo event. However, the coronagraph data was unavailable during the event. We temporally fitted the flux rope mesh between 17:24 and 18:24 on 12 July 2012, when the full CME was visible clearly in the coronagraph FOV. The 3D reconstruction of the CME was done considering the following aspects noticed in the observations:
\begin{itemize}
    \item The streamer is observed to be bent from the COR2A perspective, and there is movement through the streamer. However, there are no such signatures from the COR2B perspective, which means that the CME interacted with the streamer on its western limb during the evolution. Hence, the southern part of the CME is fitted in such a way that it looks like it fits part of the streamer, but it is the flux rope passing through it.
    \item From COR2A FOV, we fit the angular half-width of the CME in the northern limb to not fit the northern streamer completely, which is on the way of the CME and is then deflected. 
    \item For fitting the leading edge, we fit the COR2A observation of the magnetic cloud (darker region) as it is brighter than the COR2B counterpart. Physically, the CME front is closer to the COR2A's perspective, so it is better to rely on it. The leading edge looks underestimated from the COR2B perspective.
    \item The stand-off distance between the shock and the CME is insignificant, making it difficult to distinguish the CME parts for reconstruction. These observations point to a very twisted flux rope with local deformations that are not straightforward to capture with the smooth global mesh structure of the existing reconstruction models.
    \item Flux rope models like FRi3D or the Torus model represent a more realistic geometry of the CME represented by the reconstruction shown in Fig.~\ref{fig:20120712_recon_fri3d}. Whereas to constrain parameters for a spheromak model, the CME volume is overestimated and does not represent an extended flux rope geometry (see, for example, Fig.~4 of \citet{Scolini2019}.)
\end{itemize}


\subsection{Event 2: September 10, 2014}\label{app:event2}
The CME2 was tracked between 18:54 and 19:54 on 10 September 2014, when all the features of the CME were visible in the white light images recorded by LASCO C3 and STB COR2B. The 3D reconstruction of this event is based on the following distinct features:

\begin{itemize}
    \item The CME interacts with two streamers during its evolution in the coronagraph FOV. From the perspective of LASCO C3, the CME pushes through the streamer visible on the north-western limb (Streamer~1) and the other close to the north pole (Streamer~2). The CME shock and the magnetic cloud interact with the streamers as distinct brightening of the streamers, which in turn fades out when the CME has completely passed. We fit the parts of the Streamer~2 on the northern flank, and part of the Streamer~1 lies within the southern extent of the CME on its western limb. 
    \item It is the front part of the CME which interacts with Streamer~1. Although the shock is observed to disturb the streamer from the COR 2A perspective, the bulk of the CME moves away from the streamer root. As it is a back-sided halo event from COR 2A, we fit the features as per C3 features. Hence, it might seem that the features are ill-fitted in COR 2A FOV. 
    \item Streamer~2 is brightened throughout the complete propagation of the CME, which means that the streamer is present on the front side of the Sun.
    \item Both the shock front and the twisted magnetic cloud structure have a wavy morphology, which we are trying to fit with a smooth croissant-like structure. The brighter structure seen close to the CME nose in COR 2 FOV could be the twisted extension at the CME back, and hence, it is less bright from the C3 perspective. However, it isn't very easy to do justice to fit all the features of the CME with simplified reconstruction models. Hence, we fit the features that are best visible from the Earth-directed FOV. 
\end{itemize}

\begin{figure}[ht!]
    \centering
    {\includegraphics[width=0.5\textwidth,trim={0.5cm 0cm 0cm 0cm},clip=]{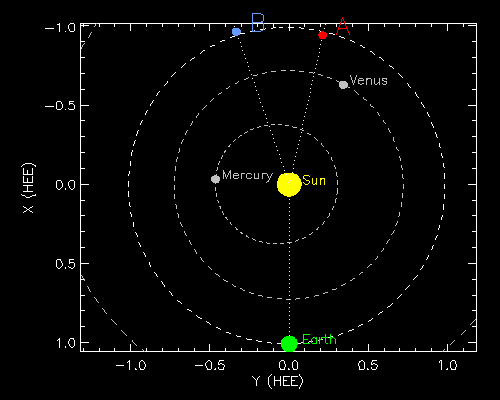}}
    \caption{The position of the STEREO-A and STEREO-B spacecraft during the CME eruption on September 10, 2014. The image was created using the freely available `Where is STEREO' tool (\url{https://stereo-ssc.nascom.nasa.gov/cgi-bin/make_where_gif}).}
    \label{fig:20140910_recon_position}
\end{figure}

\begin{figure}[ht!]
    \centering
    {\includegraphics[width=0.55\textwidth,trim={4.5cm 3.cm 4.5cm 3.cm},clip=]{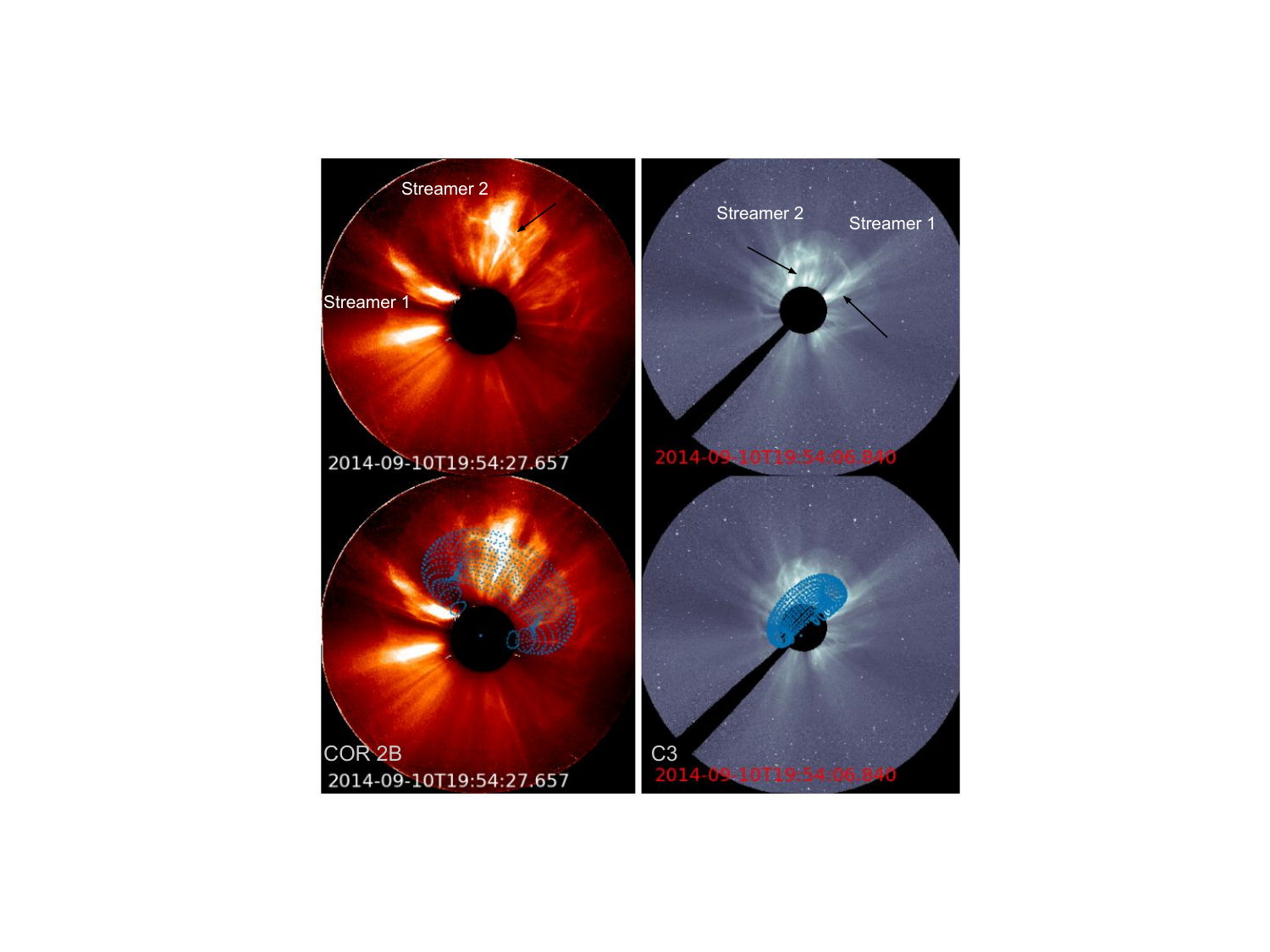}}
    \caption{The white light images observed from COR2-B and C3 and the same images overlaid with FRi3D reconstructed wire frame in the bottom panels. The arrows show different projected white light features used as proxies to reconstruct the 3D structure, as discussed in the text.}
    \label{fig:20140910_recon_fri3d}
\end{figure}

\begin{figure}[ht!]
    \centering
    {\includegraphics[width=0.55\textwidth,trim={4.5cm 3.cm 4.5cm 3.cm},clip=]{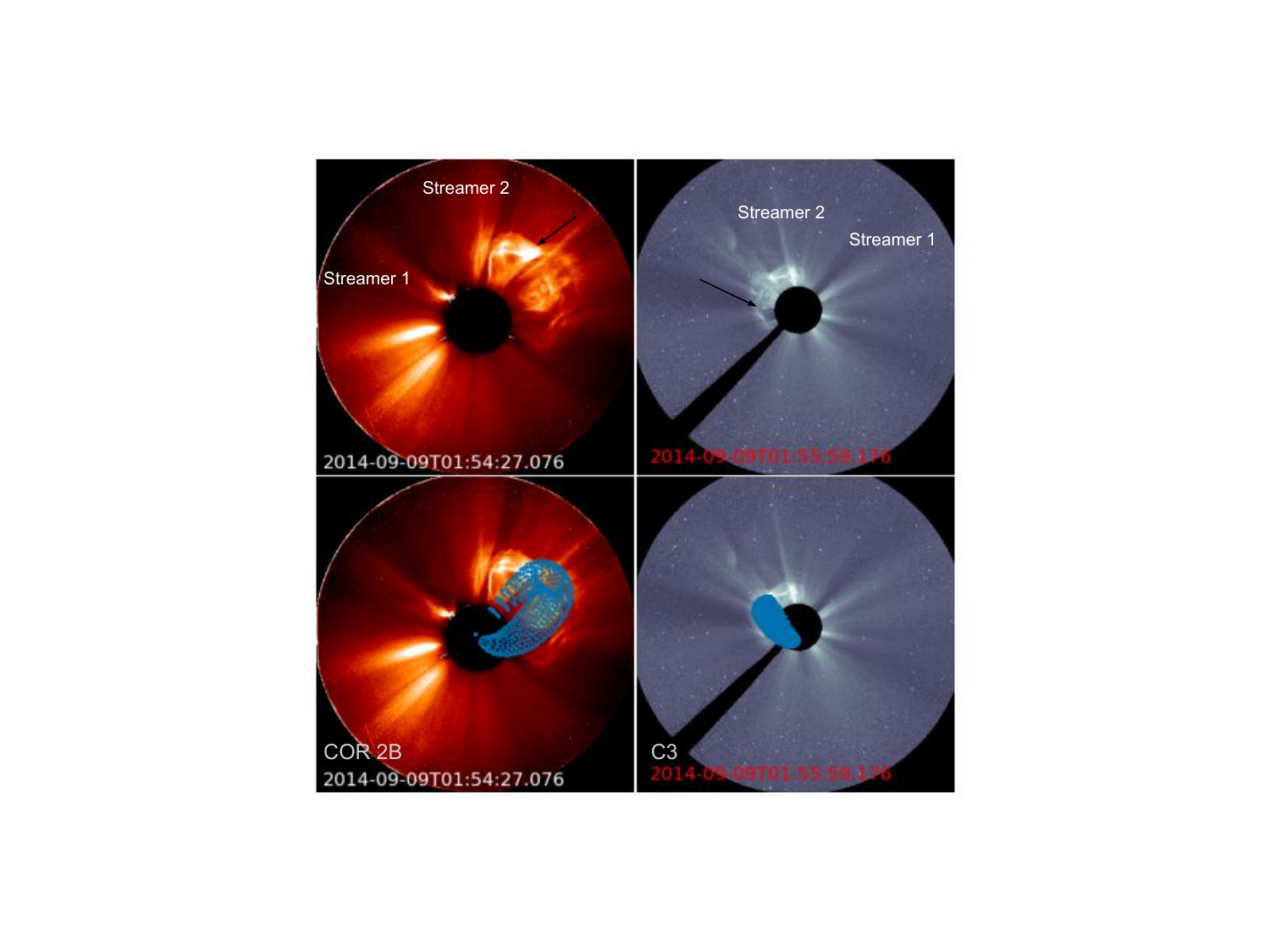}}
    \caption{The white light images observed from COR2-B and C3 and the same images overlaid with FRi3D reconstructed wire frame in the bottom panels. The arrows show different projected white light features used as proxies to reconstruct the 3D structure, as discussed in the text.}
    \label{fig:20140910_recon_fri3d_cme1}
\end{figure}

We also performed the reconstruction of CME1, which erupted more than a day before CME2, with the FRi3D model and the results are presented in Fig.~\ref{fig:20140910_recon_fri3d_cme1}.

\end{document}